\documentclass[onecolumn]{IEEEtran}

\usepackage{amssymb}
\usepackage{amsfonts}
\usepackage{amsmath}
\usepackage{array}
\usepackage{cite}
\usepackage[T1]{fontenc}
\usepackage{graphicx}

\def\hb{\boldsymbol{h}}
\def\ellb{\boldsymbol{\ell}}
\def\kappab{\mathbf{\kappa}}
\def\lambdab{\mathbf{\lambda}}
\def\xb{\boldsymbol{x}}
\def\tb{\boldsymbol{t}}

\def\zb{\boldsymbol{z}}

\def\diff{\, \mathrm{d}}
\newcommand{\abs}[1]{\left\lvert#1\right\rvert}
\newcommand{\indicf}[2]{\mathbb{I}_{#1}\left(#2\right)}
\DeclareMathOperator*{\argmax}{arg\,max}

\input{tcilatex}

\begin{document}

\title{A Bayesian Lower Bound for Parameter Estimation \\
of Poisson Data Including Multiple Changes (extended) }
\author{Lucien~Bacharach, Mohammed~Nabil~El~Korso, Alexandre~Renaux, and
Jean-Yves Tourneret\thanks{%
L. Bacharach and A. Renaux are with Laboratory of Signals and Systems (L2S),
Universit\'{e} Paris-Sud, 91192 Gif-sur Yvette, France.}\thanks{%
M. N. El Korso is with Laboratory Energetics, Mechanics and Electromagnetism
(LEME), Universit\'{e} Paris-Ouest, 91410 Ville d'Avray, France.}\thanks{%
Jean-Yves Tourneret is with the IRIT/INP-ENSEEIHT/T\'{e}SA, University of
Toulouse, 31071 Toulouse, France.}}
\maketitle

\begin{abstract}
This paper derives lower bounds for the mean square errors of parameter
estimators in the case of Poisson distributed data subjected to multiple
abrupt changes. Since both change locations (discrete parameters) and
parameters of the Poisson distribution (continuous parameters) are unknown,
it is appropriate to consider a mixed Cram\'{e}r-Rao/Weiss-Weinstein bound
for which we derive closed-form expressions and illustrate its tightness by
numerical simulations.
\end{abstract}



\section{Introduction}

\label{sect/intro}

Parameter estimation in the context of discrete Poisson time series
submitted to multiple abrupt changes is of practical interest in many
applications, such as the segmentation of multivariate astronomical time
series \cite{Sca98,JSBAAGGSTT05,DTS07}. In this context, the observed data
(distributed according to a Poisson distribution) are subjected to abrupt
changes whose locations are unknown. The values of the Poisson parameters
associated with each interval are also unknown quantities that need to be
estimated. Several strategies have been investigated in the literature \cite%
{Dju94,Lav98,TDL03,PADF02,Fea05}. However, to the best of our knowledge,
lower bounds for the mean square error of the resulting simulators have been
derived only in a few specific cases. For instance, the case of a single
change-point in the observation window was studied in \cite{FT03}. The case
of multiple changes was considered in \cite{LRNM10}. The difficulty of
deriving bounds for the parameters of piece-wise stationary data is mainly
due to the discrete nature of changepoint locations for which classical
bounds such as the Cram\'{e}r-Rao bound (CRB) are not appropriate anymore.
In \cite{FT03,LRNM10}, the authors considered a lower bound for the mean
square error (MSE) that does not require the differentiability of the
log-likelihood function. Specifically, deterministic bounds, such as the
Chapman-Robbins bound, have been derived for a single change-point in \cite%
{FT03}, and then extended to multiple changes in \cite{LRNM10}, with the
strong assumptions that the Poisson parameters are known. On the other hand,
and in order to improve the tightness of the resulting bound, we proposed in 
\cite{BREC15,BREC16} the use of the Weiss-Weinstein bound (WWB), which is
known to be one of the tightest bound in the family of the Bayesian bounds.
Nevertheless, these analyses were limited to the case of known Poisson
parameters both in the single \cite{BREC15} and multiple \cite{BREC16}
changepoint scenarios.

In this paper, we fill this gap by proposing and deriving a new Bayesian
lower bound for the global MSE (GMSE) of the parameters of Poisson
distributed data subjected to multiple changepoints. The proposed bound is a
mixed Cram\'{e}r-Rao (CR)/Weiss-Weinstein (WW) bound adapted to the fact
that some unknown parameters are discrete (the WW part of the mixed bound is
associated with the change locations) and that the other parameters are
continuous (the CR part of the mixed bound is associated with the Poisson
parameters). The idea of combining these two bounds had already been
introduced in \cite{BV06}, under a recursive form. Of course, using a WWB for
both discrete and continuous parameters would be theoretically possible.
However, the WW bound is expressed as the supremum of a set of matrices whose
computation is infeasible in our scenario. Thus, the mixed Cram\'{e}r-Rao/%
Weiss-Weinstein bound is the appropriate alternative, whose computation can be
achieved using a convex optimization procedure to compute this supremum based
on the computation of the minimum volume ellipsoid covering a union of derived
ellipsoids.

\section{Multiple Change-Points in Poisson Time-Series: problem formulation}

\label{sect/pb}

We consider an independent discrete Poisson time series subjected to
multiple changes. The resulting observation vector $\boldsymbol{x}=\left[
x_{1},\ldots ,x_{T}\right]^T $ (of length $T$) is defined as%
\begin{equation}
\begin{cases}
x_{t} \sim \mathcal{P}\left( \lambda _{1}\right) , & \text{for }t=1,\ldots ,
t_{1} \\ 
x_{t} \sim \mathcal{P}\left( \lambda _{2}\right) , & \text{for }t=t_{1}+1,
\ldots ,t_{2} \\ 
\qquad \vdots & \vdots \\ 
x_{t} \sim \mathcal{P}\left( \lambda _{K+1}\right) , & \text{for }t=t_{K}+1,
\ldots ,T,%
\end{cases}
\label{eqn/model}
\end{equation}%
where $\mathcal{P}\left( \lambda _{k}\right) $, for $k=1,\ldots ,K+1$,
denotes the Poisson distribution of parameter $\lambda _{k}$ on the $k$-th
segment, i.e., $\Pr (x_t = \kappa_t ) = \lambda_k^{\kappa_t} \exp\{
-\lambda_k\}/(\kappa_t!)$, $K$ denotes the total number of changes (assumed
to be known), and $t_{k}$ denotes the $k$-th change location, i.e., the
sample point after which the parameter $\lambda _{k}$ of the current segment
switches to $\lambda _{k+1}$. The segmentation problem addressed in this
work consists of i) segmenting the time series $\boldsymbol{x}$, i.e.,
estimating the locations of the changes $t_{k}$, and ii) estimating the
Poisson parameters $\lambda _{k}$ on each segment. The resulting unknown
parameter vector is $\mathbf{\theta }=\bigl[\mathbf{\lambda }^{T},%
\boldsymbol{t}^{T}\bigr] ^{T} $, with $\mathbf{\lambda }\overset{\Delta }{=}%
\left[ \lambda _{1},\ldots ,\lambda _{K+1}\right]^{T} $ and $\boldsymbol{t}%
\overset{\Delta }{=}\left[ t_{1},\ldots ,t_{K}\right]^{T} $. This unknown
parameter vector lies in the parameter space $\Theta =\mathbb{R}%
_{+}^{K+1}\times \left\{1,\ldots ,T\right\} ^{K}$, where $\mathbb{R}_{+}$
denotes the set of real positive numbers. Using a Bayesian framework, we
consider that both vectors $\mathbf{\lambda }$ and $\boldsymbol{t}$ are
assigned a known prior. More precisely, the Poisson parameters $\lambda _{k}$
are assumed to be independent and identically distributed (i.i.d.), and are
assigned the conjugate gamma distributions with parameters $\alpha_k $ and $%
\beta $, leading to the following prior%
\begin{equation}
f\left( \boldsymbol{\lambda }\right) =\prod_{k=1}^{K+1}\frac{\beta
^{\alpha_k }}{\Gamma \left( \alpha_k \right) }\lambda _{k}^{\alpha_k -1}\exp
\left( -\beta \lambda _{k}\right) \mathbb{I}_{\mathbb{R}_{+}}\left( \lambda
_{k}\right)  \label{eqn/prior_lambda}
\end{equation}%
in which $\mathbb{I}_{\mathcal{E}}\left( .\right) $ denotes the indicator
function on the set $\mathcal{E}$, and $\Gamma (.)$ denotes the usual gamma
function, i.e., for $\alpha >0$, $\Gamma (\alpha) = \int_0^{+\infty}
x^{\alpha- 1}\exp(-x)\mathrm{d} x$. On the other hand, we assume that each
change location $t_{k}$, for $k=1,\ldots ,K$, is defined as the following
random walk $t_{k}=t_{k-1}+\varepsilon _{k} $ where $\varepsilon _{k}$ are
i.i.d. variables following a discrete uniform distribution on the set of
integers $\left\{ 1,\ldots ,\tau \right\} $, and $t_0 = 0$. The value of $%
\tau $ is chosen so that the final change $t_{K}$ is at least located before
the last observation, i.e., the maximum possible value for $\tau $ is $\tau
_{\max }=\left\lfloor \left( T-1\right) /K\right\rfloor $, where $%
\left\lfloor .\right\rfloor $ denotes the floor function. Consequently, we
obtain the following prior distribution for the unknown vector $\boldsymbol{t%
}$%
\begin{equation}
\Pr \left( \boldsymbol{t}=\boldsymbol{\ell}\right) =\frac{1}{\tau ^{K}}%
\prod_{k=1}^{K}\mathbb{I}_{\left\{ \ell _{k-1}+1,\ldots ,\ell _{k-1}+\tau
\right\} }\left( \ell _{k}\right)  \label{eqn/prior_t}
\end{equation}%
with $\ell_0 = 0$. Since vectors $\mathbf{\lambda }$ and $\boldsymbol{t}$
are independent, the joint prior for $\mathbf{\lambda}$ and $\boldsymbol{t}$
is expressed as $f\left( \boldsymbol{\lambda },\boldsymbol{t}=\boldsymbol{%
\ell }\right) =f\left( \boldsymbol{\lambda }\right) \Pr \left( \boldsymbol{t}%
=\boldsymbol{\ell}\right) $.

From the model (\ref{eqn/model}), the likelihood of the observations can be
written as%
\begin{equation}
f\left( \boldsymbol{x}=\mathbf{\kappa }|\boldsymbol{\lambda },\boldsymbol{t}=%
\boldsymbol{\ell}\right) =\prod_{k=1}^{K+1}\prod_{t=\ell _{k-1}+1}^{\ell
_{k}}\frac{\lambda _{k}^{\kappa _{t}}}{\kappa _{t}!}\exp \left\{ -\lambda
_{k}\right\} .  \label{eqn/likelihood}
\end{equation}

The aim of the present paper is to study the estimation performance of the
vector $\mathbf{\theta }$ by deriving a lower bound on the mean square error
(MSE) of any Bayesian estimator $\hat{\mathbf{\theta}}(\boldsymbol{x} )$ of $%
\mathbf{\theta }$. Both subvectors $\mathbf{\lambda}$ and $\boldsymbol{t}$
of $\mathbf{\theta}$ have to be estimated simultaneously. However, as
already mentioned in Section \ref{sect/intro}, the Cram\'{e}r-Rao bound is
not suited for changepoint analysis, since $\boldsymbol{\ell}$ is a vector
of discrete parameters. Thus, the idea is to use two different lower bounds
w.r.t. each subvector of $\mathbf{\theta }$, resulting in a ``mixed''
Bayesian bound that corresponds to the Bayesian CR bound for the first
subvector $\mathbf{\lambda}$ of $\mathbf{\theta }$, and that corresponds to
the so-called WW bound for the second subvector $\boldsymbol{t}$ of $\mathbf{%
\theta}$. As already mentioned, the use of such a combined CR/WW lower bound
was initiated in \cite{BV06} in a target tracking context. Since our framework
is different, the next section is devoted to the presentation of this bound,
which we will refer to as the ``Bayesian Cram\'{e}r-Rao/Weiss-Weinstein
bound'' (BCRWWB). It is in fact a special case of a general family of lower
bounds exposed in \cite{WW88}.

\section{Bayesian Cram\'{e}r-Rao/Weiss-Weinstein bound}
\label{sect/theory}

We are interested in studying the estimation performance of a
parameter vector $\mathbf{\theta }$ that lies in a parameter space
$\Theta = \mathbb{R}^{K+1}\times \mathbb{N}^K$. As explained in the previous
section, this parameter vector can be split into two subvectors,
$\mathbf{\lambda }\in\Theta _{\mathbf{\lambda }}= \mathbb{R}_{+}^{K+1}$
and $\boldsymbol{t}\in \Theta _{\boldsymbol{t}}= \mathbb{N}^{K}$,
so that $\mathbf{\theta }=%
\bigl[ 
\mathbf{\lambda }^{T},\boldsymbol{t}^{T}\bigr]^{T}$ and $\Theta =\Theta _{%
\mathbf{\lambda }}\times \Theta _{\boldsymbol{t}}$. 
From a set of observations $\boldsymbol{x}\in \Omega $, the vector $\mathbf{%
\theta }$ can be estimated by using any Bayesian estimation scheme, leading
to an estimator $\hat{\mathbf{\theta}}\left( \boldsymbol{x}\right) =\bigl[ 
\hat{\mathbf{\lambda}}\left( \boldsymbol{x}\right) ^{T},\hat{\boldsymbol{t}}%
\left( \boldsymbol{x}\right) ^{T}\bigr] ^{T}$. Let us recall that we aim at
obtaining a lower bound on the \emph{global} mean square error (GMSE) of
this estimator, which corresponds to the Bayesian CR bound w.r.t. $\mathbf{%
\lambda}$, and which corresponds to the WW bound w.r.t. $\boldsymbol{t}$.
The GMSE of $\hat{\mathbf{\theta}}\left( \boldsymbol{x}\right)$ is
defined as the $(2K+1)\times (2K+1)$ matrix%
\begin{equation}
\mathbf{GMSE}\bigl( \mathbf{\hat{\theta}}\bigr) =\mathbb{E}_{\boldsymbol{x},%
\boldsymbol{\theta }}\left\{ \bigl[ \boldsymbol{\theta }-\boldsymbol{\hat{%
\theta}}\left( \boldsymbol{x}\right) \bigr] \bigl[ \boldsymbol{\theta }-%
\boldsymbol{\hat{\theta}}\left( \boldsymbol{x}\right) \bigr] ^{T}\right\}
\label{eqn/def_GMSE}
\end{equation}%
in which $\mathbb{E}_{\boldsymbol{x},\boldsymbol{\theta }}\left\{ .\right\} $
denotes the expectation operation w.r.t. the joint distribution $f\left( 
\boldsymbol{x},\boldsymbol{\theta }\right) $ which depends on both the
observations and the parameters. Based on \cite{WW88}, by appropriately
choosing some real-valued measurable functions $\psi _{k}\left( \boldsymbol{x%
},\boldsymbol{\theta }\right) $, $k=1,\ldots ,2K+1$, defined on $\Omega
\times \Theta $ such that the following integrals exist and satisfy $%
\int_{\Theta }\!\psi _{k}\left( \boldsymbol{x},\boldsymbol{\theta }\right)
f\left( \boldsymbol{x},\boldsymbol{\theta }\right) \mathrm{d}\boldsymbol{%
\theta }=0$ for almost every (a.e.) $\boldsymbol{x} \in \Omega$ and for $%
k=1,\ldots , 2K+1$, the following matrix inequality holds%
\begin{equation}
\mathbf{GMSE}\bigl( \boldsymbol{\hat{\theta}}\bigr) \succeq \boldsymbol{VP}%
^{-1}\boldsymbol{V}^{T}  \label{eqn/general_matrix_ineq}
\end{equation}%
in which $\boldsymbol{V}$ is a $(2K+1)\times (2K+1)$ matrix whose elements
are given by%
\begin{equation}
\left[ \boldsymbol{V}\right] _{k,l}=\mathbb{E}_{\boldsymbol{x},\boldsymbol{%
\theta }}\left\{ \theta _{k}\psi _{l}\left( \boldsymbol{x},\boldsymbol{%
\theta }\right) \right\}  \label{eqn/Vmn_general}
\end{equation}%
and $\boldsymbol{P}$ is a $(2K+1)\times (2K+1)$ symmetric matrix, whose
elements are given by%
\begin{equation}
\left[ \boldsymbol{P}\right] _{k,l}=\mathbb{E}_{\boldsymbol{x},\boldsymbol{%
\theta }}\left\{ \psi _{k}\left( \boldsymbol{x},\boldsymbol{\theta }\right)
\psi _{l}\left( \boldsymbol{x},\boldsymbol{\theta }\right) \right\} \text{.}
\label{eqn/Pmn_general}
\end{equation}%
Note that the matrix inequality (\ref{eqn/general_matrix_ineq}) means that
the difference between its left and its right hand sides is a nonnegative
definite matrix. One key point in the theory developed in \cite{WW88} is the
choice of the measurable functions $\psi _{k}$. For $k$ restricted to $%
\left\{ 1,\ldots ,K+1\right\} $ (continuous Poisson parameters), we define
these functions as for the CR bound, i.e.,%
\begin{equation}
\psi _{k}\left( \boldsymbol{x},\boldsymbol{\theta }\right) =%
\begin{cases}
\frac{\partial \ln f\left( \boldsymbol{x},\boldsymbol{\theta }\right) }{%
\partial \lambda_{k}}, & \text{if }\boldsymbol{\theta }\in \Theta ^{\prime }
\\ 
0, & \text{if }\boldsymbol{\theta }\notin \Theta ^{\prime }%
\end{cases}
\label{eqn/psim_CR}
\end{equation}%
where $\Theta ^{\prime }=\left\{ \boldsymbol{\theta }\in \Theta :f\left( 
\boldsymbol{x},\boldsymbol{\theta }\right) >0\text{ a.e. }\boldsymbol{x}\in
\Omega \right\} $. Conversely, for $k$ restricted to $\left\{
1,\ldots,K\right\} $ (changepoint locations), we define these measurable
functions as for the WW bound, i.e.,%
\begin{equation}
\psi _{K+1+k}\left( \boldsymbol{x},\boldsymbol{\theta }\right) =\sqrt{\frac{%
f\left( \boldsymbol{x},\boldsymbol{\theta }+\boldsymbol{h}_{k}\right) }{%
f\left( \boldsymbol{x},\boldsymbol{\theta }\right) }}-\sqrt{\frac{f\left( 
\boldsymbol{x},\boldsymbol{\theta }-\boldsymbol{h}_{k}\right) }{f\left( 
\boldsymbol{x},\boldsymbol{\theta }\right) }}  \label{eqn/psim_WW}
\end{equation}%
where $\boldsymbol{h}_{k}$ is any vector of size $2K+1$ of the form $%
\boldsymbol{h}_{k}=\bigl[ \mathbf{0}_{K+1}^{T},$ $\mathbf{0}_{k-1}^{T},h_{k},%
\mathbf{0}_{K-k}^{T}\bigr] ^{T}$, for $k=1,\ldots ,K$, in which $\boldsymbol{%
0}_{k}$ denotes the zero vector of length $k$. Note that the value of $h_k$
can be arbitrarily chosen by the user as far as it allows the invertibility
of $\boldsymbol{P}$.

The next step in our analysis is to derive the matrix V. Denote as $%
\boldsymbol{V}_{22}$ the $K\times K$ diagonal matrix whose elements are, for
any $k \in \left\{ 1,\ldots ,K\right\}$%
\begin{equation}
\left[ \boldsymbol{V}_{22}\right] _{k,k}=-h_{k}\mathbb{E}_{\boldsymbol{x},%
\boldsymbol{\theta }}\left\{ \sqrt{\frac{f\left( \boldsymbol{x},\boldsymbol{%
\theta }+\boldsymbol{h}_{k}\right) }{f\left( \boldsymbol{x},\boldsymbol{%
\theta }\right) }}\right\} \text{.}  \label{eqn/V22_mm}
\end{equation}%
Substituting (\ref{eqn/psim_CR}) and (\ref{eqn/psim_WW}) into (\ref%
{eqn/Vmn_general}), we obtain%
\begin{equation}
\boldsymbol{V}= 
\begin{bmatrix}
- \boldsymbol{I}_{K+1} & \boldsymbol{0}_{(K+1)\times K} \\ 
\boldsymbol{0}_{K\times (K+1)} & \boldsymbol{V}_{22}%
\end{bmatrix}
\label{eqn/Vblocks}
\end{equation}%
where $\boldsymbol{I}_{K+1}$ denotes the $(K+1)\times (K+1)$ identity matrix
and $\boldsymbol{0}_{(K+1)\times K}$ is the $(K+1)\times K$ zero matrix,
provided the following conditions are satisfied

\begin{enumerate}
\item $f\left( \boldsymbol{x},\boldsymbol{\theta }\right) $ is absolutely
continuous w.r.t. $\lambda _{k}$, $k=1,\ldots ,K+1$, a.e. $x\in \Omega $;

\item $\lim_{\lambda _{k}\rightarrow 0 }\lambda _{k}f\left( \boldsymbol{x},%
\boldsymbol{\theta }\right) =\lim_{\lambda _{k}\rightarrow +\infty
}\lambda_{k}f\left( \boldsymbol{x},\boldsymbol{\theta }\right) =0$, $%
k=1,\ldots ,K+1$, a.e. $x\in \Omega $.
\end{enumerate}

Note that these two conditions correspond to the necessary and usual
regularity conditions for the derivation of the Bayesian CR bound.

Similarly, by plugging (\ref{eqn/psim_CR}) and (\ref{eqn/psim_WW}) into (\ref%
{eqn/Pmn_general}), we obtain the expression of the matrix $\boldsymbol{P}$,
which can be split into four blocks as follows%
\begin{equation}
\boldsymbol{P}=%
\begin{bmatrix}
\boldsymbol{P}_{11} & \boldsymbol{P}_{12} \\ 
\boldsymbol{P}_{12}^{T} & \boldsymbol{P}_{22}%
\end{bmatrix}
\label{eqn/Pblocks}
\end{equation}%
in which $\boldsymbol{P}_{11}$ is the $(K+1)\times (K+1)$ matrix whose
elements are%
\begin{equation}
\left[ \boldsymbol{P}_{11}\right] _{k,l}=\mathbb{E}_{\boldsymbol{x},%
\boldsymbol{\theta }}\left\{ \frac{\partial \ln f\left( \boldsymbol{x},%
\boldsymbol{\theta }\right) }{\partial \lambda _{k}}\frac{\partial \ln
f\left( \boldsymbol{x},\boldsymbol{\theta }\right) }{\partial \lambda _{l}}%
\right\}  \label{eqn/P11_mn}
\end{equation}%
$\boldsymbol{P}_{12}$ is the $(K+1)\times K$ matrix whose elements are%
\begin{equation}
\hspace{-1em}\left[ \boldsymbol{P}_{12}\right] _{k,l}=\mathbb{E}_{%
\boldsymbol{x},\boldsymbol{\theta }}\left\{ \frac{\partial \ln f\left( 
\boldsymbol{x},\boldsymbol{\theta }\right) }{\partial \lambda _{k}}\left( 
\sqrt{\frac{f\left( \boldsymbol{x},\boldsymbol{\theta }+\boldsymbol{h}%
_{l}\right) }{f\left( \boldsymbol{x},\boldsymbol{\theta }\right) }}-\sqrt{%
\frac{f\left( \boldsymbol{x},\boldsymbol{\theta }-\boldsymbol{h}_{l}\right) 
}{f\left( \boldsymbol{x},\boldsymbol{\theta }\right) }}\right) \right\}
\label{eqn/P12_mn}
\end{equation}%
and finally $\boldsymbol{P}_{22}$ is the $K\times K$ matrix whose elements
are%
\begin{equation}
\hspace{-1em}\left[ \boldsymbol{P}_{22}\right] _{k,l}\!=\mathbb{E}_{%
\boldsymbol{x},\boldsymbol{\theta }}\left\{ \left( \sqrt{\frac{f\left( 
\boldsymbol{x},\boldsymbol{\theta }+\boldsymbol{h}_{k}\right) }{f\left( 
\boldsymbol{x},\boldsymbol{\theta }\right) }}-\sqrt{\frac{f\left( 
\boldsymbol{x},\boldsymbol{\theta }-\boldsymbol{h}_{k}\right) }{f\left( 
\boldsymbol{x},\boldsymbol{\theta }\right) }}\right) \left( \sqrt{\frac{%
f\left( \boldsymbol{x},\boldsymbol{\theta }+\boldsymbol{h}_{l}\right) }{%
f\left( \boldsymbol{x},\boldsymbol{\theta }\right) }}-\sqrt{\frac{f\left( 
\boldsymbol{x},\boldsymbol{\theta }-\boldsymbol{h}_{l}\right) }{f\left( 
\boldsymbol{x},\boldsymbol{\theta }\right) }}\right) \right\} \text{.}
\label{eqn/P22_mn}
\end{equation}

Note that the same bound can be obtained by defining all the functions $%
\psi_k$ for $k\in \left\{ 1,\ldots ,2K+1\right\}$ as in (\ref{eqn/psim_WW}),
with $\boldsymbol{h}_k = \bigl[ \mathbf{0}_{k-1}^T,h_k, \mathbf{0}_{2K+1-k}^T%
\bigr]^T$, by letting $h_k$ tend to $0$ for $k = 1,\ldots , K+1$, and by
using a Taylor expansion. Finally, the tightest lower bound is obtained by
maximizing the right hand side of (\ref{eqn/general_matrix_ineq}) w.r.t. $%
h_1,\ldots ,h_{K}$.

\section{Application to multiple change-points in Poisson Time-Series}

\label{sect/bound_results}

This section presents the main results about the derivation of the lower
bound presented in Section \ref{sect/theory} for the problem formulated in
Section \ref{sect/pb}. The full calculation details are provided in the
appendices. The joint distribution of the observation and parameter vectors
can be expressed as 
\begin{equation}
f\left( \boldsymbol{x}=\mathbf{\kappa },\mathbf{\theta }\right) =f\left( 
\boldsymbol{x}=\mathbf{\kappa }|\mathbf{\lambda },\boldsymbol{t}=\boldsymbol{%
\ell }\right) f\left( \mathbf{\lambda },\boldsymbol{t}=\boldsymbol{\ell }%
\right) .  \label{eqn/joint_law}
\end{equation}%
After plugging (\ref{eqn/prior_lambda}), (\ref{eqn/prior_t}) and (\ref%
{eqn/likelihood}) into (\ref{eqn/joint_law}), we can deduce the expressions
of $f\left( \boldsymbol{x},\mathbf{\theta }+\boldsymbol{h}_{k}\right) $ for $%
k=1,\ldots ,K$, and $\partial f\left( \boldsymbol{x},\mathbf{\theta }\right)
/\partial \lambda _{k}$ for $k=1,\ldots ,K+1$. Let us first introduce some
useful notations for the following mathematical functions. We first define
the function $\varphi _{h_{k}}(\boldsymbol{y})$ of the vector $\boldsymbol{y}%
=[y_{1},y_{2}]^{T}\in \mathbb{R}_{+}^{2}$ as 
\begin{equation}
\varphi _{h_{k}}\left( \boldsymbol{y}\right) =y_{1}^{\alpha
_{k}-1}y_{2}^{\alpha _{k+1}-1}\exp \Biggl\{{-}\beta \left(
y_{1}+y_{2}\right) -\left\vert h_k\right\vert \frac{\left( \sqrt{y_{2}}-\sqrt{%
y_{1}}\right) ^{2}}{2}\Biggr\}\mspace{-18mu}  \label{eqn/def_varphi}
\end{equation}%
and the following integral as 
\begin{equation}
\Phi \left( h_{k}\right) =\frac{\beta ^{\alpha _{k}+\alpha _{k+1}}}{\Gamma
\left( \alpha _{k}\right) \Gamma \left( \alpha _{k+1}\right) }\int_{\mathbb{R%
}_{+}^{2}}\varphi _{h_{k}}\left( \boldsymbol{y}\right) \mathrm{d}\boldsymbol{%
y}.  \label{eqn/def_Phi}
\end{equation}%
We also define the function $\phi _{h_{k},h_{k+1}}(\boldsymbol{z})$ of the
vector $\boldsymbol{z}=[z_{1},z_{2},z_{3}]^{T}\in \mathbb{R}_{+}^{3}$ as the
trivariate version of $\varphi _{h_{k}}$, i.e., 
\begin{equation}
\phi _{h_{k},h_{k+1}}\left( \boldsymbol{z}\right) =z_{1}^{\alpha
_{k}-1}z_{2}^{\alpha _{k+1}-1}z_{3}^{\alpha _{k+2}-1}\exp \Biggl\{-\beta
\left( z_{1}+z_{2}+z_{3}\right) -\left\vert h_{k}\right\vert \frac{\left( 
\sqrt{z_{2}}-\sqrt{z_{1}}\right) ^{2}}{2}-\left\vert h_{k+1}\right\vert 
\frac{\left( \sqrt{z_{3}}-\sqrt{z_{2}}\right) ^{2}}{2}\Biggr\}.\mspace{-18mu}
\label{eqn/def_phi}
\end{equation}%
We finally define the three functions $u$, $v$ and $w$ as follows 
\begin{gather}
u\left( \tau ,h_{k}\right) =%
\begin{cases}
\multicolumn{1}{c@{}}{\frac{(\tau -\left\vert h_{k}\right\vert )^{2}}{\tau
^{2}}}&\text{ if } k\leq K-1 \text{ and } \abs{h_k}\leq \tau  \\ 
\multicolumn{1}{c@{}}{\frac{\tau -\left\vert h_{K}\right\vert }{\tau }
}& \text{ if } k = K \text{ and } \abs{h_K}\leq \tau  \\ 
\multicolumn{1}{c@{}}{0}&\text{ if } \abs{h_k} > \tau %
\end{cases}
\label{eqn/def_u} \\
v\left( \tau ,h_{k},h_{k+1}\right) =%
\begin{cases}
\frac{\left( \tau -\left\vert h_{k}\right\vert \right) (\tau -|h_{k+1}|)}{%
\tau ^{3}}\text{ if }k\leq K-1 \\ 
\multicolumn{1}{r@{}}{\text{and }\max \left( \left\vert h_{k}\right\vert
,\left\vert h_{k+1}\right\vert \right) \leq \tau \text{;}} \\ 
\multicolumn{1}{r@{}}{\frac{\tau -\left\vert h_{K}\right\vert }{\tau ^{2}}%
\quad \text{ if }k=K\text{ and }\left\vert h_{K}\right\vert \leq \tau \text{;%
}} \\ 
\multicolumn{1}{r@{}}{0\hspace{1.5em}\text{ if }\max \left( \left\vert
h_{k}\right\vert ,\lvert h_{k+1}\rvert \right) >\tau \text{,}}%
\end{cases}
\label{eqn/def_v} \\
\begin{split}
w\left( \boldsymbol{z},\tau ,h_{k},h_{k+1}\right)
= 2\max \left( \tau -\lvert h_{k}\rvert -\lvert h_{k+1}\rvert ,0\right)
& -\max \bigl(\tau -\max \left( \lvert h_{k}\rvert ,\lvert
h_{k+1}\rvert \right) ,0\bigr) \\
& -\max \left( \tau -\lvert h_{k}\rvert -\lvert h_{k+1}\rvert
+1,0\right)
+\dfrac{1-r^{1-\min (\lvert h_{k}\rvert ,\lvert h_{k+1}\rvert
)}\left( \boldsymbol{z}\right) }{1-r\left( \boldsymbol{z}\right) }
\label{eqn/def_w}
\end{split}
\end{gather}%
in which $r\left( \boldsymbol{z}\right) =\exp \left\{ -z_{2}+\sqrt{z_{1}z_{2}%
}+\sqrt{z_{2}z_{3}}-\sqrt{z_{1}z_{3}}\right\} $. Using these functions, we
now give the expressions of the matrix blocks composing $\boldsymbol{V}$ and 
$\boldsymbol{P}$, i.e., $\boldsymbol{V}_{22}$, $\boldsymbol{P}_{11}$, $%
\boldsymbol{P}_{12}$, $\boldsymbol{P}_{21}$ and $\boldsymbol{P}_{22}$ which
were introduced in Section \ref{sect/theory}.


After plugging (\ref{eqn/joint_law}) into (\ref{eqn/V22_mm}) and computing
the expectations, we obtain, for $k=1,\ldots ,K$ 
\begin{equation}
[\boldsymbol{V}_{22}]_{k,k} = -h_{k} \: u\left( \tau ,h_{k}\right) \Phi
\left( h_{k} \right) .  \label{eqn/def_V22kk}
\end{equation}


The expression of $\boldsymbol{P}_{11}$ is obtained by substituting (\ref%
{eqn/joint_law}) into (\ref{eqn/P11_mn}), which leads to a diagonal matrix
whose elements have the following form (for $\alpha_k > 2$) 
\begin{equation}
[\boldsymbol{P}_{11}]_{k,k}=\left(\frac{\beta \left( \tau +1\right) }{%
2\left( \alpha_k -1\right) }+\frac{\beta^2}{\alpha_k - 2}\right).
\label{eqn/P11kk}
\end{equation}
Similarly, the expressions of $\boldsymbol{P}_{12}$ (of size $(K+1)\times K$%
) and $\boldsymbol{P}_{21}$ (of size $K\times (K+1)$) can be obtained after
plugging (\ref{eqn/joint_law}) into (\ref{eqn/P12_mn}), which leads to 
\begin{equation}
\boldsymbol{P}_{12}=\boldsymbol{P}_{21}^{T}= 
\begin{bmatrix}
A_{1,1} & 0       & \cdots & 0        \\ 
A_{2,1} & A_{2,2} & \ddots & \vdots   \\ 
0       & A_{3,2} & \ddots & 0        \\ 
\vdots  & \ddots  & \ddots & A_{K,K}  \\ 
0       & \cdots  & 0      & A_{K+1,K}
\end{bmatrix}%
\label{eqn/P12}
\end{equation}%
where, for $k=1,\ldots ,K$, and $l=k$ or $l=k-1$,%
\begin{equation}
A_{k,l}
= \pm h_{l} \,
  u\left( \tau ,h_{l}\right)
  \frac{\beta^{\alpha_{l} +\alpha_{l+1}}}
       {\Gamma (\alpha_{l}) \Gamma (\alpha_{l+1})}
  \int_{\mathbb{R}_{+}^{2}}
  \Biggl(
  \sqrt{\left(\frac{y_1}{y_2}\right)^{\pm 1}}
  -1
  \Biggr)
  \varphi_{h_{l}} \left( \boldsymbol{y}\right)
  \diff\boldsymbol{y}  \label{eqn/def_Akl}
\end{equation}
where both ``$\pm$'' signs are ``$+$'' signs if $l=k-1$, and they are
``$-$'' signs if $l=k$.


Finally, by substituting (\ref{eqn/joint_law}) into (\ref{eqn/P22_mn}), the
matrix $\boldsymbol{P}_{22}$ can be written as the following symmetric
tridiagonal matrix 
\begin{equation}
\boldsymbol{P}_{22}= 
\begin{bmatrix}
B_{1}  & C_{1}  & 0      & \cdots  & 0       \\ 
C_{1}  & B_{2}  & C_{2}  & \ddots  & \vdots  \\ 
0      & C_{2}  & B_{3}  & \ddots  & 0       \\ 
\vdots & \ddots & \ddots & \ddots  & C_{K-1} \\ 
0      & \cdots & 0      & C_{K-1} & B_{K}
\end{bmatrix}%
\label{eqn/P22_final}
\end{equation}%
where, for $k=1,\ldots ,K$,
\begin{equation}
B_{k}=2\bigl( u(\tau ,h_{k})
      -u\left( \tau ,2h_{k}\right)\Phi \left( 2h_{k}\right) \bigr)
\label{eqn/def_Bk}
\end{equation}
and, for $k=1,\ldots ,K-1$,
\begin{equation}
C_{k}=v\left( \tau ,h_{k},h_{k+1}\right) \frac{\beta ^{\alpha_{k}+
\alpha_{k+1}+\alpha_{k+2} }}{\Gamma \left( \alpha_{k} \right)\Gamma \left(
\alpha_{k+1} \right) \Gamma \left( \alpha_{k+2} \right) } {} \int_{\mathbb{R}%
_{+}^{3}} \phi _{h_k,h_{k+1} }\left( \boldsymbol{z}\right) w(\boldsymbol{z}%
,\tau ,h_k,h_{k+1}) \mathrm{d}\boldsymbol{z}.  \label{eqn/def_Ck}
\end{equation}

\subsection{Practical computation of the bound}

The lower bound given by the right-hand side of (\ref%
{eqn/general_matrix_ineq}), that we will denote by $\boldsymbol{R} $ can be
computed using the previous formulas, i.e., from (\ref{eqn/def_V22kk}) to (%
\ref{eqn/def_Ck}). It can be noticed that some integrals (in (\ref%
{eqn/def_V22kk}), (\ref{eqn/def_Akl}) and (\ref{eqn/def_Ck})), do not have
any closed-form expression requiring some numerical scheme for their
computation. In this paper, we have used the adaptive quadrature method \cite%
{Sha08} that proved efficient for our computations.

In addition, we would like to stress that, even if it does not appear
explicitly with the adopted notations, the matrix $\boldsymbol{R} $ actually
depends on the parameters $\alpha_1,\ldots,\alpha_{K+1},\beta,\tau,h_1,%
\ldots,h_{K}$ (only the dependency on $h_1,\ldots,h_K$ has been mentioned
from (\ref{eqn/def_varphi}) to (\ref{eqn/def_w})). Since each vector $%
\boldsymbol{h}=(h_1,\ldots,h_K)$ leads to a lower bound $\boldsymbol{R} (%
\boldsymbol{h})$, one obtains a finite set of lower bounds $\mathcal{W} = \{ 
\boldsymbol{R} (\boldsymbol{h} ) \mid \boldsymbol{h} \in \mathcal{H}\}$, in
which $\mathcal{H}$ is the set of all possible values of $\boldsymbol{h}$.
As already mentioned, the proper Cram\'{e}r-Rao/Weiss-Weinstein lower bound
is the tightest value of $\boldsymbol{R} (\boldsymbol{h})$, namely the
supremum of $\mathcal{W}$, that we denote by $\boldsymbol{B}=\sup (\mathcal{W%
}) = \sup_{h_1,\ldots,h_K} \boldsymbol{R}(\boldsymbol{h})$. The supremum
operation has to be taken w.r.t. the Loewner partial ordering, denoted by ``$%
\preceq$'' \cite{BL00}. This ordering implies that a unique supremum in the
finite set $\mathcal{W}$ might not exist. However, it is possible to
approximate this supremum by computing a minimal upper-bound $\boldsymbol{B}%
^{*}$ of the set $\mathcal{W}$: this bound is such that, for all $%
\boldsymbol{h} \in \mathcal{H}$, $\boldsymbol{B}^{*}\succeq \boldsymbol{R}(%
\boldsymbol{h})$, and there is no smaller matrix $\boldsymbol{B}^{\prime}
\preceq \boldsymbol{B}^{*}$ that also verifies $\boldsymbol{B}%
^{\prime}\succeq \boldsymbol{R}(\boldsymbol{h})$, $\forall \boldsymbol{h}
\in \mathcal{H}$. It has been shown in \cite{BV04,LRNM10} that finding $%
\boldsymbol{B}^{*}$ is equivalent to finding the minimum volume
hyper-ellipsoid $\varepsilon(\boldsymbol{B}^{*}) = \{\boldsymbol{x}\in%
\mathbb{R}^K\mid \boldsymbol{x}^T\boldsymbol{B}^{*}\boldsymbol{x} \leq 1\}$
that covers the union of hyper-ellipsoids $\varepsilon\bigl(\boldsymbol{R}(%
\boldsymbol{h})\bigr) = \{\boldsymbol{x}\in\mathbb{R}^K\mid \boldsymbol{x}^T%
\boldsymbol{R}(\boldsymbol{h})\boldsymbol{x} \leq 1\}$. The search of this
ellipsoid can actually be formulated as the following convex optimization
problem \cite{BV04}: 
\begin{align}
& \text{minimize} & & \log \left( \det \bigl( \boldsymbol{B}^{1/2} \bigr)%
\right)  \label{eqn/opt_pb} \\
& \text{subject to} & & 
\begin{cases}
\: b_1 \geq 0, b_2\geq 0,\ldots ,b_{N_{\boldsymbol{h}}}\geq 0, \\ 
\begin{bmatrix}
\boldsymbol{B}^{-1}-b_n \bigl(\boldsymbol{R}(\boldsymbol{h})_n\bigr)^{-1} & 
\mathbf{0}_{(2K+1) \times 1} \\ 
\mathbf{0}_{1\times (2K+1)} & b_n -1%
\end{bmatrix}
\preceq \mathbf{0}_{2K+2} \\ 
\hfill (n = 1,\ldots ,N_{\boldsymbol{h}})%
\end{cases}
\notag
\end{align}
in which $N_{\boldsymbol{h}}$ denotes the number of elements of the set $%
\mathcal{H}$, and $\boldsymbol{R}(\boldsymbol{h})_n \in \mathcal{W}$ is an
indexed version of $\boldsymbol{R}(\boldsymbol{h})$ (i.e., when $n$ varies
from $1$ to $N_{\boldsymbol{h}}$, $\boldsymbol{h}$ runs through all the
possible combinations of $h_1,\ldots ,h_K$, and $\boldsymbol{R}(\boldsymbol{h%
})_n$ runs through all the elements of $\mathcal{W}$). The problem (\ref%
{eqn/opt_pb}) can be solved efficiently using a semidefinite programming
tool, such as the one provided in the CVX package \cite{GB08}.

\section{Numerical results}

\begin{figure}[!t]
\centering
\includegraphics[width=0.5%
\columnwidth]{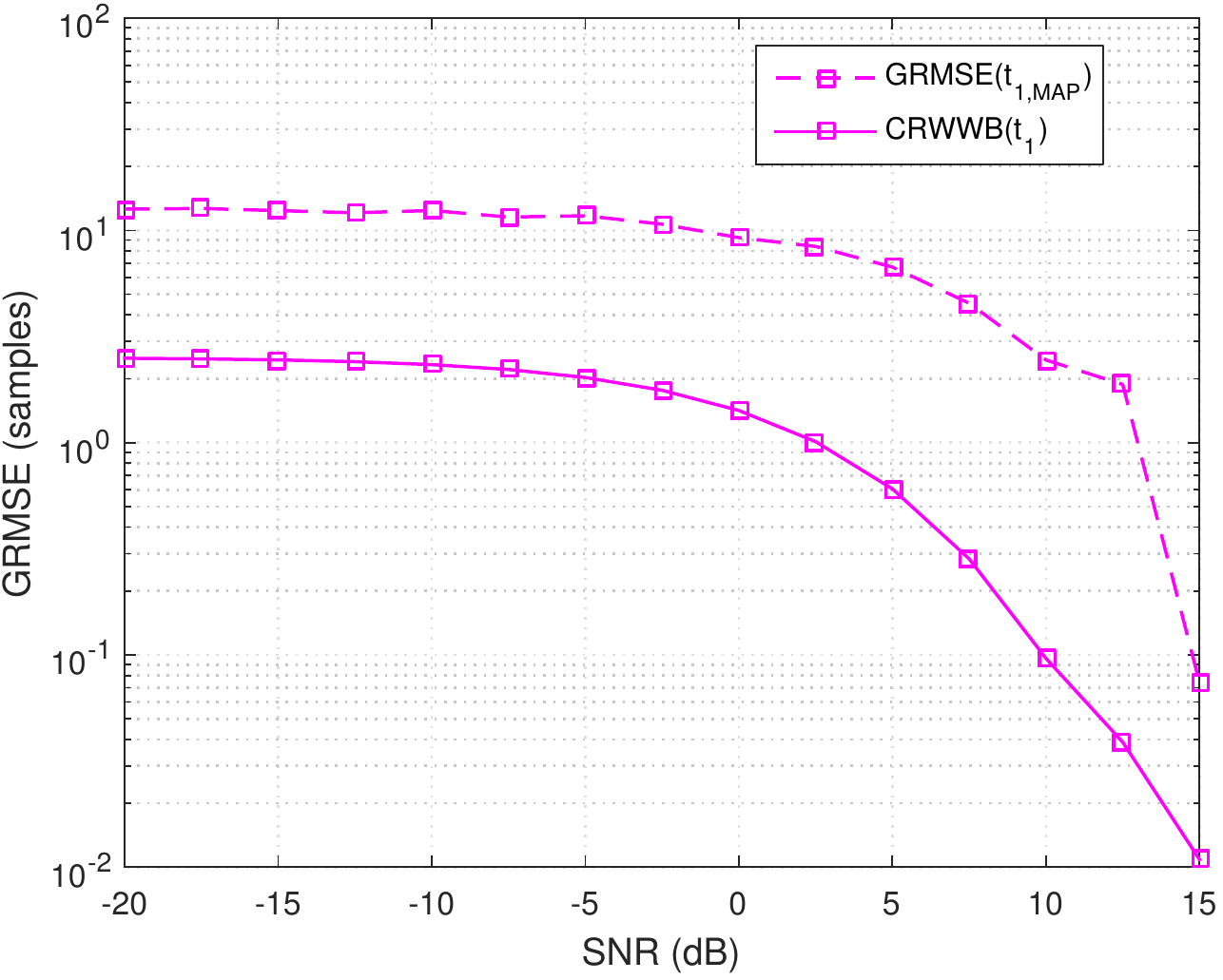}
\caption{Estimated GRMSE and proposed lower bound w.r.t. the change-point $%
t_1$, versus SNR, with $T=80$ snapshots and $K=1$ change in the mean rate of
a Poisson time series.}
\label{fig/sim1}
\end{figure}

This section analyzes the evolution of the proposed bound as a function of a
parameter that is classically used for changepoint estimation performance.
This parameter is either referred to as ``amount of change'' \cite{FJK10},
``magnitude of change'' or ``signal-to-noise ratio'' (SNR): 
in \cite{FT03,LRNM10,BREC15}, for Poisson distributed data, the SNR is
defined as $\nu = (\lambda_{k+1}-\lambda_k)^2/\lambda_k^2 $, for $k =
1,\ldots ,K+1$. In our context, since each $\lambda_k$ is a random variable
with a gamma distribution of parameters $\alpha_k$ and $\beta$ (as stated in
(\ref{eqn/prior_lambda})), this leads to a lower bound $\boldsymbol{B}$ that
does not depend on $\lambda_1,\ldots ,\lambda_{K+1}$, and \textit{a fortiori}
on $\nu$. However, the bound depends upon the parameters $\alpha_k$ and $%
\beta$, which can then be used to drive the average $(\lambda_k)_{_{\mathsf{%
mean}}} = \alpha_k/\beta$ generated by the Gamma prior. Thus, by
substituting $(\lambda_k)_{_{\mathsf{mean}}}$ with $\lambda_k$ in the
definition of $\nu$, we obtain $\bar{\nu} =
(\alpha_{k+1}-\alpha_k)^2/\alpha_k^2$. Such a definition implies that the
higher $\bar{\nu}$, the higher the amount of change between two consecutive
segments, \emph{on average}.

In this study, we present some simulation results obtained for $T=80$
observations, $K=1$ change, and with $\bar{\nu}$ ranging from $-20$~dB to $%
15 $~dB. Such a choice for the value of $K$ is justified by the fact that it
yields less complex expressions of the estimators given in (\ref%
{eqn/MAP_lambda}) and (\ref{eqn/MAP_t}). We chose $\alpha_1 = 3$, and the
subsequent $\alpha_2$ is given by $\alpha_2=\alpha_1 (1+\sqrt{\bar{\nu}})$.
We compare the proposed bound with the estimated global mean square error
(GMSE) of the maximum a posteriori (MAP) estimator of $\boldsymbol{\theta}
=[\lambda_1,\lambda_2,t_1]^T$. It is worth mentioning that, given the
posterior density $f(\mathbf{\lambda},\boldsymbol{t}|\boldsymbol{x} = 
\mathbf{\kappa})$ (that is proportional to (\ref{eqn/joint_law})), there is
a closed-form expression of the MAP estimator of $\mathbf{\lambda}$, for a
given $\boldsymbol{\ell}$, that is, for $k = 1,\ldots ,K+1$: 
\begin{equation}
\hat{\lambda}_k^{\mathsf{MAP}}\left(\ell_{k-1},\ell_{k}\right) = \frac{%
\alpha_k + \left(\sum\limits_{t= \ell_{k-1}+1}^{\ell_k}\kappa_t \right)-1}{%
\beta +(\ell_{k} - \ell_{k-1})}.  \label{eqn/MAP_lambda}
\end{equation}
This closed form expression is then used to obtain the MAP estimator of $%
\boldsymbol{t}$ 
\begin{equation}
\hat{\boldsymbol{t}}^{\mathsf{MAP}} = \argmax_{\boldsymbol{\ell}} \ln f\left(%
\hat{\mathbf{\lambda}}^{\mathsf{MAP}}(\boldsymbol{\ell}), \boldsymbol{t}=%
\boldsymbol{\ell}|\boldsymbol{x} = \mathbf{\kappa}\right).  \label{eqn/MAP_t}
\end{equation}
The estimated global root mean square error (GRMSE) of $\hat{\boldsymbol{t}}%
^{\mathsf{MAP}}$ computed using 1000 Monte-Carlo runs and the associated
lower bound are compared in Fig. \ref{fig/sim1}. Even if there exists a gap
between the GRMSE and the bound, the difference decreases as $\bar{\nu}$
increases: at $\bar{\nu}=10$~dB, the difference in terms of number of
samples is no more than $3$~samples; at $\bar{\nu}=15$~dB, it is less than $%
0.1$~samples. The MAP behavior even seems to be closer to the bound for $%
\bar{\nu} \geq 15$~dB. However, it could not be displayed for numerical
reasons, the GRMSE tending steeply to zero. Finally, the derived bound
provides a fair approximation of the changepoint estimation behavior, in
this context of Poisson data when the Poisson parameters $\lambda_k$ are
unknown.

\appendix[Details about the derivation of the bound]

In this section, we give all the calculation details leading to the bound
given in Section \ref{sect/bound_results}.

\subsection{Derivation of $\boldsymbol{V}_{22}$ and $\boldsymbol{P}_{22}$}

Let us first remark that%
\begin{equation}
\left[ \boldsymbol{V}_{22}\right] _{k,k}=-h_{k}\zeta \left( \boldsymbol{h}%
_{k},\boldsymbol{0}_{2K+1}\right) \qquad \text{and}\qquad \left[ \boldsymbol{%
P}_{22}\right] _{k,l}=\zeta \left( \boldsymbol{h}_{k},\boldsymbol{h}%
_{l}\right) +\zeta \left( -\boldsymbol{h}_{k},-\boldsymbol{h}_{l}\right)
-\zeta \left( -\boldsymbol{h}_{k},\boldsymbol{h}_{l}\right) -\zeta \left( 
\boldsymbol{h}_{k},-\boldsymbol{h}_{l}\right)  \label{eqn/V22andP22zeta}
\end{equation}%
in which $\zeta \left( \boldsymbol{h}_{k},\boldsymbol{h}_{l}^{\prime
}\right) $, for $k\in \{1,\ldots ,K\}$, denotes%
\begin{align}
\zeta \left( \boldsymbol{h}_{k},\boldsymbol{h}_{l}^{\prime }\right) &=%
\mathbb{E}_{\boldsymbol{x},\mathbf{\lambda },\boldsymbol{t}}\left\{ \frac{%
\sqrt{f\left( \boldsymbol{x}=\boldsymbol{\kappa },\mathbf{\lambda },%
\boldsymbol{t}=\boldsymbol{\ell}+\boldsymbol{h}_{k}\right) f\left( 
\boldsymbol{x}=\boldsymbol{\kappa },\mathbf{\lambda },\boldsymbol{t}=%
\boldsymbol{\ell}+\boldsymbol{h}_{l}^{\prime }\right) }}{f\left( \boldsymbol{%
x}=\boldsymbol{\kappa },\mathbf{\lambda },\boldsymbol{t}=\boldsymbol{\ell}%
\right) }\right\}  \notag \\
&=\sum_{\boldsymbol{\ell}\in \mathbb{Z}^{K}}\int_{\mathbb{R}_{+}^{K+1}}\sum_{%
\boldsymbol{\kappa }\in \mathbb{N}^{T}}\sqrt{f\left( \boldsymbol{x}=%
\boldsymbol{\kappa },\mathbf{\lambda },\boldsymbol{t}=\boldsymbol{\ell}+%
\boldsymbol{h}_{k}\right) f\left( \boldsymbol{x}=\boldsymbol{\kappa },%
\mathbf{\lambda },\boldsymbol{t}=\boldsymbol{\ell}+\boldsymbol{h}%
_{l}^{\prime }\right) }\mathrm{d}\mathbf{\lambda } \label{eqn/def_zeta}
\end{align}%
with vectors $\boldsymbol{h}_{k}$ and $\boldsymbol{h}_{l}^{\prime }$ of the
form $\boldsymbol{h}_{k}=\bigl[\mathbf{0}_{k-1}^{T},h_{k},\mathbf{0}%
_{K-k}^{T}\bigr]^{T}$and $\boldsymbol{h}_{l}^{\prime }=\bigl[\mathbf{0}%
_{l-1}^{T},h_{l}^{\prime },\mathbf{0}_{K-l}^{T}\bigr]^{T}$, both with size $%
K $. Note that function $\zeta$, even though it is written as a function of
two vectors $\hb_k$ and $\hb_l'$, is actually a function of the two scalars
$h_k$ and $h'_l$ (which are the non zero components of the two aforementioned
vectors). Throughout the following developments, we will either use
$\zeta ( \hb_{k},\hb_{l}')$ or $\zeta ( h_{k},h_{l}')$, depending on the
convenience.

Since $\boldsymbol{V}_{22}$ is diagonal and $\boldsymbol{P}_{22}$ is
symmetric, we can assume $l\geq k$, without loss of generality.

Developing each probability density function (p.d.f.) in (\ref{eqn/def_zeta}%
) from (\ref{eqn/joint_law}), we have 
\begin{align}
\zeta (\boldsymbol{h}_{k},\boldsymbol{h}_{l}^{\prime })& =\sum_{\boldsymbol{%
\ell }\in \mathbb{Z}^{K}}\int_{\mathbb{R}_{+}^{K+1}}\left[ \sqrt{f(\mathbf{%
\lambda },\boldsymbol{t}=\boldsymbol{\ell }+\boldsymbol{h}_{k})f(\mathbf{%
\lambda },\boldsymbol{t}=\boldsymbol{\ell }+\boldsymbol{h}_{l}^{\prime })}%
\sum_{\mathbf{\kappa }\in \mathbb{N}^{T}}\sqrt{f(\boldsymbol{x}=\mathbf{%
\kappa }|\mathbf{\lambda },\boldsymbol{t}=\boldsymbol{\ell }+\boldsymbol{h}%
_{k})f(\boldsymbol{x}=\mathbf{\kappa }|\mathbf{\lambda },\boldsymbol{t}=%
\boldsymbol{\ell }+\boldsymbol{h}_{l}^{\prime })}\right] \mathrm{d}\mathbf{%
\lambda }  \notag \\
& =\sum_{\boldsymbol{\ell }\in \mathbb{Z}^{K}}\int_{\mathbb{R}_{+}^{K+1}}\pi
(\mathbf{\lambda },\boldsymbol{\ell },\boldsymbol{h}_{k},\boldsymbol{h}%
_{l}^{\prime })\;\acute{\zeta}(\mathbf{\lambda },\boldsymbol{\ell },%
\boldsymbol{h}_{k},\boldsymbol{h}_{l}^{\prime })\,\mathrm{d}\mathbf{\lambda }
\label{eqn/zeta_pi_zetacute}
\end{align}%
in which 
\begin{equation}
\pi (\mathbf{\lambda },\boldsymbol{\ell },\boldsymbol{h}_{k},\boldsymbol{h}%
_{l}^{\prime })=\sqrt{f(\mathbf{\lambda },\boldsymbol{t}=\boldsymbol{\ell }+%
\boldsymbol{h}_{k})f(\mathbf{\lambda },\boldsymbol{t}=\boldsymbol{\ell }+%
\boldsymbol{h}_{l}^{\prime })}  \label{eqn/def_pi}
\end{equation}%
and 
\begin{equation}
\acute{\zeta}(\mathbf{\lambda },\boldsymbol{\ell },\boldsymbol{h}_{k},%
\boldsymbol{h}_{l}^{\prime })=\sum_{\mathbf{\kappa }\in \mathbb{N}^{T}}\sqrt{%
f(\boldsymbol{x}=\mathbf{\kappa }|\mathbf{\lambda },\boldsymbol{t}=%
\boldsymbol{\ell }+\boldsymbol{h}_{k})f(\boldsymbol{x}=\mathbf{\kappa }|%
\mathbf{\lambda },\boldsymbol{t}=\boldsymbol{\ell }+\boldsymbol{h}%
_{l}^{\prime })}.  \label{eqn/def_zetacute}
\end{equation}%
We first calculate (\ref{eqn/def_pi}), and then (\ref{eqn/def_zetacute}).

\subsubsection{Derivation of $\protect\pi(\mathbf{\protect\lambda},%
\boldsymbol{\ell},\boldsymbol{h}_k,\boldsymbol{h}_l^{\prime})$}

From (\ref{eqn/prior_lambda}) and (\ref{eqn/prior_t}), one can deduce 
\begin{align}
\pi (\mathbf{\lambda },\boldsymbol{\ell },\boldsymbol{h}_{k},\boldsymbol{h}%
_{l}^{\prime })& =\prod_{i=1}^{K+1}\frac{\beta ^{\alpha _{i}}}{\Gamma \left(
\alpha _{i}\right) }\lambda _{i}^{\alpha _{i}-1}\exp \left( -\beta \lambda
_{i}\right) \mathbb{I}_{\mathbb{R}_{+}}\left( \lambda _{i}\right)   \notag \\
& \quad \times \frac{1}{\tau ^{K}}\left( \prod_{i=1}^{K}\mathbb{I}_{\left\{ \ell _{i-1}+1,\ldots ,\ell _{i-1}+\tau \right\}
}\left( \ell _{i}\right) \right) \mathbb{I}_{\left\{ \ell
_{k-1}-h_{k}+1,\ldots ,\ell _{k-1}-h_{k}+\tau \right\} }\left( \ell
_{k}\right) \mathbb{I}_{\left\{ \ell _{k}+h_{k}+1,\ldots ,\ell
_{k}+h_{k}+\tau \right\} }\left( \ell _{k+1}\right)   \notag \\
& \quad \times \phantom{\frac{1}{\tau ^{K}}}\left( \prod_{i=1}^{K}\mathbb{I}_{\left\{ \ell _{i-1}+1,\ldots ,\ell _{i-1}+\tau
\right\} }\left( \ell _{i}\right) \right) \mathbb{I}_{\left\{ \ell
_{l-1}-h_{l}^{\prime }+1,\ldots ,\ell _{l-1}-h_{l}^{\prime }+\tau \right\}
}\left( \ell _{l}\right) \mathbb{I}_{\left\{ \ell _{l}+h_{l}^{\prime
}+1,\ldots ,\ell _{l}+h_{l}^{\prime }+\tau \right\} }\left( \ell
_{l+1}\right)   \notag \\
& =\prod_{i=1}^{K+1}\frac{\beta ^{\alpha _{i}}}{\Gamma \left( \alpha
_{i}\right) }\lambda _{i}^{\alpha _{i}-1}\exp \left( -\beta \lambda
_{i}\right) \mathbb{I}_{\mathbb{R}_{+}}\left( \lambda _{i}\right) \cdot 
\frac{1}{\tau ^{K}}\mathbb{I}_{\mathcal{J}_{h_{k},h_{l}^{\prime }}}\left( 
\boldsymbol{\ell }\right)   \label{eqn/pi_expanded}
\end{align}%
in which $\mathcal{J}_{h_{k},h_{l}^{\prime }}$, for $l>k+1$, denotes the
following set (that is a subset of $\mathbb{N}^{K}$) 
\begin{equation}
\mathcal{J}_{h_{k},h_{l}^{\prime }}=\left( \prod_{\substack{ i=1 \\ i\neq
k,k+1 \\ i\neq l,l+1}}^{K}\left\{ \ell _{i-1}+1,\ldots ,\ell _{i-1}+\tau
\right\} \right) \!%
\begin{array}[t]{@{}l}
\times \bigl(\left\{ \ell _{k-1}+1,\ldots ,\ell _{k-1}+\tau \right\} \cap
\left\{ \ell _{k-1}-h_{k}+1,\ldots ,\ell _{k-1}-h_{k}+\tau \right\} \bigr)
\\ 
\times \bigl(\left\{ \ell _{k}+1,\ldots ,\ell _{k}+\tau \right\} \cap
\left\{ \ell _{k}+h_{k}+1,\ldots ,\ell _{k}+h_{k}+\tau \right\} \bigr) \\ 
\times \bigl(\left\{ \ell _{l-1}+1,\ldots ,\ell _{l-1}+\tau \right\} \cap
\left\{ \ell _{l-1}-h_{l}^{\prime }+1,\ldots ,\ell _{l-1}-h_{l}^{\prime
}+\tau \right\} \bigr) \\ 
\times \bigl(\left\{ \ell _{l}+1,\ldots ,\ell _{l}+\tau \right\} \cap
\left\{ \ell _{l}+h_{l}^{\prime }+1,\ldots ,\ell _{l}+h_{l}^{\prime }+\tau
\right\} \bigr),%
\end{array}
\label{eqn/J_UT}
\end{equation}%
for $l=k+1$, it is 
\begin{equation}
\mathcal{J}_{h_{k},h_{k+1}^{\prime }}=\left( \prod_{\substack{ i=1 \\ i\neq
k,k+1,k+2}}^{K}\hspace{-1.5em}\left\{ \ell _{i-1}+1,\ldots ,\ell _{i-1}+\tau
\right\} \! \right) \!\!%
\begin{array}[t]{@{}l}
\times \bigl(\left\{ \ell _{k-1}+1,\ldots ,\ell _{k-1}+\tau \right\} \cap
\left\{ \ell _{k-1}-h_{k}+1,\ldots ,\ell _{k-1}-h_{k}+\tau \right\} \bigr)
\\ 
\begin{split}
\! \times \bigl(\left\{ \ell _{k}+1,\ldots ,\ell _{k}+\tau \right\} & \cap
\left\{ \ell _{k}+h_{k}+1,\ldots ,\ell _{k}+h_{k}+\tau \right\}  \\
& \cap \left\{ \ell _{k}-h_{k+1}^{\prime }+1,\ldots ,\ell
_{k}-h_{k+1}^{\prime }+\tau \right\} \bigr)
\end{split}
\\
\times
\begin{array}[t]{@{}l@{}}
\bigl(\left\{ \ell _{k+1}+1,\ldots ,\ell _{k+1}+\tau \right\} \\
\phantom{\bigl(\left\{ \ell _{k}+1,\ldots ,\ell _{k}+\tau \right\}}
\cap \left\{ \ell _{k+1}+h'_{k+1} +1,\ldots ,\ell _{k+1}+h'_{k+1} +\tau
\right\} \bigr),%
\end{array}
\end{array}
\mspace{-10mu}
\label{eqn/J_FSD}
\end{equation}%
and for $l=k$, it is 
\begin{equation}
\mathcal{J}_{h_{k},h_{k}^{\prime }}=\left( \prod_{\substack{ i=1 \\ i\neq
k,k+1}}^{K}\left\{ \ell _{i-1}+1,\ldots ,\ell _{i-1}+\tau \right\} \right) \!%
\begin{array}[t]{@{}l}
\begin{array}[t]{@{}r@{}l}
\times \bigl(\left\{ \ell _{k-1}+1,\ldots ,\ell _{k-1}+\tau \right\}  & 
\:\cap \left\{ \ell _{k-1}-h_{k}+1,\ldots ,\ell _{k-1}-h_{k}+\tau \right\} 
\\ 
& \:\cap \left\{ \ell _{k-1}-h_{k}^{\prime }+1,\ldots ,\ell
_{k-1}-h_{k}^{\prime }+\tau \right\} \bigr)%
\end{array}
\\ 
\begin{split}
\!\times \bigl(\left\{ \ell _{k}+1,\ldots ,\ell _{k}+\tau \right\} & \cap
\left\{ \ell _{k}+h_{k}+1,\ldots ,\ell _{k}+h_{k}+\tau \right\}  \\
& \cap \left\{ \ell _{k}+h_{k}^{\prime }+1,\ldots ,\ell _{k}+h_{k}^{\prime
}+\tau \right\} \bigr),
\end{split}%
\end{array}
\label{eqn/J_D}
\end{equation}%
with $\ell _{0}=0$.

To summarize this, and in an effort to make it more explicit, the set $%
\mathcal{J}_{h_k,h_l^{\prime}}\subset \mathbb{N}^K$ can be written as the
cartesian product of $K$ sets $\mathcal{I}_i\subset \mathbb{N}$ of
consecutive integers, with $i=1,\ldots K$: 
\begin{equation}
\mathcal{J}_{h_k,h_l^{\prime}}=\prod_{i=1}^{K} \mathcal{I}_i.
\end{equation}
The lowest elements $(\ell_i)_{\min}=\min \mathcal{I}_i$ and the greatest
elements $(\ell_i)_{\max}=\max \mathcal{I}_i$ of the sets $\mathcal{I}_i$
for $i=1,\ldots ,K$ are given in tables \ref{tab/ranges_elli_UT}, \ref%
{tab/ranges_elli_FSD}, and \ref{tab/ranges_elli_D} for the three cases i) $%
l>k+1$ (case ``UT'', for ``upper triangle''), ii) $l = k+1$ (case ``FSD'',
for ``first subdiagonal''), and iii) $l = k$ (case ``D'', for ``diagonal''),
respectively.

\begin{table}[t]
\caption{Lowest elements $(\ell_i)_{\min}$ and greatest elements $%
(\ell_i)_{\max}$ of the sets $\mathcal{I}_i$ for $i=1,\ldots ,K$, in the
case UT ($l>k+1$).}
\label{tab/ranges_elli_UT}\renewcommand{\arraystretch}{1.5} \centering%
\begin{tabular}{|c||c|c|}
\hline
$i$ & $(\ell_{i})_{\min}$ & $(\ell_{i})_{\max}$ \\ \hline
$k$ & $\ell_{k-1}+1+\max(-h_{k} ,0)$ & $\ell_{k-1}+\tau-\max( h_{k} ,0)$ \\ 
$k+1$ & $\ell_{k} +1+\max( h_{k} ,0)$ & $\ell_{k} +\tau-\max(-h_{k} ,0)$ \\ 
$l$ & $\ell_{l-1}+1+\max(-h^{\prime}_{l},0)$ & $\ell_{l-1}+\tau-\max(
h^{\prime}_{l},0)$ \\ 
$l+1$ (if $l+1 \leq K$) & $\ell_{l} +1+\max( h^{\prime}_{l},0)$ & $\ell_{l}
+\tau -\max(-h^{\prime}_{l},0)$ \\ 
$i\neq k,k+1,l,l+1$ & $\ell_{i-1}+1$ & $\ell_{i-1}+\tau $ \\ \hline
\end{tabular}%
\end{table}

\begin{table}[t]
\caption{Lowest elements $(\ell_i)_{\min}$ and greatest elements $%
(\ell_i)_{\max}$ of the sets $\mathcal{I}_i$ for $i=1,\ldots ,K$, in the
case FSD ($l=k+1$).}
\label{tab/ranges_elli_FSD}\renewcommand{\arraystretch}{1.5} \centering%
\begin{tabular}{|c||c|c|}
\hline
$i$ & $(\ell_{i})_{\min }$ & $( \ell_{i})_{\max }$ \\ \hline
$k$ & $\ell_{k-1}+1+\max ( -h_{k},0) $ & $\ell_{k-1}+\tau -\max (h_{k},0) $
\\ 
$k+1$ & $\ell_{k}+1+\max ( h_{k},-h^{\prime}_{k+1},0) $ & $\ell_{k}+\tau
-\max (-h_{k},h^{\prime}_{k+1},0) $ \\ 
$k+2$ (if $k+2 \leq K$) & $\ell_{k+1}+1+\max ( h^{\prime}_{k+1},0) $ & $%
\ell_{k+1}+\tau -\max (-h^{\prime}_{k+1},0) $ \\ 
$i\neq k,k+1,k+2$ & $k_{i-1}+1$ & $k_{i-1}+\tau $ \\ \hline
\end{tabular}%
\end{table}

\begin{table}[!t]
\caption{Lowest elements $(\ell_i)_{\min}$ and greatest elements $%
(\ell_i)_{\max}$ of the sets $\mathcal{I}_i$ for $i=1,\ldots ,K$, in the
case D ($l=k$).}
\label{tab/ranges_elli_D}\renewcommand{\arraystretch}{1.5} \centering%
\begin{tabular}{|c||c|c|}
\hline
$i$ & $( \ell_{i})_{\min }$ & $( \ell_{i})_{\max }$ \\ \hline
$k$ & $\ell_{k-1}+1+\max ( -h_{k},-h^{\prime}_{k},0) $ & $\ell_{k-1}+\tau
-\max (h_{k},h^{\prime}_{k},0) $ \\ 
$k+1$ (if $k+1 \leq K$) & $\ell_{k}+1+\max ( h_{k},h^{\prime}_{k},0) $ & $%
\ell_{k}+\tau -\max (-h_{k},-h^{\prime}_{k},0) $ \\ 
$i\neq k,k+1$ & $\ell_{i-1}+1$ & $\ell_{i-1}+\tau $ \\ \hline
\end{tabular}%
\end{table}

\subsubsection{Derivation of $\acute{\protect\zeta}(\mathbf{\protect\lambda},%
\boldsymbol{\ell},\boldsymbol{h}_k,\boldsymbol{h}_l^{\prime})$}

It follows directly from (\ref{eqn/likelihood}) that 
\begin{multline}
\hspace{-1em}f(\boldsymbol{x}=\mathbf{\kappa }| \mathbf{\lambda },%
\boldsymbol{t}=\boldsymbol{\ell }+\boldsymbol{h}_{k}) = \mspace{-12mu} \prod
_{\substack{ i=1 \\ i\neq k,k+1}}^{K+1} \left[ \prod_{t=\ell_{i-1}+1}^{%
\ell_i} \sqrt{\frac{\lambda_i^{\kappa_t}}{\kappa_t !}\exp(-\lambda_i)} %
\right] \prod_{t=\ell_{k-1}+1}^{\ell_k+h_k} \sqrt{\frac{\lambda_k^{\kappa_t}%
}{\kappa_t !}\exp(-\lambda_k)} \prod_{t=\ell_{k}+h_k+1}^{\ell_{k+1}} \sqrt{%
\frac{\lambda_{k+1}^{\kappa_t}}{\kappa_t !}\exp(-\lambda_{k+1})}
\label{eqn/likelihood_hk}
\end{multline}
and 
\begin{multline}
\hspace{-1em}f(\boldsymbol{x}=\mathbf{\kappa }| \mathbf{\lambda },%
\boldsymbol{t}=\boldsymbol{\ell }+\boldsymbol{h}_{l}^{\prime}) = %
\mspace{-12mu} \prod_{\substack{ i=1 \\ i\neq l,l+1}}^{K+1} \left[
\prod_{t=\ell_{i-1}+1}^{\ell_i} \sqrt{\frac{\lambda_i^{\kappa_t}}{\kappa_t !}%
\exp(-\lambda_i)} \right] \prod_{t=\ell_{l-1}+1}^{\ell_l+h_l^{\prime}} \sqrt{%
\frac{\lambda_l^{\kappa_t}}{\kappa_t !}\exp(-\lambda_l)} \prod_{t=%
\ell_{l}+h_l^{\prime}+1}^{\ell_{l+1}} \sqrt{\frac{\lambda_{l+1}^{\kappa_t}}{%
\kappa_t !}\exp(-\lambda_{l+1})}  \label{eqn/likelihood_hl}
\end{multline}
with $\ell_0=0$ and $\ell_{K+1}=T$. By plugging (\ref{eqn/likelihood_hk})
and (\ref{eqn/likelihood_hl}) into (\ref{eqn/def_zetacute}), one obtains the
expression of $\acute{\zeta}(\mathbf{\lambda},\boldsymbol{\ell},\boldsymbol{h%
}_k,\boldsymbol{h}_l^{\prime})$, whose writing actually depends on the three
aforementioned cases, i.e., i) UT ($l>k+1$), ii) FSD ($l=k+1$), and iii) D ($%
l = k$).

\paragraph*{i) Case UT ($l>k+1$)}

In this case, by plugging (\ref{eqn/likelihood_hk}) and (\ref%
{eqn/likelihood_hl}) into (\ref{eqn/def_zetacute}), one obtains: 
\begin{align}
\acute{\zeta}(\mathbf{\lambda},\boldsymbol{\ell},\boldsymbol{h}_k,%
\boldsymbol{h}_l^{\prime}) &= \sum_{\kappa_1=0}^{+\infty} \ldots
\sum_{\kappa_T=0}^{+\infty} \left( \prod_{\substack{ i=1 \\ i\neq k,k+1  \\ %
i\neq l,l+1}}^{K+1} \left[ \prod_{t=\ell_{i-1}+1}^{\ell_i} \frac{%
\lambda_i^{\kappa_t}}{\kappa_t !}\exp(-\lambda_i) \right] \prod_{t=%
\ell_{k-1}+1}^{\ell_k-\max(-h_k,0)} \frac{\lambda_k^{\kappa_t}}{\kappa_t !}%
\exp(-\lambda_k) \right.  \notag \\
& \quad \hphantom{ \sum_{\kappa_1 \in \mathbb{N}} \ldots \sum_{\kappa_T\in
\mathbb{N}} \left( \vphantom{\prod_{\substack{i=1\\ i\neq k,k+1 \\ i\neq
l,l+1}}^{K+1}} \right. } {} \times
\prod_{t=\ell_{k}-\max(-h_k,0)+1}^{\ell_{k}+\max(h_k,0)} \frac{\sqrt{%
\lambda_k\lambda_{k+1}}^{\kappa_t}
}{\kappa_t !} \exp\left\{-\frac{%
\lambda_k+\lambda_{k+1}}{2}\right\}
\prod_{t=\ell_{k}+\max(h_k,0)+1}^{\ell_{k+1}} \frac{\lambda_{k+1}^{\kappa_t}%
}{\kappa_t !}\exp(-\lambda_{k+1})  \notag \\
& \quad \hphantom{ \sum_{\kappa_1 \in \mathbb{N}} \ldots \sum_{\kappa_T\in
\mathbb{N}} \left( \vphantom{\prod_{\substack{i=1\\ i\neq k,k+1 \\ i\neq
l,l+1}}^{K+1}} \right. } {} \times
\prod_{t=\ell_{l-1}+1}^{\ell_l-\max(-h_l^{\prime},0)} \frac{%
\lambda_l^{\kappa_t}}{\kappa_t !}\exp(-\lambda_l)
\prod_{t=\ell_{l}-\max(-h_l^{\prime},0)+1}^{\ell_{l}+\max(h_l^{\prime},0)} 
\frac{\sqrt{\lambda_l\lambda_{l+1}}^{\kappa_t}}{\kappa_t !} \exp\left\{-%
\frac{\lambda_l+\lambda_{l+1}}{2}\right\}  \notag \\
& \quad \hphantom{ \sum_{\kappa_1 \in \mathbb{N}} \ldots \sum_{\kappa_T\in
\mathbb{N}} \left( \vphantom{\prod_{\substack{i=1\\ i\neq k,k+1 \\ i\neq
l,l+1}}^{K+1}} \right. } {} \times \left.
\prod_{t=\ell_{l}+\max(h_l^{\prime},0)+1}^{\ell_{l+1}} \frac{%
\lambda_{l+1}^{\kappa_t}}{\kappa_t !}\exp(-\lambda_{l+1}) %
\vphantom{\prod_{\substack{i=1\\ i\neq k,k+1 \\ i\neq l,l+1}}^{K+1}} \right)
\label{eqn/zetacute_expanded}
\end{align}
Note that the sums indices $\kappa_1,\kappa_2,\ldots ,\kappa_{\ell_1},
\kappa_{\ell_1+1},\ldots ,\kappa_{\ell_2},\ldots,\kappa_{\ell_K},\ldots ,
\kappa_T$ are separated in (\ref{eqn/zetacute_expanded}). This implies that
the sums of products become products of sums, and since 
\begin{equation}
\sum_{\kappa_t = 0}^{+\infty} \frac{\lambda_i^{\kappa_t}}{\kappa_t !}%
\exp(-\lambda_i) = 1,
\end{equation}
then (\ref{eqn/zetacute_expanded}) becomes 
\begin{align}
\acute{\zeta}(\mathbf{\lambda},\boldsymbol{\ell},\boldsymbol{h}_k,%
\boldsymbol{h}_l^{\prime}) &=
\prod_{t=\ell_{k}-\max(-h_k,0)+1}^{\ell_k+\max(h_k,0)} \left( \exp\left\{-%
\frac{\lambda_k+\lambda_{k+1}}{2}\right\} \sum_{\kappa_t = 0}^{+\infty} 
\frac{\sqrt{\lambda_k\lambda_{k+1}}^{\kappa_t}}{\kappa_t !} \right)  \notag
\\
& \quad \hspace{4em} \times
\prod_{t=\ell_{l}-\max(-h_l^{\prime},0)+1}^{\ell_l+\max(h_l^{\prime},0)}
\left( \exp\left\{-\frac{\lambda_l+\lambda_{l+1}}{2}\right\} \sum_{\kappa_t
= 0}^{+\infty} \frac{\sqrt{\lambda_l\lambda_{l+1}}^{\kappa_t}}{\kappa_t !}
\right)  \notag \\
& = \exp\left\{-\vert h_k\vert\frac{\lambda_k+\lambda_{k+1}}{2}\right\}
\left( \sum_{\kappa = 0}^{+\infty} \frac{\sqrt{\lambda_k\lambda_{k+1}}%
^{\kappa}}{\kappa !} \right)^{\vert h_k\vert} \exp\left\{-\vert
h_l^{\prime}\vert\frac{\lambda_l+\lambda_{l+1}}{2}\right\} \left(
\sum_{\kappa = 0}^{+\infty} \frac{\sqrt{\lambda_l\lambda_{l+1}}^{\kappa}}{%
\kappa !} \right)^{\vert h_l^{\prime}\vert}  \notag \\
& = \exp\left\{-\vert h_k\vert\frac{\lambda_k+\lambda_{k+1}}{2}\right\}
\exp\left\{\vert h_k\vert\sqrt{\lambda_k\lambda_{k+1}}\right\}
\exp\left\{-\vert h_l^{\prime}\vert\frac{\lambda_l+\lambda_{l+1}}{2}\right\}
\exp\left\{\vert h_l^{\prime}\vert\sqrt{\lambda_l\lambda_{l+1}}\right\} 
\notag \\
& = \exp\left\{-\vert h_k\vert\frac{\left(\sqrt{\lambda_{k+1}} -\sqrt{%
\lambda_k}\right)^2}{2}\right\} \exp\left\{-\vert h_l^{\prime}\vert\frac{%
\left(\sqrt{\lambda_{l+1}} -\sqrt{\lambda_l}\right)^2}{2}\right\}  \notag \\
&= \rho^{\left| h_k\right|} (\lambda_k ,\lambda_{k+1}) \rho^{\left|
h^{\prime}_l\right|}(\lambda_l ,\lambda_{l+1})  \label{eqn/zetacute_UT}
\end{align}
in which, for $k=1,\ldots, K$, 
\begin{equation}
\rho (\lambda_k ,\lambda_{k+1}) = \exp\left\{-\frac{\left(\sqrt{\lambda_{k+1}%
} -\sqrt{\lambda_k}\right)^2}{2}\right\}.
\end{equation}
Note that $\acute{\zeta}(\mathbf{\lambda},\boldsymbol{\ell},\boldsymbol{h}_k,%
\boldsymbol{h}_l^{\prime})$ does actually not depend on $\boldsymbol{\ell}$
in this case.

\paragraph*{ii) Case FSD ($l=k+1$)}

This case is more complicated than the two others, because the writing of (%
\ref{eqn/def_zetacute}) depends on whether $\ell_{k+1}+h^{\prime}_{k+1}%
\gtrless\ell_k+h_k$ (note that in the case UT ($l>k+1$), one always has $%
\ell_{l}+h^{\prime}_{l}>\ell_k+h_k$ ; this can be seen by analyzing table %
\ref{tab/ranges_elli_UT}). The case $\ell_{k+1}+h^{\prime}_{k+1}<\ell_k+h_k$
is often referred to as ``overlap case'' in the following. It is first of
interest to determine when this case occurs.

By analyzing the line ``$i=k+1$'' in table \ref{tab/ranges_elli_FSD}, we can
first remark that this case is possible only if $h_k>0$ and $%
h^{\prime}_{k+1}<0$. In addition, since the set $\mathcal{I}_{k+1}$ depends
on $\ell_k$, the formal condition for the overlap case to occur can be
written as: 
\begin{equation}
\exists \ell_{k+1} \in \mathcal{I}_{k+1},\,
\ell_{k+1}+h^{\prime}_{k+1}<\ell_k+h_k,
\label{eqn/overlap_condition_formal}
\end{equation}
which is equivalent to 
\begin{equation}
(\ell_{k+1})_{\min } + h_{k+1}<\ell_k+h_k.
\end{equation}
Using the expression of $(\ell_{k+1})_{\min }$ from table \ref%
{tab/ranges_elli_FSD}, we obtain that the condition for the overlap case is
finally 
\begin{equation}
h_k>0 \quad \text{and} \quad h^{\prime}_{k+1}<0 \quad \text{and} \quad \min
(\left| h_k\right|,\left| h^{\prime}_{k+1}\right|) \geq 2.
\label{eqn/condition_overlap}
\end{equation}

In the case whithout overlap, i.e., when the condition (\ref%
{eqn/condition_overlap}) is not met, the derivation of $\acute{\zeta}(%
\mathbf{\lambda},\boldsymbol{\ell},\boldsymbol{h}_k,\boldsymbol{h}%
_l^{\prime})$ is exactly the same as in the case UT, which means we obtain 
\begin{equation}
\acute{\zeta}(\mathbf{\lambda},\boldsymbol{\ell},\boldsymbol{h}_k,%
\boldsymbol{h}_{k+1}^{\prime})= \rho^{\left| h_k\right|} (\lambda_k
,\lambda_{k+1}) \rho^{\left| h^{\prime}_{k+1}\right|}(\lambda_{k+1}
,\lambda_{k+2}).  \label{eqn/zetacute_FSD_nooverlap}
\end{equation}

On the other hand, when there is overlap, i.e., when the condition (\ref%
{eqn/condition_overlap}) is met, the derivation of (\ref{eqn/def_zetacute})
is done in the following way: 
\begin{align}
\acute{\zeta}(\mathbf{\lambda },\boldsymbol{\ell },\boldsymbol{h}_{k},%
\boldsymbol{h}_{k+1}^{\prime })& =\sum_{\kappa _{1}=0}^{+\infty }\ldots
\sum_{\kappa _{T}=0}^{+\infty }\left( \prod_{\substack{ i=1 \\ i\neq k+1,k+2
}}^{K+1}\left[ \prod_{t=\ell _{i-1}+1}^{\ell _{i}}\frac{\lambda _{i}^{\kappa
_{t}}}{\kappa _{t}!}\exp (-\lambda _{i})\right] \prod_{t=\ell _{k}+1}^{\ell
_{k+1}+h_{k+1}^{\prime }}\frac{\sqrt{\lambda _{k}\lambda _{k+1}}^{\kappa
_{t}}}{\kappa _{t}!}\exp \left\{ -\frac{\lambda _{k}+\lambda _{k+1}}{2}%
\right\} \right.   \notag \\
& \quad \hphantom{ \sum_{\kappa_1 \in \mathbb{N}} \ldots \sum_{\kappa_T\in
\mathbb{N}} \left( \vphantom{\prod_{\substack{i=1\\ i\neq k,k+1,k+2}}^{K+1}}
\right. } {}\times \prod_{t=\ell _{k+1}+h_{k+1}^{\prime }+1}^{\ell
_{k}+h_{k}}\frac{\sqrt{\lambda _{k}\lambda _{k+2}}^{\kappa _{t}}}{\kappa
_{t}!}\exp \left\{ -\frac{\lambda _{k}+\lambda _{k+2}}{2}\right\}   \notag \\
& \quad \hphantom{ \sum_{\kappa_1 \in \mathbb{N}} \ldots \sum_{\kappa_T\in
\mathbb{N}} \left( \vphantom{\prod_{\substack{i=1\\ i\neq k,k+1,k+2}}^{K+1}}
\right. } {}\times \left. \prod_{t=\ell _{k}+h_{k}+1}^{\ell _{k+1}}\frac{%
\sqrt{\lambda _{k+1}\lambda _{k+2}}^{\kappa _{t}}}{\kappa _{t}!}\exp \left\{
-\frac{\lambda _{k+1}+\lambda _{k+2}}{2}\right\} \vphantom{\prod_{%
\substack{i=1\\ i\neq k,k+1,k+2}}^{K+1}}\right)   \notag \\
& =\frac{\rho ^{(\ell _{k+1}+h_{k+1}^{\prime })-\ell _{k}}(\lambda
_{k},\lambda _{k+1})\rho ^{\ell _{k+1}-(\ell _{k}+h_{k})}(\lambda
_{k+1},\lambda _{k+2})}{\rho ^{(\ell _{k+1}+h_{k+1}^{\prime })-(\ell
_{k}+h_{k})}(\lambda _{k},\lambda _{k+2})}  \label{eqn/zetacute_FSD_overlap}
\end{align}%
using the same manipulations as those leading to (\ref{eqn/zetacute_UT}).
Note that in this case, $\acute{\zeta}(\mathbf{\lambda },\boldsymbol{\ell },%
\boldsymbol{h}_{k},\boldsymbol{h}_{k+1}^{\prime })$ effectively depends on $%
\boldsymbol{\ell }$. We introduce the following definition, where the notation
$\lambdab_{k:k+2}$ denotes the truncated vector
$[\lambda_k,\lambda_{k+1},\lambda_{k+2}]^T$ of $\lambdab$
\begin{align}
r(\lambdab_{k:k+2}) &= \frac{\rho(\lambda_{k}  ,\lambda_{k+1})
                             \rho(\lambda_{k+1},\lambda_{k+2})}
                            {\rho(\lambda_{k}  ,\lambda_{k+2})}
                       \label{eqn/def_r_general} \\
                    &= \exp \left\{
                            -\lambda_{k+1}+\sqrt{\lambda_{k}  \lambda_{k+1}}
                                          +\sqrt{\lambda_{k+1}\lambda_{k+2}}
                                          -\sqrt{\lambda_{k}  \lambda_{k+2}}
                            \right\}.
                       \label{eqn/def_r_poiss}
\end{align}
Then we retrieve the function $r(.)$ used in (\ref{eqn/def_w}). Using this
definition, we can rewrite (\ref{eqn/zetacute_FSD_overlap}) as
\begin{equation}
\acute{\zeta}(\lambdab,\ellb,\hb_{k},\hb'_{k+1})
= \rho^{h_{k}}     (\lambda_{k}  ,\lambda_{k+1}) \,
  \rho^{-h'_{k+1}} (\lambda_{k+1},\lambda_{k+2}) \,
  r^{(\ell_{k+1}+h'_{k+1})-(\ell_{k}+h_{k})}(\lambdab_{k:k+2})
\label{eqn/zetacute_FSD_overlap_r}
\end{equation}

\paragraph*{iii) Case D ($l=k$)}

In this case, we have to consider the fact that $\boldsymbol{h}^{\prime}_k$
can either take three values: either $\boldsymbol{h}^{\prime}_k=\boldsymbol{h%
}_k$, or $\boldsymbol{h}^{\prime}_k=-\boldsymbol{h}_k$, or $\boldsymbol{h}%
^{\prime}_k=\mathbf{0}_{2K+1}$, according to (\ref{eqn/V22andP22zeta}).

If $\boldsymbol{h}_{k}^{\prime }=\boldsymbol{h}_{k}$, (\ref{eqn/def_zetacute}%
) gives directly 
\begin{align}
\acute{\zeta}(\mathbf{\lambda },\boldsymbol{\ell },\boldsymbol{h}_{k},%
\boldsymbol{h}_{k})& =\sum_{\mathbf{\kappa }\in \mathbb{N}^{T}}f(\boldsymbol{%
x}=\mathbf{\kappa }|\mathbf{\lambda },\boldsymbol{t}=\boldsymbol{\ell }+%
\boldsymbol{h}_{k})  \notag \\
& =1  \label{eqn/zetacute_D1}
\end{align}%
since it is the sum of a probability distribution over its whole domain.

If $\boldsymbol{h}_{k}^{\prime }=-\boldsymbol{h}_{k}$, equation (\ref%
{eqn/zetacute_expanded}) becomes 
\begin{align}
\acute{\zeta}(\mathbf{\lambda },\boldsymbol{\ell },\boldsymbol{h}_{k},%
\boldsymbol{h}_{l}^{\prime })& =\sum_{\kappa _{1}=0}^{+\infty }\ldots
\sum_{\kappa _{T}=0}^{+\infty }\left( \prod_{\substack{ i=1 \\ i\neq k,k+1}}%
^{K+1}\left[ \prod_{t=\ell _{i-1}+1}^{\ell _{i}}\frac{\lambda _{i}^{\kappa
_{t}}}{\kappa _{t}!}\exp (-\lambda _{i})\right] \prod_{t=\ell
_{k-1}+1}^{\ell _{k}-\left\vert h_{k}\right\vert }\frac{\lambda _{k}^{\kappa
_{t}}}{\kappa _{t}!}\exp (-\lambda _{k})\right.   \notag \\
& \quad \hphantom{ \sum_{\kappa_1 \in \mathbb{N}} \ldots \sum_{\kappa_T\in
\mathbb{N}} \left( \vphantom{\prod_{\substack{i=1\\ i\neq k,k+1}}^{K+1}}
\right. } {}\times \prod_{t=\ell _{k}-\left\vert h_{k}\right\vert +1}^{\ell
_{k}+\left\vert h_{k}\right\vert }\frac{\sqrt{\lambda _{k}\lambda _{k+1}}%
^{\kappa _{t}}}{\kappa _{t}!}\exp \left\{ -\frac{\lambda _{k}+\lambda _{k+1}%
}{2}\right\}   \notag \\
& \quad \hphantom{ \sum_{\kappa_1 \in \mathbb{N}} \ldots \sum_{\kappa_T\in
\mathbb{N}} \left( \vphantom{\prod_{\substack{i=1\\ i\neq k,k+1}}^{K+1}}
\right. } {}\times \left. \prod_{t=\ell _{k}+\left\vert h_{k}\right\vert
+1}^{\ell _{k+1}}\frac{\lambda _{k+1}^{\kappa _{t}}}{\kappa _{t}!}\exp
(-\lambda _{k+1})\vphantom{\prod_{\substack{i=1\\ i\neq k,k+1}}^{K+1}}%
\right)   \notag \\
& =\rho ^{2\left\vert h_{k}\right\vert }(\lambda _{k},\lambda _{k+1})
\label{eqn/zetacute_D2}
\end{align}%
using the same manipulations as those leading to (\ref{eqn/zetacute_UT}).

At last, if $\boldsymbol{h}_{k}^{\prime }=\mathbf{0}_{2K+1}$, then equation (%
\ref{eqn/zetacute_expanded}) becomes: 
\begin{align}
\acute{\zeta}(\mathbf{\lambda },\boldsymbol{\ell },\boldsymbol{h}_{k},%
\mathbf{0}_{2K+1})& =\sum_{\kappa _{1}=0}^{+\infty }\ldots \sum_{\kappa
_{T}=0}^{+\infty }\left( \prod_{\substack{ i=1 \\ i\neq k,k+1}}^{K+1}\left[
\prod_{t=\ell _{i-1}+1}^{\ell _{i}}\frac{\lambda _{i}^{\kappa _{t}}}{\kappa
_{t}!}\exp (-\lambda _{i})\right] \prod_{t=\ell _{k-1}+1}^{\ell _{k}-\max
(-h_{k},0)}\frac{\lambda _{k}^{\kappa _{t}}}{\kappa _{t}!}\exp (-\lambda
_{k})\right.   \notag \\
& \quad \hphantom{ \sum_{\kappa_1 \in \mathbb{N}} \ldots \sum_{\kappa_T\in
\mathbb{N}} \left( \vphantom{\prod_{\substack{i=1\\ i\neq k,k+1}}^{K+1}}
\right. } {}\times \prod_{t=\ell _{k}-\max (-h_{k},0)+1}^{\ell _{k}+\max
(h_{k},0)}\frac{\sqrt{\lambda _{k}\lambda _{k+1}}^{\kappa _{t}}}{\kappa _{t}!%
}\exp \left\{ -\frac{\lambda _{k}+\lambda _{k+1}}{2}\right\}   \notag \\
& \quad \hphantom{ \sum_{\kappa_1 \in \mathbb{N}} \ldots \sum_{\kappa_T\in
\mathbb{N}} \left( \vphantom{\prod_{\substack{i=1\\ i\neq k,k+1}}^{K+1}}
\right. } {}\times \left. \prod_{t=\ell _{k}+\max (h_{k},0)+1}^{\ell _{k+1}}%
\frac{\lambda _{k+1}^{\kappa _{t}}}{\kappa _{t}!}\exp (-\lambda _{k+1})%
\vphantom{\prod_{\substack{i=1\\ i\neq k,k+1}}^{K+1}}\right)   \notag \\
& =\rho ^{\left\vert h_{k}\right\vert }(\lambda _{k},\lambda _{k+1})
\label{eqn/zetacute_D3}
\end{align}%
using once again the same manipulations as those leading to (\ref%
{eqn/zetacute_UT}).

\subsubsection{Derivation of $\protect\zeta (\boldsymbol{h}_{k},\boldsymbol{h%
}_{l}^{\prime })$ and final expressions of $\boldsymbol{V}_{22}$ and $%
\boldsymbol{P}_{22}$}

In order to obtain closed-form expressions of $\zeta (\boldsymbol{h}_{k},%
\boldsymbol{h}_{l}^{\prime })$, we use the expressions obtained for $\pi (%
\mathbf{\lambda },\boldsymbol{\ell },\boldsymbol{h}_{k},\boldsymbol{h}%
_{l}^{\prime })$ and $\acute{\zeta}(\mathbf{\lambda },\boldsymbol{\ell },%
\boldsymbol{h}_{k},\boldsymbol{h}_{l}^{\prime })$ in the three cases

\begin{description}
\item[\textit{i)}] UT, using (\ref{eqn/pi_expanded}) with $\mathcal{J}%
_{h_{k},h_{l}^{\prime }}$ given by (\ref{eqn/J_UT}), and (\ref%
{eqn/zetacute_UT}),

\item[\textit{ii)}] FSD, using (\ref{eqn/pi_expanded}) with $\mathcal{J}%
_{h_{k},h_{k+1}^{\prime }}$ given by (\ref{eqn/J_FSD}), and either (\ref%
{eqn/zetacute_FSD_overlap}) or (\ref{eqn/zetacute_FSD_nooverlap}), depending
on whether there is overlap or not, respectively (see condition (\ref%
{eqn/condition_overlap})),

\item[\textit{iii)}] D, using (\ref{eqn/pi_expanded}), with $\mathcal{J}%
_{h_{k},h_{k}^{\prime }}$ given by (\ref{eqn/J_D}), and either (\ref%
{eqn/zetacute_D1}), (\ref{eqn/zetacute_D2}) or (\ref{eqn/zetacute_D3})
depending on whether $h_{k}^{\prime }=h_{k}$, $h_{k}^{\prime }=-h_{k}$ or $%
h_{k}^{\prime }=0$.
\end{description}

We then plug them into (\ref{eqn/zeta_pi_zetacute}) to obtain the final
expression of $\zeta (\boldsymbol{h}_{k},\boldsymbol{h}_{l}^{\prime })$.
Finally, the expressions of $\boldsymbol{V}_{22}$ and $\boldsymbol{P}_{22}$
are obtained by using (\ref{eqn/V22andP22zeta}). Let us give these details
in each of the three aforementioned cases UT, FSD and D.

\paragraph*{i) Case UT ($l>k+1$) and derivation of the upper-triangle terms
of $\boldsymbol{P}_{22}$}

Notice first that this case enable us to derive the upper-triangle terms of $%
\boldsymbol{P}_{22}$, according to (\ref{eqn/V22andP22zeta}). As just
explained, by plugging (\ref{eqn/pi_expanded}) (in which $\mathcal{J}%
_{h_{k},h_{l}^{\prime }}$ is given by (\ref{eqn/J_UT})) and (\ref%
{eqn/zetacute_UT}) into (\ref{eqn/zeta_pi_zetacute}), we obtain%
\begin{align}
\zeta (\boldsymbol{h}_{k},\boldsymbol{h}_{l}^{\prime })& =\sum_{\boldsymbol{%
\ell }\in \mathbb{Z}^{K}}\int_{\mathbb{R}_{+}^{K+1}}\left( \prod_{i=1}^{K+1}%
\frac{\beta ^{\alpha _{i}}}{\Gamma \left( \alpha _{i}\right) }\lambda
_{i}^{\alpha _{i}-1}\exp \left( -\beta \lambda _{i}\right) \right) \frac{1}{%
\tau ^{K}}\mathbb{I}_{\mathcal{J}_{h_{k},h_{l}^{\prime }}}\left( \boldsymbol{%
\ell }\right) \rho ^{\left\vert h_{k}\right\vert }(\lambda _{k},\lambda
_{k+1})\rho ^{\left\vert h_{l}^{\prime }\right\vert }(\lambda _{l},\lambda
_{l+1})\mathrm{d}\mathbf{\lambda }  \notag \\
& =\frac{1}{\tau ^{K}}\left( \sum_{\boldsymbol{\ell }\in \mathbb{Z}^{K}}%
\mathbb{I}_{\mathcal{J}_{h_{k},h_{l}^{\prime }}}\left( \boldsymbol{\ell }%
\right) \right) \frac{\beta ^{\alpha _{k}+\alpha _{k+1}}}{\Gamma \left(
\alpha _{k}\right) \Gamma \left( \alpha _{k+1}\right) }
\frac{\beta ^{\alpha _{l}+\alpha _{l+1}}}{\Gamma \left(
\alpha _{l}\right) \Gamma \left( \alpha _{l+1}\right) } \notag \\
& \quad \times \left( \int_{\mathbb{R%
}_{+}^{2}}\lambda _{k}^{\alpha _{k}-1}\lambda _{k+1}^{\alpha _{k+1}-1}\exp
\left\{ -\beta \left( \lambda _{k}+\lambda _{k+1}\right) -|h_{k}|\frac{%
\left( \sqrt{\lambda _{k+1}}-\sqrt{\lambda _{k}}\right) ^{2}}{2}\right\} 
\mathrm{d}\lambda _{k}\mathrm{d}\lambda _{k+1}\right)   \notag \\
& \quad \times \left( \int_{\mathbb{R%
}_{+}^{2}}\lambda _{l}^{\alpha _{l}-1}\lambda _{l+1}^{\alpha _{l+1}-1}\exp
\left\{ -\beta \left( \lambda _{l}+\lambda _{l+1}\right) -|h_{l}^{\prime }|%
\frac{\left( \sqrt{\lambda _{l+1}}-\sqrt{\lambda _{l}}\right) ^{2}}{2}%
\right\} \mathrm{d}\lambda _{l}\mathrm{d}\lambda _{l+1}\right)   \notag \\
& =\frac{\mathrm{Card}\left( \mathcal{J}_{h_{k},h_{l}^{\prime }}\right) }{%
\tau ^{K}}\Phi \left( h_{k}\right) \Phi \left( h_{l}^{\prime }\right) 
\label{eqn/zeta_UT}
\end{align}%
in which $\Phi \left( .\right) $ is defined in (\ref{eqn/def_Phi}), and $%
\mathrm{Card}\left( \mathcal{J}_{h_{k},h_{l}^{\prime }}\right) $ denotes the
cardinality of the set $\mathcal{J}_{h_{k},h_{l}^{\prime }}$. By analyzing Table \ref{tab/ranges_elli_UT}, we can show that
\begin{align}
\frac{\mathrm{Card}\left( \mathcal{J}_{h_{k},h_{l}^{\prime }}\right) }{\tau
^{K}} &=u\left( \tau ,h_{k}\right) u\left( \tau ,h_{l}\right)   \notag \\
&=\frac{\mathrm{Card}\left( \mathcal{J}_{-h_{k},-h_{l}}\right) }{\tau ^{K}}=%
\frac{\mathrm{Card}\left( \mathcal{J}_{-h_{k},h_{l}}\right) }{\tau ^{K}}=%
\frac{\mathrm{Card}\left( \mathcal{J}_{h_{k},-h_{l}}\right) }{\tau ^{K}}
\label{eqn/CardJtauK_UT}
\end{align}%
where $u\left(\tau, .\right) $ is defined in (\ref{eqn/def_u}). Then, plugging
(\ref{eqn/zeta_UT}) into (\ref{eqn/V22andP22zeta}), we obtain%
\begin{align}
\left[ \boldsymbol{P}_{22}\right] _{k,l} &=\zeta \left( \boldsymbol{h}_{k},%
\boldsymbol{h}_{l}\right) +\zeta \left( -\boldsymbol{h}_{k},-\boldsymbol{h}%
_{l}\right) -\zeta \left( -\boldsymbol{h}_{k},\boldsymbol{h}_{l}\right)
-\zeta \left( \boldsymbol{h}_{k},-\boldsymbol{h}_{l}\right)   \notag \\
&=0  \label{eqn/P22UT}
\end{align}%
since neither $\Phi \left( h_{k}\right) $ nor $\mathrm{Card}\left( \mathcal{J}_{h_{k},h_{l}^{\prime }}\right)$ depend on the signs of $h_{k}$ and $h'_{l}$.
Equation (\ref{eqn/P22UT}) finally gives us the tridiagonal structure of matrix $\boldsymbol{P}_{22}$, that appears in (\ref{eqn/P22_final}).

\paragraph*{ii) Case FSD ($l=k+1$) and derivation of the first superdiagonal
                terms of $\boldsymbol{P}_{22}$}

Let us first remark that the expression of $[\boldsymbol{P}_{22}]_{k,k+1}$ in
(\ref{eqn/V22andP22zeta}) can be rewritten as
\begin{equation}
[\boldsymbol{P}_{22}]_{k,k+1} = \mathrm{sign}(h_k h_{k+1})
                                \bigl[
                                \zeta (\abs{h_k},\abs{h_{k+1}})
                                + \zeta (-\abs{h_k},-\abs{h_{k+1}})
                                - \zeta (-\abs{h_k},\abs{h_{k+1}})
                                - \zeta (\abs{h_k},-\abs{h_{k+1}})
                                \bigr].
\label{eqn/P22FSD}
\end{equation}
This writing enables us to dispose of considerations on the signs of $h_k$ and
$h_{k+1}$ in order to determine in which term of (\ref{eqn/V22andP22zeta}),
a possible overlap has to be taken into account. Here, this has to be done
only for the last term $\zeta (\abs{h_k},-\abs{h_{k+1}})$, whatever the signs
of $h_k$ and $h_{k+1}$.

Let us first derive $\zeta (h_k,h'_{k+1})$ in the case without overlap, i.e.,
when the condition (\ref{eqn/condition_overlap}) is \emph{not} met. This will
give us the expression for the three first terms of (\ref{eqn/P22FSD}), and
possibly the fourth if $\min(\abs{h_k},\abs{h_{k+1}}) = 1$. As already
explained, by plugging (\ref{eqn/pi_expanded}) (in which $\mathcal{J}%
_{h_{k},h_{l}^{\prime }}$ is given by (\ref{eqn/J_FSD})) and (\ref%
{eqn/zetacute_FSD_nooverlap}) into (\ref{eqn/zeta_pi_zetacute}), we obtain, in
the same way as in the case UT:
\begin{align}
\zeta(\hb_k,\hb'_{k+1})
&= \sum_{\ellb \in \mathbb{Z}^{K}} \int_{\mathbb{R}_{+}^{K+1}}
   \left(
   \prod_{i=1}^{K+1}
   \frac{\beta^{\alpha_i}}{\Gamma(\alpha_i)}
   \lambda_i^{\alpha_i-1}\exp(-\beta \lambda_i)
   \right)
   \frac{1}{\tau^K}\indicf{\mathcal{J}_{h_k,h'_{k+1}}}{\ellb}
   \rho^{\abs{h_k}}(\lambda_k,\lambda_{k+1})
   \rho^{\abs{h'_{k+1}}}(\lambda_{k+1},\lambda_{k+2})
   \diff\lambdab \notag \\
&= \frac{1}{\tau^K}
   \left(
   \sum_{\ellb \in \mathbb{Z}^{K}}
   \indicf{\mathcal{J}_{h_k,h'_{k+1}}}{\ellb}
   \right)
   \frac{\beta^{\alpha_k+\alpha_{k+1}+\alpha_{k+2}}}
        {\Gamma(\alpha_k)\Gamma(\alpha_{k+1})\Gamma(\alpha_{k+2})}
   \notag \\
&  \begin{aligned}
   \quad \: \times
   \int_{\mathbb{R}_{+}^{3}}
   \lambda_k^{\alpha_k-1}
   \lambda_{k+1}^{\alpha_{k+1}-1}
   \lambda_{k+2}^{\alpha_{k+2}-1}
   \exp\Biggl\{
       &{-}\beta (\lambda_k+\lambda_{k+1}+\lambda_{k+2})
        -\abs{h_k}\frac{(\sqrt{\lambda_{k+1}}-\sqrt{\lambda_k})^2}{2}
        \notag \\
       &{-} \abs{h'_{k+1}}
            \frac{(\sqrt{\lambda_{k+2}}-\sqrt{\lambda_{k+1}})^2}{2}
       \Biggr\}
   \diff\lambda_k\diff\lambda_{k+1}\diff\lambda_{k+2}
   \end{aligned}
   \notag \\
&= \frac{\mathrm{Card}\left(\mathcal{J}_{h_k,h'_{k+1}}\right)}{\tau^K}
   \frac{\beta^{\alpha_k+\alpha_{k+1}+\alpha_{k+2}}}
        {\Gamma(\alpha_k)\Gamma(\alpha_{k+1})\Gamma(\alpha_{k+2})}
   \int_{\mathbb{R}_{+}^{3}}
   \phi_{h_k,h'_{k+1}}(\lambdab_{k:k+2})
   \diff\lambda_k\diff\lambda_{k+1}\diff\lambda_{k+2}
\label{eqn/zeta_FSD_nooverlap_0}
\end{align}
where $\phi_{h_k,h'_{k+1}}(.)$ is defined in (\ref{eqn/def_phi}). By reading
Table \ref{tab/ranges_elli_FSD}, we can deduce that
\begin{multline}
\hspace{-1em}
\frac{\mathrm{Card}\left(\mathcal{J}_{h_k,h'_{k+1}}\right)}{\tau^K}
= \\
\begin{cases}
\multicolumn{1}{c}{
\dfrac{(\tau-\abs{h_k})
       \bigl(\tau-\max(-h_k,h'_{k+1},0)-\max(h_k,-h'_{k+1},0)\bigr)
       (\tau-\abs{h'_{k+1}})}
     {\tau^3},} & \begin{minipage}[t]{0.3\textwidth}
                if  $k+1\leq K-1$ \\
                and $\max(\abs{h_k},\abs{h'_{k+1}})\leq \tau -1$,
                \end{minipage} \\
\multicolumn{1}{c}{
\dfrac{(\tau-\abs{h_{K-1}})
       \bigl(\tau-\max(-h_{K-1},h'_{K},0)-\max(h_{K-1},-h'_{K},0)\bigr)}
     {\tau^2},} & \begin{minipage}[t]{0.3\textwidth}
                if $k+1 = K$ \\
                and $\max(\abs{h_{K-1}},\abs{h'_{K}})\leq \tau -1$,
                \end{minipage} \\
\multicolumn{1}{c}{0, \vphantom{\dfrac{\bigl(}{\tau^2}}}
                & \text{if } \max(\abs{h_{K-1}},\abs{h'_{K}})\geq \tau.
\end{cases}
\end{multline}
Finally, the results for the first three terms in (\ref{eqn/P22FSD}) are,
for $k<K-1$:
\begin{equation}
\begin{split}
\zeta(\abs{h_k},\abs{h_{k+1}})
= \zeta(-{\abs{h_k}},-{\abs{h_{k+1}}})
&= \frac{(\tau-\abs{h_k})
        (\tau-\abs{h_k}-\abs{h_{k+1}})
        (\tau-\abs{h_{k+1}})}
       {\tau^3}
  \\
& \quad \: \times
  \frac{\beta^{\alpha_k+\alpha_{k+1}+\alpha_{k+2}}}
        {\Gamma(\alpha_k)\Gamma(\alpha_{k+1})\Gamma(\alpha_{k+2})}
  \int_{\mathbb{R}_{+}^{3}}
  \phi_{\abs{h_k},\abs{h_{k+1}}}(\lambdab_{k:k+2})
  \diff\lambdab_{k:k+2}
\end{split}
\label{eqn/zeta_FSD_nooverlap_kltK-1_1} 
\end{equation}
\begin{equation}
\begin{split}
\zeta(-{\abs{h_k}},\abs{h_{k+1}})
&= \frac{(\tau-\abs{h_k})
        \bigl(\tau-\max(\abs{h_k},\abs{h_{k+1}})\bigr)
        (\tau-\abs{h_{k+1}})}
       {\tau^3} \\
& \quad \: \times
  \frac{\beta^{\alpha_k+\alpha_{k+1}+\alpha_{k+2}}}
        {\Gamma(\alpha_k)\Gamma(\alpha_{k+1})\Gamma(\alpha_{k+2})}
  \int_{\mathbb{R}_{+}^{3}}
  \phi_{\abs{h_k},\abs{h_{k+1}}}(\lambdab_{k:k+2})
  \diff\lambdab_{k:k+2}
\end{split}
\label{eqn/zeta_FSD_nooverlap_kltK-1_2}
\end{equation}
and for $k=K-1$, (\ref{eqn/zeta_FSD_nooverlap_kltK-1_1}) and
(\ref{eqn/zeta_FSD_nooverlap_kltK-1_2}) become respectively
\begin{equation}
\begin{split}
\zeta(\abs{h_{K-1}},\abs{h_{K}})
= \zeta(-{\abs{h_{K-1}}},-{\abs{h_{K}}})
&= \frac{(\tau-\abs{h_{K-1}})
        (\tau-\abs{h_{K-1}}-\abs{h_{K}})}
       {\tau^2}
  \\
& \quad \: \times
  \frac{\beta^{\alpha_{K-1}+\alpha_{K}+\alpha_{K+1}}}
        {\Gamma(\alpha_{K-1})\Gamma(\alpha_{K})\Gamma(\alpha_{K+1})}
  \int_{\mathbb{R}_{+}^{3}}
  \phi_{\abs{h_{K-1}},\abs{h_{K}}}(\lambdab_{K-1:{K+1}})
  \diff\lambdab_{K-1:K+1} 
\end{split}
\raisetag{36pt}
\label{eqn/zeta_FSD_nooverlap_keqK-1_1} 
\end{equation}
and
\begin{equation}
\begin{split}
\zeta(-{\abs{h_{K-1}}},\abs{h_{K}})
&= \frac{(\tau-\abs{h_{K-1}})
        \bigl(\tau-\max(\abs{h_{K-1}},\abs{h_{K}})\bigr)}
       {\tau^2} \\
& \quad \: \times
  \frac{\beta^{\alpha_{K-1}+\alpha_{K}+\alpha_{K+1}}}
        {\Gamma(\alpha_{K-1})\Gamma(\alpha_{K})\Gamma(\alpha_{K+1})}
  \int_{\mathbb{R}_{+}^{3}}
  \phi_{\abs{h_{K-1}},\abs{h_{K}}}(\lambdab_{K-1:K+1})
  \diff\lambdab_{K-1:K+1} 
\end{split}
\label{eqn/zeta_FSD_nooverlap_keqK-1_2}
\end{equation}
If there is no overlap (i.e., $\min(\abs{h_k},\abs{h_{k+1}}) = 1$), the
fourth term in (\ref{eqn/P22FSD}) is also given by
(\ref{eqn/zeta_FSD_nooverlap_kltK-1_2}) or
(\ref{eqn/zeta_FSD_nooverlap_keqK-1_2}), i.e., we have
\begin{equation}
\zeta(\abs{h_k},-{\abs{h_{k+1}}})=\zeta(-{\abs{h_k}},\abs{h_{k+1}}).
\end{equation}

    This term has a different writing if there is an overlap, i.e., when
$\min(\abs{h_k},\abs{h_{k+1}}) > 1$. By referring to the overlap condition~%
(\ref{eqn/overlap_condition_formal}), it appears that for values of
$\ell_{k+1}$ that satisfy $\ell_{k+1}<\ell_k+\abs{h_k}+\abs{h_{k+1}}$,
the quantity $\acute{\zeta}(\lambdab,\ellb,\abs{\hb_k},-{\abs{\hb_{k+1}}})$
has to be written according
to (\ref{eqn/zetacute_FSD_overlap}), whereas for values of $\ell_{k+1}$ such
that $\ell_{k+1}\geq\ell_k+\abs{h_k}+\abs{h_{k+1}}$, the quantity
$\acute{\zeta}(\lambdab,\ellb,\abs{\hb_k},-{\abs{\hb_{k+1}}})$
has to be written according to (\ref{eqn/zetacute_FSD_nooverlap}).
Then, the sum in
(\ref{eqn/zeta_pi_zetacute}) w.r.t. $\ell_{k+1}$ has to be split into two
parts, the first of which contains the terms with overlap, i.e., for
$\ell_{k+1}\in\{\ell_k+\max(\abs{h_k},\abs{h_{k+1}})+1,
                \ldots,
                \ell_k+\abs{h_k}+\abs{h_{k+1}}-1\}$, whereas the second
part contains the terms without overlap, i.e., for
$\ell_{k+1}\in\{\ell_k+\abs{h_k}+\abs{h_{k+1}},\ldots,\ell_k+\tau\}$. More
explicitly, we rewrite (\ref{eqn/zeta_pi_zetacute}) in the following way
\begin{align}
\zeta (\abs{h_k},-{\abs{h_{k+1}}}) &=
\sum_{\boldsymbol{\ell }\in \mathbb{Z}^{K}}\int_{\mathbb{R}_{+}^{K+1}}\pi
(\mathbf{\lambda },\boldsymbol{\ell },\boldsymbol{h}_{k},\hb'_{k+1})\;\acute{\zeta}(\mathbf{\lambda },\boldsymbol{\ell },%
\boldsymbol{h}_{k},\hb'_{k+1})\,\mathrm{d}\mathbf{\lambda }
\notag \\
&=
\int_{\mathbb{R}_{+}^{K+1}}
\left[
\prod_{i=1}^{K+1}
\frac{\beta ^{\alpha _{i}}}
     {\Gamma ( \alpha_{i}) }
     \lambda _{i}^{\alpha _{i}-1}
     \exp ( -\beta \lambda_{i})
\left(
\sum_{\ellb_{1:k}\in\mathbb{Z}^{k}}
\bigl[
S_1 (\lambdab,\ellb_{1:k},h_k,h_{k+1})
+ S_2 (\lambdab,\ellb_{1:k},h_k,h_{k+1})
\bigr]
\right)
\right]
\diff\lambdab
\label{eqn/zeta_FSD_overlapS1S2}
\end{align}
in which, for $i,j = 1,\ldots, K$, $\ellb_{i:j}$ denotes the truncated
vector $[\ell_i,\ldots ,\ell_j]^T$,
the quantity $S_1(\lambdab,\ellb_{1:k},h_k,h_{k+1})$ is
\begin{equation}
S_1(\lambdab,\ellb_{1:k},h_k,h_{k+1}) =
\sum_{\ell_{k+1} = (\ell_{k+1})_{\min}}
    ^{\ell_k+\abs{h_k}+\abs{h_{k+1}}-1}
\left[
\sum_{\ellb_{k+2:K}\in\mathbb{Z}^{K-k-1}}
\frac{1}{\tau ^{K}}
\mathbb{I}_{\mathcal{J}_{\vert h_{k}\vert,-{\vert h_{k+1}\vert}}}(\ellb ) \;
\acute{\zeta}(\lambdab,\ellb,\abs{h_{k}},-{\abs{h_{k+1}}})
\right]
\label{eqn/def_S1}
\end{equation}
in which $\acute{\zeta}(\lambdab,\ellb,\abs{h_{k}},-{\abs{h_{k+1}}})$ is
expressed according to (\ref{eqn/zetacute_FSD_overlap_r}), and
$S_2 (\lambdab,\ellb_{1:k},h_k,h_{k+1})$ denotes
\begin{equation}
S_2(\lambdab,\ellb_{1:k},h_k,h_{k+1}) =
\sum_{\ell_{k+1} = \ell_k+\abs{h_k}+\abs{h_{k+1}}}
    ^{\ell_k+\tau}
\left[
\sum_{\ellb_{k+2:K}\in\mathbb{Z}^{K-k-1}}
\frac{1}{\tau ^{K}}
\mathbb{I}_{\mathcal{J}_{\vert h_{k}\vert,-{\vert h_{k+1}\vert}}}(\ellb ) \;
\acute{\zeta}(\lambdab,\ellb,\abs{h_{k}},-{\abs{h_{k+1}}})
\right]
\label{eqn/def_S2}
\end{equation}
in which $\acute{\zeta}(\lambdab,\ellb,\abs{h_{k}},-{\abs{h_{k+1}}})$ is
expressed according to (\ref{eqn/zetacute_FSD_nooverlap}).

For the subsequent derivation, we will use the following rewriting of
$\mathcal{J}_{\vert h_{k}\vert,-{\vert h_{k+1}\vert}}$,
for $k<K-1$:
\begin{equation}
\mathcal{J}_{\vert h_{k}\vert,-{\vert h_{k+1}\vert}}
= \tilde{\mathcal{J}}_{1:k  ,\vert h_{k  }\vert}
  \times
  \{\ell_{k}+\max(\abs{h_{k}},\abs{h_{k+1}})+1,\ldots ,\ell_k +\tau \}
  \times
  \tilde{\mathcal{J}}_{k+2:K,-\vert h_{k+1}\vert}
\label{eqn/Jsplit_kltK-1}
\end{equation}
with
\begin{equation}
\tilde{\mathcal{J}}_{1:k  ,\vert h_{k  }\vert}
= \left(
  \prod_{i=1}^{k-1}
  \{ \ell _{i-1}+1,\ldots ,\ell _{i-1}+\tau\}
  \right)
  \times
  \{ \ell _{k-1}+1,\ldots ,\ell _{k-1}-\abs{h_{k}}+\tau\}
\label{eqn/def_Jtilde1}
\end{equation}
and
\begin{equation}
\tilde{\mathcal{J}}_{k+2:K,-\vert h_{k+1}\vert}
= \{ \ell_{k+1}+1,\ldots ,\ell_{k+1}-\abs{h_{k+1}}+\tau\}
  \times
  \left(
  \prod_{i=k+3}^{K}
  \{ \ell _{i-1}+1,\ldots ,\ell _{i-1}+\tau\}
  \right).
\label{eqn/def_Jtilde2}
\end{equation}
In the same manner, for $k=K-1$, we will rewrite
$\mathcal{J}_{\vert h_{k}\vert,-{\vert h_{k+1}\vert}}$
as
\begin{equation}
\mathcal{J}_{\vert h_{K-1}\vert,-{\vert h_{K}\vert}}
= \tilde{\mathcal{J}}_{1:K-1  ,\vert h_{K-1}\vert}
  \times
  \{\ell_{K-1}+\max(\abs{h_{K-1}},\abs{h_{K}})+1,\ldots ,\ell_{K-1} +\tau \}.
\label{eqn/Jsplit_keqK-1}
\end{equation}

Let us first develop (\ref{eqn/def_S1}), for $k < K-1$, by splitting
$\mathcal{J}_{\vert h_{k}\vert,-{\vert h_{k+1}\vert}}$
according to (\ref{eqn/Jsplit_kltK-1})
\begin{align}
S_1(\lambdab,\ellb_{1:k},h_k,h_{k+1})
&= \frac{1}{\tau ^{K}}
   \frac{\rho^{\abs{h_{k}}  } (\lambda_{k}  ,\lambda_{k+1}) \,
         \rho^{\abs{h_{k+1}}} (\lambda_{k+1},\lambda_{k+2})}
        {\bigl(r(\lambdab_{k:k+2})\bigr)^{\ell_k+\abs{h_{k+1}}+\abs{h_{k}}}}
   \notag \\
&  \quad \: \times
   \sum_{\ell_{k+1}=\ell_k+\max(\abs{h_k},\abs{h_{k+1}})+1}
       ^{\ell_k+\abs{h_k}+\abs{h_{k+1}}-1}
   \left(
   \bigl(r(\lambdab_{k:k+2})\bigr)^{\ell_{k+1}}
   \sum_{\ellb_{k+2:K}\in\mathbb{Z}^{K-k-1}}
   \mathbb{I}_{\mathcal{J}_{\vert h_{k}\vert,-{\vert h_{k+1}\vert}}}(\ellb )
   \right)
   \notag \\
&= \frac{\mathrm{Card}
         \left( \tilde{\mathcal{J}}_{k+2:K,-\vert h_{k+1}\vert}\right)}
        {\tau ^{K}}
   \frac{\rho^{\abs{h_{k}}  } (\lambda_{k}  ,\lambda_{k+1}) \,
         \rho^{\abs{h_{k+1}}} (\lambda_{k+1},\lambda_{k+2})}
        {\bigl(r(\lambdab_{k:k+2})\bigr)^{\ell_k+\abs{h_{k+1}}+\abs{h_{k}}}}\,
   \mathbb{I}_{\tilde{\mathcal{J}}_{1:k,\vert h_{k}\vert}}(\ellb_{1:k})
   \notag \\
&  \quad \: \times
   \sum_{\ell_{k+1}=\ell_k+\max(\abs{h_k},\abs{h_{k+1}})+1}
       ^{\ell_k+\abs{h_k}+\abs{h_{k+1}}-1}
   \hspace{-2em}
   \bigl(r(\lambdab_{k:k+2})\bigr)^{\ell_{k+1}}
   \notag \\
&= \frac{\mathrm{Card}
         \left( \tilde{\mathcal{J}}_{k+2:K,-\vert h_{k+1}\vert}\right)}
        {\tau ^{K}}
   \frac{\rho^{\abs{h_{k}}  } (\lambda_{k}  ,\lambda_{k+1}) \,
         \rho^{\abs{h_{k+1}}} (\lambda_{k+1},\lambda_{k+2})}
        {\bigl(
         r(\lambdab_{k:k+2})
         \bigr)^{\ell_k+\abs{h_{k+1}}+\abs{h_{k}}}} \,
   \mathbb{I}_{\tilde{\mathcal{J}}_{1:k,\vert h_{k}\vert}}(\ellb_{1:k}) \,
   \bigl(r(\lambdab_{k:k+2})\bigr)^{\ell_k+\max(\abs{h_k},\abs{h_{k+1}})+1}
   \notag \\
&  \quad \: \times
   \frac{1-\bigl(r(\lambdab_{k:k+2})\bigr)^{\abs{h_k}+\abs{h_{k+1}}
                                            -\max(\abs{h_k},\abs{h_{k+1}})-1}}
        {1-r(\lambdab_{k:k+2})}, \qquad
   \begin{minipage}[t]{0.35\textwidth}
   (by summing the terms of the geometric series with common ratio
   $r(\lambdab_{k:k+2})$)
   \end{minipage}
   \notag \\
&= \frac{\mathrm{Card}
         \left( \tilde{\mathcal{J}}_{k+2:K,-\vert h_{k+1}\vert}\right)}
        {\tau ^{K}}
   \rho^{\abs{h_{k}}  } (\lambda_{k}  ,\lambda_{k+1}) \,
   \rho^{\abs{h_{k+1}}} (\lambda_{k+1},\lambda_{k+2}) \,
   \bigl(r(\lambdab_{k:k+2})\bigr)^{1-\min(\abs{h_k},\abs{h_{k+1}})}
   \notag \\
&  \quad \: \times
   \frac{1-\bigl(r(\lambdab_{k:k+2})\bigr)^{\min(\abs{h_k},\abs{h_{k+1}})-1}}
        {1-r(\lambdab_{k:k+2})} \,
   \mathbb{I}_{\tilde{\mathcal{J}}_{1:k,\vert h_{k}\vert}}(\ellb_{1:k}).
   \label{eqn/S1_expanded}
\end{align}
According to (\ref{eqn/def_Jtilde2}), we have
\begin{equation}
\mathrm{Card}
\left( \tilde{\mathcal{J}}_{k+2:K,-\vert h_{k+1}\vert}\right)
= \begin{cases}
  (\tau - \abs{h_{k+1}})\tau^{K-k-2}, & \text{if } \abs{h_{k+1}}\leq\tau-1 \\
  0,                                  & \text{if } \abs{h_{k+1}}\geq\tau
  \end{cases}
\end{equation}
so $S_1$ is given, for $k<K-1$, and provided that
$\abs{h_{k+1}}\leq\tau-1$, by
\begin{equation}
\mspace{-1mu}
S_1(\lambdab,\ellb_{1:k},h_k,h_{k+1})
= \frac{\bigl(\tau \!-\!\abs{h_{k+1}}\bigr)}{\tau ^{k+2}}
  \rho^{\abs{h_{k}}  } (\lambda_{k}  ,\lambda_{k+1}) \,
  \rho^{\abs{h_{k+1}}} (\lambda_{k+1},\lambda_{k+2}) \,
  \frac{\bigl(r(\lambdab_{k:k+2})\bigr)^{1-\min(\abs{h_k},\abs{h_{k+1}})}
        \!\!-1}
       {1-r(\lambdab_{k:k+2})}\,
  \indicf{\tilde{\mathcal{J}}_{1:k,\vert h_{k}\vert\!\!}}{\ellb_{1:k}}
  \mspace{-12mu}
\label{eqn/S1_final_1}
\end{equation}
and if $k=K-1$, by splitting
$\mathcal{J}_{\vert h_{K-1}\vert,-{\vert h_{K}\vert}}$
according to (\ref{eqn/Jsplit_keqK-1}),
we obtain, by similar manipulations as those leading to
(\ref{eqn/S1_expanded})
\begin{equation}
\begin{split}
S_1(\lambdab,\ellb_{1:K-1},h_{K-1},h_{K})
&= \frac{1}{\tau^{K}}
   \rho^{\abs{h_{K-1}}} (\lambda_{K-1},\lambda_{K}) \,
   \rho^{\abs{h_{K}}  } (\lambda_{K  },\lambda_{K+1}) \,
   \frac{\bigl(r(\lambdab_{K-1:K+1})\bigr)^{1-\min(\abs{h_{K-1}},\abs{h_{K}})}
         \!\!-1}
        {1-r(\lambdab_{K-1:K+1})}\\
&  \quad \times \:
   \indicf{\tilde{\mathcal{J}}_{1:K-1,\vert h_{K-1}\vert\!\!}}{\ellb_{1:K-1}}.
\end{split}
\label{eqn/S1_final_2}
\end{equation}

Let us now develop $S_2$
from (\ref{eqn/def_S2}) where, as already mentioned,
$\acute{\zeta}(\lambdab,\ellb,\abs{h_{k}},-{\abs{h_{k+1}}})$ is given by
(\ref{eqn/zetacute_FSD_nooverlap}). For $k<K-1$,
\begin{align}
S_2(\lambdab,\ellb_{1:k},h_k,h_{k+1})
&= \sum_{\ell_{k+1} = \ell_k+\abs{h_k}+\abs{h_{k+1}}}
   ^{\ell_k+\tau}
   \left[
   \sum_{\ellb_{k+2:K}\in\mathbb{Z}^{K-k-1}}
   \frac{1}{\tau ^{K}}
   \mathbb{I}_{\mathcal{J}_{\vert h_{k}\vert,-{\vert h_{k+1}\vert}}}(\ellb)\;
   \rho^{\left| h_k\right|} (\lambda_k,\lambda_{k+1})
   \rho^{\left| h_{k+1}\right|}(\lambda_{k+1},\lambda_{k+2})
   \right] \notag \\
&= \frac{1}{\tau ^{K}}
   \rho^{\left| h_k\right|} (\lambda_k,\lambda_{k+1})
   \rho^{\left| h_{k+1}\right|}(\lambda_{k+1},\lambda_{k+2})
   \sum_{\ell_{k+1} = \ell_k+\abs{h_k}+\abs{h_{k+1}}}^{\ell_k+\tau}
   \left[
   \sum_{\ellb_{k+2:K}\in\mathbb{Z}^{K-k-1}}
   \mathbb{I}_{\mathcal{J}_{\vert h_{k}\vert,-{\vert h_{k+1}\vert}}}(\ellb)
   \right] \notag \\
&= \frac{\mathrm{Card}
         \left( \tilde{\mathcal{J}}_{k+2:K,-\vert h_{k+1}\vert}\right)
         \max\bigl( \tau-\abs{h_k}-\abs{h_{k+1}}+1,0 \bigr)}
        {\tau ^{K}}
   \rho^{\left| h_k\right|} (\lambda_k,\lambda_{k+1})
   \rho^{\left| h_{k+1}\right|}(\lambda_{k+1},\lambda_{k+2})
   \indicf{\tilde{\mathcal{J}}_{1:k,\vert h_{k}\vert}}{\ellb_{1:k}}
   \notag \\
&= \frac{\bigl( \tau-\abs{h_{k+1}}             \bigr)
         \bigl( \tau-\abs{h_k}-\abs{h_{k+1}}+1 \bigr)}
        {\tau^{k+2}}
   \rho^{\left| h_k\right|} (\lambda_k,\lambda_{k+1})
   \rho^{\left| h_{k+1}\right|}(\lambda_{k+1},\lambda_{k+2})
   \indicf{\tilde{\mathcal{J}}_{1:k,\vert h_{k}\vert}}{\ellb_{1:k}},
\label{eqn/S2_final_1}
\end{align}
provided that $\tau-\abs{h_k}-\abs{h_{k+1}}+1 > 0$ and
$\abs{h_{k+1}}\leq\tau-1$.

If $k=K-1$, we obtain in the same way
\begin{equation}
S_2(\lambdab,\ellb_{1:K-1},h_{K-1},h_{K})
= \frac{\max\bigl( \tau-\abs{h_{K-1}}-\abs{h_{K}}+1,0 \bigr)}
       {\tau^K}
  \rho^{\left| h_{K-1}\right|} (\lambda_{K-1},\lambda_{K})
  \rho^{\left| h_{K  }\right|}(\lambda_{K},\lambda_{K+1})
  \indicf{\tilde{\mathcal{J}}_{1:K-1,\vert h_{K-1}\vert}}{\ellb_{1:K-1}}.
\label{eqn/S2_final_2}
\end{equation}
We then plug (\ref{eqn/S1_final_1}) and (\ref{eqn/S2_final_1}) into
(\ref{eqn/zeta_FSD_overlapS1S2}) to obtain, for $k<K-1$
\begin{align}
\zeta (\abs{h_k},-{\abs{h_{k+1}}})
&= \int_{\mathbb{R}_{+}^{K+1}}
   \left[
   \prod_{i=1}^{K+1}
   \frac{\beta ^{\alpha _{i}}}
        {\Gamma ( \alpha_{i}) }
   \lambda _{i}^{\alpha _{i}-1}
   \exp ( -\beta \lambda_{i})
   \left(
   \sum_{\ellb_{1:k}\in\mathbb{Z}^{k}}
   \frac{\bigl( \tau-\abs{h_{k+1}} \bigr)}
        {\tau^{k+2}}
   \rho^{\left| h_{k}  \right|}(\lambda_{k}  ,\lambda_{k+1})
   \rho^{\left| h_{k+1}\right|}(\lambda_{k+1},\lambda_{k+2})
   \right.
   \right.
   \notag \\
&  \phantom{= \int_{\mathbb{R}_{+}^{K+1}}\left[
              \vphantom{\prod_{i=1}^{K+1}
                        \left(
                        \sum_{\ellb_{1:k}\in\mathbb{Z}^{k}}
                        \right)}
                        \right.}
   \left.
   \left.
   \times
   \left(
   \frac{\bigl(r(\lambdab_{k:k+2})\bigr)^{1-\min(\abs{h_k},\abs{h_{k+1}})}
         -1}
       {1-r(\lambdab_{k:k+2})}
   + \max(\tau-\abs{h_k}-\abs{h_{k+1}}+1,0)
   \right)
   \indicf{\tilde{\mathcal{J}}_{1:k,\vert h_{k}\vert}}{\ellb_{1:k}}
   \vphantom{\sum_{\ellb_{1:k}\in\mathbb{Z}^{k}}}
   \!
   \right)
   \vphantom{\prod_{i=1}^{K+1}}
   \!
   \right]
   \diff\lambdab
   \notag \\
&= \frac{\mathrm{Card}
         \left( \tilde{\mathcal{J}}_{k+2:K,-\vert h_{k+1}\vert}\right)}
        {\tau^{K}}
   \int_{\mathbb{R}_{+}^{K+1}}
   \left[
   \prod_{i=1}^{K+1}
   \frac{\beta ^{\alpha _{i}}}
        {\Gamma ( \alpha_{i}) }
   \lambda _{i}^{\alpha _{i}-1}
   \exp ( -\beta \lambda_{i})
   \rho^{\left| h_{k}  \right|}(\lambda_{k}  ,\lambda_{k+1})
   \rho^{\left| h_{k+1}\right|}(\lambda_{k+1},\lambda_{k+2})
   \vphantom{\sum_{\ellb_{1:k}\in\mathbb{Z}^{k}}}
   \right.
   \notag \\
&  \phantom{= \frac{\mathrm{Card}
                    \left(
                    \tilde{\mathcal{J}}_{k+2:K,-\vert h_{k+1}\vert}
                    \right)}
                   {\tau^{K}}
              \int_{\mathbb{R}_{+}^{K+1}}\left[
              \vphantom{\prod_{i=1}^{K+1}
                        \left(
                        \sum_{\ellb_{1:k}\in\mathbb{Z}^{k}}
                        \right)}
                        \right.}
   \times
   \left(
   \frac{\bigl(r(\lambdab_{k:k+2})\bigr)^{1-\min(\abs{h_k},\abs{h_{k+1}})}
         -1}
       {1-r(\lambdab_{k:k+2})}
   + \max(\tau-\abs{h_k}-\abs{h_{k+1}}+1,0)
   \right)
   \notag \\
&  \phantom{= \frac{\mathrm{Card}
                    \left(
                    \tilde{\mathcal{J}}_{k+2:K,-\vert h_{k+1}\vert}
                    \right)}
                   {\tau^{K}}
              \int_{\mathbb{R}_{+}^{K+1}}\left[
              \vphantom{\prod_{i=1}^{K+1}
                        \left(
                        \sum_{\ellb_{1:k}\in\mathbb{Z}^{k}}
                        \right)}
                        \right.}
   \left.
   \times
   \sum_{\ellb_{1:k}\in\mathbb{Z}^{k}}
   \indicf{\tilde{\mathcal{J}}_{1:k,\vert h_{k}\vert}}{\ellb_{1:k}}
   \right]
   \diff \lambdab
   \notag \\
&= \frac{\mathrm{Card}
         \left( \tilde{\mathcal{J}}_{k+2:K,-\vert h_{k+1}\vert}\right)
         \mathrm{Card}
         \left( \tilde{\mathcal{J}}_{1:k,\vert h_{k}\vert}\right)}
        {\tau^{K}}
   \notag \\
&  \quad \: \times
   \int_{\mathbb{R}_{+}^{K+1}}
   \left[
   \prod_{i=1}^{K+1}
   \frac{\beta ^{\alpha _{i}}}
        {\Gamma ( \alpha_{i}) }
   \lambda _{i}^{\alpha _{i}-1}
   \exp ( -\beta \lambda_{i})
   \rho^{\left| h_{k}  \right|}(\lambda_{k}  ,\lambda_{k+1})
   \rho^{\left| h_{k+1}\right|}(\lambda_{k+1},\lambda_{k+2})
   \vphantom{\sum_{\ellb_{1:k}\in\mathbb{Z}^{k}}}
   \right.
   \notag \\
&  \phantom{\quad \: \times
            \int_{\mathbb{R}_{+}^{K+1}}
            \left[
            \vphantom{\left(
                      \sum_{\ellb_{1:k}\in\mathbb{Z}^{k}}
                      \right)} \right.}
   \left.
   \times
   \left(
   \frac{\bigl(r(\lambdab_{k:k+2})\bigr)^{1-\min(\abs{h_k},\abs{h_{k+1}})}
         -1}
       {1-r(\lambdab_{k:k+2})}
   + \max\bigl(\tau-\abs{h_k}-\abs{h_{k+1}}+1,0\bigr)
   \right)
   \vphantom{\left(
             \sum_{\ellb_{1:k}\in\mathbb{Z}^{k}}
             \right)}
   \right]
   \diff \lambdab
   \label{eqn/zeta_FSD_overlap_expanded}
\end{align}
where $\mathrm{Card}\left( \tilde{\mathcal{J}}_{1:k,\vert h_{k}\vert}\right)$
is given, according to (\ref{eqn/def_Jtilde1}), by
\begin{equation}
\mathrm{Card}\left( \tilde{\mathcal{J}}_{1:k,\vert h_{k}\vert}\right)
= \begin{cases}
  \tau^{k-1} \bigl(\tau -\abs{h_{k}}\bigr),
                                     & \text{if } \abs{h_{k}}\leq\tau-1 ; \\
  0,                                 & \text{if } \abs{h_{k}}\geq\tau .
  \end{cases}
\end{equation}
Then, after integrating out w.r.t. variables $\lambda_1,\ldots ,
\lambda_{k-1},\lambda_{k+3},\ldots ,\lambda_{K}$ in
(\ref{eqn/zeta_FSD_overlap_expanded}), and provided that
$\max\bigl( \abs{h_k},\abs{h_{k+1}}\bigr) \leq \tau -1$,
we finally obtain
\begin{align}
\zeta (\abs{h_k},-{\abs{h_{k+1}}})
&= \frac{\bigl( \tau-\abs{h_{k  }} \bigr)
         \bigl( \tau-\abs{h_{k+1}} \bigr)}
        {\tau^{3}}
   \int_{\mathbb{R}_{+}^{3}}
   \phi_{\abs{h_k},\abs{h_{k+1}}}(\lambdab_{k:k+2})
   \left(
   \frac{\bigl(r(\lambdab_{k:k+2})\bigr)^{1-\min(\abs{h_k},\abs{h_{k+1}})}
         -1}
        {1-r(\lambdab_{k:k+2})}
   \right.
   \notag \\
&  \phantom{= \frac{\bigl( \tau-\abs{h_{k+1}} \bigr)
                    \bigl( \tau-\abs{h_{k  }} \bigr)}
                   {\tau^{3}}
              \int_{\mathbb{R}_{+}^{3}}
              \phi_{\abs{h_k},\abs{h_{k+1}}}(\lambdab_{k:k+2})
              \left(
              \vphantom{\frac{\bigl(
                              r(\lambdab_{k:k+2})
                              \bigr)^{1-\min(\abs{h_k},\abs{h_{k+1}})}-1}
                             {1-r(\lambdab_{k:k+2})}}
              \right.}
   \left.
   + \max\bigl(\tau-\abs{h_k}-\abs{h_{k+1}}+1,0\bigr)
   \vphantom{\frac{\bigl(
                   r(\lambdab_{k:k+2})
                   \bigr)^{1-\min(\abs{h_k},\abs{h_{k+1}})}-1}
                  {1-r(\lambdab_{k:k+2})}}
   \right)\!
   \diff \lambdab_{k:k+2}
\label{eqn/zeta_FSD_overlap_kltK-1}
\end{align}
and for $k=K-1$, by plugging (\ref{eqn/S1_final_2}) and
(\ref{eqn/S2_final_2}) into (\ref{eqn/zeta_FSD_overlapS1S2}), we obtain
in the same way
\begin{align}
\zeta (\abs{h_{K-1}},-{\abs{h_{K}}}) 
&= \frac{\bigl( \tau-\abs{h_{K-1}} \bigr)}
   {\tau^{2}}
   \int_{\mathbb{R}_{+}^{3}}
   \phi_{\abs{h_{K-1}},\abs{h_{K}}}(\lambdab_{K-1:K+1})
   \left(
   \frac{\bigl(r(\lambdab_{K-1:K+1})\bigr)^{1-\min(\abs{h_{K-1},\abs{h_{K}})}}
         -1}
        {1-r(\lambdab_{K-1:K+1})}
   \right.
   \notag \\
&  \phantom{=
   \frac{\bigl( \tau-\abs{h_{K-1}} \bigr)}
        {\tau^{2}}
   \int_{\mathbb{R}_{+}^{3}}
   \phi_{\abs{h_{K-1}},\abs{h_{K}}}(\lambdab_{K-1:K+1})
   \left(
   \vphantom{
   \frac{\bigl(r(\lambdab_{K-1:K+1})\bigr)^{1-\min(\abs{h_{K-1},\abs{h_{K}})}}
         -1}
        {1-r(\lambdab_{K-1:K+1})}}
   \right.}
   \left.
   \vphantom{\frac{\bigl(
                   r(\lambdab_{K-1:K+1})\bigr)
                   ^{1-\min(\abs{h_{K-1},\abs{h_{K}})}}-1}
                  {1-r(\lambdab_{K-1:K+1})}}
   + \max(\tau-\abs{h_{K-1}}-\abs{h_{K}}+1,0)
   \right)\!
   \diff \lambdab_{K-1:K+1}.
\label{eqn/zeta_FSD_overlap_keqK-1}
\end{align}

Finally, by plugging (\ref{eqn/zeta_FSD_nooverlap_kltK-1_1}),
(\ref{eqn/zeta_FSD_nooverlap_kltK-1_2}) and
(\ref{eqn/zeta_FSD_overlap_kltK-1}) into
(\ref{eqn/P22FSD}) for $k<K-1$, and equivalently
(\ref{eqn/zeta_FSD_nooverlap_keqK-1_1}),
(\ref{eqn/zeta_FSD_nooverlap_keqK-1_2}) and
(\ref{eqn/zeta_FSD_overlap_keqK-1}) into
(\ref{eqn/P22FSD}) for $k=K-1$, we finally obtain
\begin{equation}
[\boldsymbol{P}_{22}]_{k,k+1} = C_k = 
v( \tau ,h_{k},h_{k+1})
\frac{\beta ^{\alpha_{k}+\alpha_{k+1}+\alpha_{k+2} }}
     {\Gamma \left( \alpha_{k}   \right)
      \Gamma \left( \alpha_{k+1} \right)
      \Gamma \left( \alpha_{k+2} \right) }
\int_{\mathbb{R}_{+}^{3}} \phi _{h_k,h_{k+1} }( \zb) 
                          w(\zb,\tau ,h_k,h_{k+1}) \diff\zb
\label{eqn/P22FSD_Ck}
\end{equation}
where functions $v(.)$ and $w(.)$ are defined in (\ref{eqn/def_v}) and
(\ref{eqn/def_w}), respectively.

\paragraph*{iii) Case D ($l=k$) and derivation of the diagonal
                 terms of $\boldsymbol{P}_{22}$ and of $\boldsymbol{V}_{22}$}

This case enables us to derive the diagonal terms of $\boldsymbol{P}_{22}$
(using the cases $h'_k = h_k$ and $h'_k=-h_k$), and the diagonal terms of
$\boldsymbol{V}_{22}$ (in the case $h'_k=0$), since
\begin{equation}
[\boldsymbol{P}_{22}]_{k,k}
= \zeta(\hb_k,\hb_k) + \zeta(-\hb_k,-\hb_k) - \zeta(-\hb_k,\hb_k)
  -\zeta(\hb_k,-\hb_k)
\label{eqn/P22_D_zeta}
\end{equation}
and
\begin{equation}
[\boldsymbol{V}_{22}]_{k,k}
= -h_k\zeta(\hb_k,\mathbf{0}_{2K+1}).
\label{eqn/V22_D_zeta}
\end{equation}

The two first terms in (\ref{eqn/P22_D_zeta}) can be obtained by plugging
(\ref{eqn/zetacute_D1}) and (\ref{eqn/pi_expanded}) (with
$\mathcal{J}_{h_{k},h_{k}}$ given by (\ref{eqn/J_D})) into
(\ref{eqn/zeta_pi_zetacute}), i.e.,
\begin{align}
\zeta(\hb_k,\hb_k) = \zeta(-\hb_k,-\hb_k)
&= \sum_{\ellb\in \mathbb{Z}^{K}}
   \int_{\mathbb{R}_{+}^{K+1}}
   \sum_{\kappab\in \mathbb{N}^{T}}
   f\left( \xb= \kappab,\lambdab,\tb=\ellb+\hb_{k}\right)
   \mathrm{d}\lambdab
   \notag \\
&= \sum_{\ellb\in \mathbb{Z}^{K}}
   \int_{\mathbb{R}_{+}^{K+1}}
   \biggl(
   f(\lambdab,\tb = \ellb +\hb_{k})
   \underset{=1}{
   \underbrace{
   \sum_{\kappab\in \mathbb{N}^{T}}
   f\left( \xb= \kappab|\lambdab,\tb=\ellb+\hb_{k}\right)
   }}
   \biggr)
   \mathrm{d}\lambdab
   \notag \\
&= \sum_{\ellb\in\mathbb{Z}^{K}}
   \int_{\mathbb{R}_{+}^{K+1}}
   \left(
   \prod_{i=1}^{K+1}
   \frac{\beta^{\alpha _{i}}}
        {\Gamma \left( \alpha _{i}\right) }
   \lambda_{i}^{\alpha _{i}-1}\exp \left( -\beta \lambda _{i}\right)
   \right)
   \frac{1}{\tau ^{K}}
   \indicf{\mathcal{J}_{h_{k},h_{k}}}{\ellb}
   \diff\lambdab
   \notag \\
&= \biggl(
   \frac{1}{\tau ^{K}}
   \sum_{\boldsymbol{\ell }\in \mathbb{Z}^{K}}
   \mathbb{I}_{\mathcal{J}_{h_{k},h_{k}}}\left( \ellb\right)
   \biggr)
   \prod_{i=1}^{K+1}
   \biggl(
   \underset{=1}{\underbrace{
   \frac{\beta^{\alpha _{i}}}
        {\Gamma \left( \alpha _{i}\right) }
   \int_{\mathbb{R}_{+}}
   \lambda_{i}^{\alpha _{i}-1}\exp \left( -\beta \lambda _{i}\right)
   \diff\lambda_{i}
   }}
   \biggr)
   \notag \\
&= \begin{cases}
   \frac{(\tau -\left\vert h_{k}\right\vert )^{2}}
        {\tau^{2}}
   & \text{ if } k\leq K-1 \text{ and } \abs{h_k}\leq \tau
   \\ 
   \frac{\tau -\left\vert h_{K}\right\vert }{\tau }
   & \text{ if } k = K \text{ and } \abs{h_K}\leq \tau
   \\ 
   0
   & \text{ if } \abs{h_k} > \tau
   \end{cases}
   \notag \\
&= u(\tau ,h_{k}) .
\label{eqn/zeta_D1}
\end{align}

The two other terms in (\ref{eqn/P22_D_zeta}) can be obtained by plugging
(\ref{eqn/zetacute_D2}) and (\ref{eqn/pi_expanded}) (with
$\mathcal{J}_{h_{k},-h_{k}}$ given by (\ref{eqn/J_D})) into
(\ref{eqn/zeta_pi_zetacute}), i.e.,
\begin{align}
\zeta(\hb_k,-\hb_k)
&= \sum_{\ellb\in\mathbb{Z}^{K}}
   \int_{\mathbb{R}_{+}^{K+1}}
   \left(
   \prod_{i=1}^{K+1}
   \frac{\beta^{\alpha _{i}}}
        {\Gamma \left( \alpha _{i}\right) }
   \lambda_{i}^{\alpha _{i}-1}\exp \left( -\beta \lambda _{i}\right)
   \right)
   \frac{1}{\tau ^{K}}
   \indicf{\mathcal{J}_{h_{k},-h_{k}}}{\ellb}
   \rho ^{2\left\vert h_{k}\right\vert }(\lambda _{k},\lambda _{k+1})
   \diff \lambdab
   \notag \\
&= \left(
   \frac{1}{\tau ^{K}}
   \sum_{\boldsymbol{\ell }\in \mathbb{Z}^{K}}
   \mathbb{I}_{\mathcal{J}_{h_{k},-h_{k}}}\left( \ellb\right)
   \right)
   \frac{\beta ^{\alpha_{k}+\alpha _{k+1}}}
        {\Gamma (\alpha_{k}) \Gamma ( \alpha _{k+1}) }
  \notag \\
& \quad \times
  \left( \int_{\mathbb{R}_{+}^{2}}
  \lambda _{k}^{\alpha_{k}-1}\lambda_{k+1}^{\alpha_{k+1}-1}
  \exp\left\{ -\beta \left( \lambda_{k}+\lambda_{k+1}\right)
              -|h_{k}|\left( 
                            \sqrt{\lambda_{k+1}}-\sqrt{\lambda_{k}}
                      \right)^{2}
      \right\} 
  \diff\lambda_{k}
  \diff\lambda_{k+1}\right)
  \notag \\
&= u(\tau,2h_k) \Phi(2h_k)
   \label{eqn/zeta_D2} \\
&= \zeta(-\hb_k,\hb_k) ,
\end{align}
where $u(\tau,h_k)$ is defined in (\ref{eqn/def_u}). Finally, by plugging
(\ref{eqn/zeta_D1}) and (\ref{eqn/zeta_D2}) into (\ref{eqn/P22_D_zeta}),
we obtain the diagonal terms of $[\boldsymbol{P}_{22}]_{k,k}$, for
$k = 1,\ldots ,K$
\begin{equation}
[\boldsymbol{P}_{22}]_{k,k}
= B_k
= 2\bigl(u(\tau,h_k)-u(\tau,2h_k) \Phi(2h_k)\bigr).
\label{eqn/P22D}
\end{equation}

Finally, with (\ref{eqn/P22UT}), (\ref{eqn/P22FSD_Ck}) and (\ref{eqn/P22D}),
we retrieve the results given in (\ref{eqn/P22_final}), (\ref{eqn/def_Bk})
and (\ref{eqn/def_Ck}).

To conclude this section, the expression of $\zeta(\hb_k,\mathbf{0}_{2K+1})$
can be obtained by plugging (\ref{eqn/pi_expanded}) (with
$\mathcal{J}_{h_{k},0}$ given by (\ref{eqn/J_D}) with $h'_k=0$)
and (\ref{eqn/zetacute_D3}) into (\ref{eqn/zeta_pi_zetacute}). This yields
\begin{align}
\zeta(\hb_k,\mathbf{0}_{2K+1})
&= \sum_{\ellb\in\mathbb{Z}^{K}}
   \int_{\mathbb{R}_{+}^{K+1}}
   \left(
   \prod_{i=1}^{K+1}
   \frac{\beta^{\alpha _{i}}}
        {\Gamma \left( \alpha _{i}\right) }
   \lambda_{i}^{\alpha _{i}-1}\exp \left( -\beta \lambda _{i}\right)
   \right)
   \frac{1}{\tau ^{K}}
   \indicf{\mathcal{J}_{h_{k},0}}{\ellb}
   \rho ^{\left\vert h_{k}\right\vert }(\lambda _{k},\lambda _{k+1})
   \diff \lambdab
   \notag \\
&= \frac{1}{\tau ^{K}}
   \left(
   \sum_{\boldsymbol{\ell }\in \mathbb{Z}^{K}}
   \mathbb{I}_{\mathcal{J}_{h_{k},0}}\left( \ellb\right)
   \right)
   \frac{\beta ^{\alpha_{k}+\alpha _{k+1}}}
        {\Gamma (\alpha_{k}) \Gamma ( \alpha _{k+1}) }
  \notag \\
& \quad \times
  \left( \int_{\mathbb{R}_{+}^{2}}
  \lambda _{k}^{\alpha_{k}-1}\lambda_{k+1}^{\alpha_{k+1}-1}
  \exp\left\{ -\beta \left( \lambda_{k}+\lambda_{k+1}\right)
              -|h_{k}|\frac{\left( 
                            \sqrt{\lambda_{k+1}}-\sqrt{\lambda_{k}}
                            \right)^{2}}
                           {2}
      \right\} 
  \diff\lambda_{k}
  \diff\lambda_{k+1}\right)
  \notag \\
&= u(\tau,h_k) \Phi(h_k)
   \label{eqn/zeta_D3}
\end{align}
which directly leads to the result we give in (\ref{eqn/def_V22kk}), i.e.,
\begin{equation}
[\boldsymbol{V}_{22}]_{k,k} = -h_{k} \: u\left( \tau ,h_{k}\right) \Phi
\left( h_{k} \right) .  \label{eqn/V22kk}
\end{equation}

\subsection{Derivation of $\boldsymbol{P}_{11}$}
\begin{align}
[\boldsymbol{P}_{11}]_{k,l}
&= \mathbb{E}_{\xb,\lambdab,\tb}
   \left\{
   \frac{\partial \ln f(\xb=\kappab,\lambdab,\tb=\ellb)}
        {\partial \lambda_k}
   \frac{\partial \ln f(\xb=\kappab,\lambdab,\tb=\ellb)}
        {\partial \lambda_l}
   \right\}
   \notag \\
&= \int_{\mathbb{R}_{+}^{K+1}}
   \sum_{\ellb\in\mathbb{Z}^{K}}
   f(\lambdab,\tb=\ellb)
   \Biggl[
   \sum_{\kappab\in\mathbb{N}^{T}}
   \frac{\partial \ln f(\xb=\kappab|\lambdab,\tb=\ellb)}
        {\partial \lambda_k}
   \frac{\partial \ln f(\xb=\kappab|\lambdab,\tb=\ellb)}
        {\partial \lambda_l}
   f(\xb = \kappab|\lambdab,\tb=\ellb)
   \notag \\
&  \hphantom{
   {} = \int_{\mathbb{R}_{+}^{K+1}}
        \sum_{\ellb\in\mathbb{Z}^{K}}
        f(\lambdab,\tb=\ellb)
        \Biggl[
   }
   + \frac{\partial \ln f(\lambdab,\tb=\ellb)}
          {\partial \lambda_l}
     \sum_{\kappab\in\mathbb{N}^{T}}
     \frac{\partial \ln f(\xb=\kappab|\lambdab,\tb=\ellb)}
          {\partial \lambda_k}
     f(\xb = \kappab|\lambdab,\tb=\ellb)
   \notag \\
&  \hphantom{
   {} = \int_{\mathbb{R}_{+}^{K+1}}
        \sum_{\ellb\in\mathbb{Z}^{K}}
        f(\lambdab,\tb=\ellb)
        \Biggl[
   }
   + \frac{\partial \ln f(\lambdab,\tb=\ellb)}
          {\partial \lambda_k}
     \sum_{\kappab\in\mathbb{N}^{T}}
     \frac{\partial \ln f(\xb=\kappab|\lambdab,\tb=\ellb)}
          {\partial \lambda_l}
     f(\xb = \kappab|\lambdab,\tb=\ellb)
   \notag \\
&  \hphantom{
   {} = \int_{\mathbb{R}_{+}^{K+1}}
        \sum_{\ellb\in\mathbb{Z}^{K}}
        f(\lambdab,\tb=\ellb)
        \Biggl[
   }
   + \frac{\partial \ln f(\lambdab,\tb=\ellb)}
          {\partial \lambda_k}
     \frac{\partial \ln f(\lambdab,\tb=\ellb)}
          {\partial \lambda_l}
     \sum_{\kappab\in\mathbb{N}^{T}}
     f(\xb = \kappab|\lambdab,\tb=\ellb)
     \Biggr] \diff\lambdab
\label{eqn/P11kl_expanded}
\end{align}
From (\ref{eqn/likelihood})
\begin{equation}
\frac{\partial \ln f(\xb=\kappab|\lambdab,\tb=\ellb)}
     {\partial \lambda_k}
= \ell_{k-1}-\ell_{k}
  + \frac{1}{\lambda_k}
    \sum_{t=\ell_{k-1}+1}^{\ell_k} \kappa_t
\label{eqn/deriv_likelihood}
\end{equation}
and from (\ref{eqn/prior_lambda}) and (\ref{eqn/prior_t})
\begin{equation}
\frac{\partial \ln f(\lambdab,\tb=\ellb)}
     {\partial \lambda_k}
= (\alpha_k -1)\frac{1}{\lambda_k} - \beta.
\label{eqn/deriv_prior}
\end{equation}
It is straightforward that
\begin{equation}
\sum_{\kappab\in\mathbb{N}^{T}}
     f(\xb = \kappab|\lambdab,\tb=\ellb) = 1.
\label{eqn/sum_likelihood}
\end{equation}
On the other hand,
\begin{align}
\sum_{\kappab\in\mathbb{N}^{T}}
\frac{\partial \ln f(\xb=\kappab|\lambdab,\tb=\ellb)}
     {\partial \lambda_k}
f(\xb = \kappab|\lambdab,\tb=\ellb)
&= \sum_{\kappab\in\mathbb{N}^{T}}
   \frac{\partial \ln f(\xb=\kappab|\lambdab,\tb=\ellb)}
        {\partial \lambda_k}
   \notag \\
&= \frac{\partial}{\partial \lambda_k}
   \sum_{\kappab\in\mathbb{N}^{T}}
   f(\xb = \kappab|\lambdab,\tb=\ellb)
   \notag \\
&= 0
\label{eqn/zero_terms_P11kl}
\end{align}
using (\ref{eqn/sum_likelihood}).
\begin{multline}
   \sum_{\kappab\in\mathbb{N}^{T}}
   \frac{\partial \ln f(\xb=\kappab|\lambdab,\tb=\ellb)}
        {\partial \lambda_k}
   \frac{\partial \ln f(\xb=\kappab|\lambdab,\tb=\ellb)}
        {\partial \lambda_l}
   f(\xb = \kappab|\lambdab,\tb=\ellb) \\
\begin{split}
   &= (\ell_{k-1}-\ell_{k})(\ell_{l-1}-\ell_{l})
   +
   \frac{(\ell_{l-1}-\ell_{l})}{\lambda_k}
   \sum_{\kappab\in\mathbb{N}^{T}}
   \left[
   \left(
   \sum_{t'=\ell_{k-1}+1}^{\ell_k}
   \kappa_{t'}
   \right)
   \prod_{i=1}^{K+1}
   \prod_{t=\ell_{i-1}+1}^{\ell_i}
   \frac{\lambda_i^{\kappa_t}}{\kappa_t!}
   \exp(-\lambda_i)
   \right]
   \\
&  \phantom{{}=(\ell_{k-1}-\ell_{k})(\ell_{l-1}-\ell_{l})}
   +
   \frac{(\ell_{k-1}-\ell_{k})}{\lambda_l}
   \sum_{\kappab\in\mathbb{N}^{T}}
   \left[
   \left(
   \sum_{t'=\ell_{l-1}+1}^{\ell_l}
   \kappa_{t'}
   \right)
   \prod_{i=1}^{K+1}
   \prod_{t=\ell_{i-1}+1}^{\ell_i}
   \frac{\lambda_i^{\kappa_t}}{\kappa_t!}
   \exp(-\lambda_i)
   \right]
   \\
&  \phantom{{}=(\ell_{k-1}-\ell_{k})(\ell_{l-1}-\ell_{l})}
   +
   \frac{1}{\lambda_k \lambda_l}
   \sum_{\kappab\in\mathbb{N}^{T}}
   \left[
   \left(
   \sum_{t'=\ell_{k-1}+1}^{\ell_k}
   \kappa_{t'}
   \right)
   \left(
   \sum_{t''=\ell_{l-1}+1}^{\ell_l}
   \kappa_{t''}
   \right)
   \prod_{i=1}^{K+1}
   \prod_{t=\ell_{i-1}+1}^{\ell_i}
   \frac{\lambda_i^{\kappa_t}}{\kappa_t!}
   \exp(-\lambda_i)
   \right]
   \end{split}
\label{eqn/P11kl.1}
\end{multline}
Let us first develop
\begin{align}
\left(
\sum_{t'=\ell_{k-1}+1}^{\ell_k}
\kappa_{t'}
\right)
\prod_{i=1}^{K+1}
\prod_{t=\ell_{i-1}+1}^{\ell_i}
\frac{\lambda_i^{\kappa_t}}{\kappa_t!}
\exp(-\lambda_i)
&= \left(
   \prod_{
          \substack{i=1 \\
                    i\neq k}
         }^{K+1}
   \prod_{t=\ell_{i-1}+1}^{\ell_i}
   \frac{\lambda_i^{\kappa_t}}{\kappa_t!}
   \exp(-\lambda_i)
   \right)
   \notag \\
& \quad \: \times
   \sum_{t'=\ell_{k-1}+1}^{\ell_k}
   \left[
   \left(
   \prod_{
          \substack{t=\ell_{k-1}+1 \\
                    t\neq t'}
         }^{\ell_k}
   \frac{\lambda_k^{\kappa_t}}{\kappa_t!}
   \exp(-\lambda_k)
   \right)
   \frac{\lambda_k^{\kappa_{t'}}}{(\kappa_{t'}-1)!}
   \exp(-\lambda_k)
   \right]
\label{eqn/P11kl.1/aux1_expanded}
\end{align}
Thus
\begin{multline}
\sum_{\kappab\in\mathbb{N}^{T}}
\left[
\left(
\sum_{t'=\ell_{k-1}+1}^{\ell_k}
\kappa_{t'}
\right)
\prod_{i=1}^{K+1}
\prod_{t=\ell_{i-1}+1}^{\ell_i}
\frac{\lambda_i^{\kappa_t}}{\kappa_t!}
\exp(-\lambda_i)
\right] \\
\begin{split}
&= \sum_{[\kappab_{1:\ell_{k-1}}^T,\kappab_{\ell_k+1:T}^T]^T
         \in\mathbb{N}^{T-(\ell_k-\ell_{k-1})}}
   \left(
   \prod_{
          \substack{i=1 \\
                    i\neq k}
         }^{K+1}
   \prod_{t=\ell_{i-1}+1}^{\ell_i}
   \frac{\lambda_i^{\kappa_t}}{\kappa_t!}
   \exp(-\lambda_i)
   \right)
   \\
&  \quad \: \times
   \sum_{\kappab_{\ell_{k-1}+1:\ell_{k}}\in\mathbb{N}^{\ell_k-\ell_{k-1}}}
   \left[
   \sum_{t'=\ell_{k-1}+1}^{\ell_k}
   \left[
   \left(
   \prod_{
          \substack{t=\ell_{k-1}+1 \\
                    t\neq t'}
         }^{\ell_k}
   \frac{\lambda_k^{\kappa_t}}{\kappa_t!}
   \exp(-\lambda_k)
   \right)
   \frac{\lambda_k^{\kappa_{t'}}}{(\kappa_{t'}-1)!}
   \exp(-\lambda_k)
   \right]
   \right]
\end{split}
\label{eqn/eqn/P11kl.1/sum_aux1_expanded}
\end{multline}
On the one hand
\begin{align}
\sum_{\kappab\in\mathbb{N}^{T}}
\left(
\prod_{
       \substack{i=1 \\
                 i\neq k}
      }^{K+1}
\prod_{t=\ell_{i-1}+1}^{\ell_i}
\frac{\lambda_i^{\kappa_t}}{\kappa_t!}
\exp(-\lambda_i)
\right)
&= \prod_{
          \substack{i=1 \\
                    i\neq k}
         }^{K+1}
   \prod_{t=\ell_{i-1}+1}^{\ell_i}
   \left(
   \sum_{\kappa_t\in\mathbb{N}}
   \frac{\lambda_i^{\kappa_t}}{\kappa_t!}
   \exp(-\lambda_i)
   \right)
   \notag \\
&= 1^{T-(\ell_k-\ell_{k-1})}
   \notag \\
&= 1.
\label{eqn/P11kl.1/sum_aux1_expanded.1}
\end{align}
On the other hand
\begin{multline}
\sum_{\kappab_{\ell_{k-1}+1:\ell_{k}}\in\mathbb{N}^{\ell_k-\ell_{k-1}}}
   \left[
   \sum_{t'=\ell_{k-1}+1}^{\ell_k}
   \left[
   \left(
   \prod_{
          \substack{t=\ell_{k-1}+1 \\
                    t\neq t'}
         }^{\ell_k}
   \frac{\lambda_k^{\kappa_t}}{\kappa_t!}
   \exp(-\lambda_k)
   \right)
   \frac{\lambda_k^{\kappa_{t'}}}{(\kappa_{t'}-1)!}
   \exp(-\lambda_k)
   \right]
   \right]
   \\
\begin{aligned}[b]
&= \sum_{t'=\ell_{k-1}+1}^{\ell_k}
   \left[
   \underset{=1}{
   \underbrace{
   \left(
   \sum_{\kappa\in\mathbb{N}}
   \frac{\lambda_k^{\kappa}}{\kappa!}
   \exp(-\lambda_k)
   \right)}}^{\ell_k-\ell_{k-1}-1}
   \sum_{\kappa_{t'}=1}^{+\infty}
    \frac{\lambda_k^{\kappa_{t'}}}{(\kappa_{t'}-1)!}
   \exp(-\lambda_k)
   \right]
   \\
&= \sum_{t'=\ell_{k-1}+1}^{\ell_k}
   \lambda_k
   \\
&= \lambda_k (\ell_k -\ell_{k-1}).
\label{eqn/P11kl.1/sum_aux1_expanded.2}
\end{aligned}
\end{multline}
Then, we obtain
\begin{align}
\frac{(\ell_{l-1}-\ell_{l})}{\lambda_k}
   \sum_{\kappab\in\mathbb{N}^{T}}
   \left[
   \left(
   \sum_{t'=\ell_{k-1}+1}^{\ell_k}
   \kappa_{t'}
   \right)
   \prod_{i=1}^{K+1}
   \prod_{t=\ell_{i-1}+1}^{\ell_i}
   \frac{\lambda_i^{\kappa_t}}{\kappa_t!}
   \exp(-\lambda_i)
   \right]
&= (\ell_{l-1}-\ell_{l})(\ell_k -\ell_{k-1})
\label{eqn/P11kl.1.2}\\
\intertext{and}
\frac{(\ell_{k-1}-\ell_{k})}{\lambda_l}
   \sum_{\kappab\in\mathbb{N}^{T}}
   \left[
   \left(
   \sum_{t'=\ell_{l-1}+1}^{\ell_l}
   \kappa_{t'}
   \right)
   \prod_{i=1}^{K+1}
   \prod_{t=\ell_{i-1}+1}^{\ell_i}
   \frac{\lambda_i^{\kappa_t}}{\kappa_t!}
   \exp(-\lambda_i)
   \right]
&= (\ell_{k-1}-\ell_{k})(\ell_{l}-\ell_{l-1}).
\label{eqn/P11kl.1.3}
\end{align}

The writing of the fourth term in the right hand side of
(\ref{eqn/P11kl.1}) depends on
whether $k=l$ or $k\neq l$.

\noindent $\bullet$ For $k\neq l$, using the same kind of manipulations as in
equations (\ref{eqn/P11kl.1/aux1_expanded})
to (\ref{eqn/P11kl.1/sum_aux1_expanded.2}), we find
\begin{align}
\sum_{\kappab\in\mathbb{N}^{T}}
   \left[
   \left(
   \sum_{t'=\ell_{k-1}+1}^{\ell_k}
   \kappa_{t'}
   \right)
   \left(
   \sum_{t''=\ell_{l-1}+1}^{\ell_l}
   \kappa_{t''}
   \right)
   \prod_{i=1}^{K+1}
   \prod_{t=\ell_{i-1}+1}^{\ell_i}
   \frac{\lambda_i^{\kappa_t}}{\kappa_t!}
   \exp(-\lambda_i)
   \right]
&= \left(
   \sum_{t'=\ell_{k-1}+1}^{\ell_k}\lambda_k
   \right)
   \left(
   \sum_{t''=\ell_{l-1}+1}^{\ell_l}\lambda_l
   \right)
   \notag \\
&= \lambda_k (\ell_k - \ell_{k-1})\lambda_l (\ell_l -\ell_{l-1})
\label{eqn/P11kl.1.4_kneql}
\end{align}
which finally leads, by plugging (\ref{eqn/P11kl.1.2}),
(\ref{eqn/P11kl.1.3}) and (\ref{eqn/P11kl.1.4_kneql}) into
(\ref{eqn/P11kl.1}), to
\begin{equation}
\sum_{\kappab\in\mathbb{N}^{T}}
   \frac{\partial \ln f(\xb=\kappab|\lambdab,\tb=\ellb)}
        {\partial \lambda_k}
   \frac{\partial \ln f(\xb=\kappab|\lambdab,\tb=\ellb)}
        {\partial \lambda_l}
   f(\xb = \kappab|\lambdab,\tb=\ellb)
= 0.
\label{eqn/P11kl.1_kneql}
\end{equation}

\noindent $\bullet$ Conversely, if $k=l$, then (\ref{eqn/P11kl.1.4_kneql})
becomes
\begin{multline}
\sum_{\kappab\in\mathbb{N}^{T}}
   \left[
   \left(
   \sum_{t'=\ell_{k-1}+1}^{\ell_k}
   \kappa_{t'}
   \right)^2
   \prod_{i=1}^{K+1}
   \prod_{t=\ell_{i-1}+1}^{\ell_i}
   \frac{\lambda_i^{\kappa_t}}{\kappa_t!}
   \exp(-\lambda_i)
   \right] \\
= \sum_{\kappab\in\mathbb{N}^{T}}
   \left[
   \left(
   \sum_{t'=\ell_{k-1}+1}^{\ell_k}
   \kappa_{t'}^2
   +2\sum_{u=\ell_{k-1}+1}^{\ell_k}
     \sum_{v=u+1}^{\ell_k}
   \kappa_{u}\kappa_{v}
   \right)
   \prod_{i=1}^{K+1}
   \prod_{t=\ell_{i-1}+1}^{\ell_i}
   \frac{\lambda_i^{\kappa_t}}{\kappa_t!}
   \exp(-\lambda_i)
   \right].
   \label{eqn/P11kl.1.4_k=l}
\end{multline}
On the one hand, we have
\begin{align}
\sum_{\kappab\in\mathbb{N}^{T}}
   \left[
   \left(
   \sum_{t'=\ell_{k-1}+1}^{\ell_k}
   \kappa_{t'}^2
   \right)
   \prod_{i=1}^{K+1}
   \prod_{t=\ell_{i-1}+1}^{\ell_i}
   \frac{\lambda_i^{\kappa_t}}{\kappa_t!}
   \exp(-\lambda_i)
   \right]
&= \sum_{t'=\ell_{k-1}+1}^{\ell_k}
   \sum_{\kappa_{t'}=1}^{+\infty}
   \kappa_{t'}^2
   \frac{\lambda_k^{\kappa_{t'}}}{\kappa_{t'}!}
   \exp(-\lambda_k)
   \notag \\
&= \sum_{t'=\ell_{k-1}+1}^{\ell_k}
   \left[
   \lambda_k \exp(-\lambda_k)
   \sum_{\kappa_{t'}=1}^{+\infty}
   \frac{\kappa_{t'}\lambda_k^{\kappa_{t'}-1}}{(\kappa_{t'}-1)!}
   \right]
   \notag \\
&= \sum_{t'=\ell_{k-1}+1}^{\ell_k}
   \left[
   \lambda_k \exp(-\lambda_k)
   \sum_{\kappa_{t'}=1}^{+\infty}
   \frac{\!\diff}{\!\diff\lambda_k}
   \frac{\lambda_k^{\kappa_{t'}}}{(\kappa_{t'}-1)!}
   \right]
   \notag \\
&= \sum_{t'=\ell_{k-1}+1}^{\ell_k}
   \left[
   \lambda_k \exp(-\lambda_k)
   \frac{\!\diff}{\!\diff\lambda_k}
   \left(
   \sum_{\kappa_{t'}=1}^{+\infty}
   \frac{\lambda_k^{\kappa_{t'}}}{(\kappa_{t'}-1)!}
   \right)
   \right]
   \notag \\
&= \sum_{t'=\ell_{k-1}+1}^{\ell_k}
   \left[
   \lambda_k \exp(-\lambda_k)
   \frac{\!\diff}{\!\diff\lambda_k}
   \bigl(
   \lambda_k\exp(\lambda_k)
   \bigr)
   \right]
   \notag \\
&= \sum_{t'=\ell_{k-1}+1}^{\ell_k}
   \left[
   \lambda_k (\lambda_k +1)
   \right]
   \notag \\
&= (\ell_k - \ell_{k-1})\lambda_k (\lambda_k +1)
\label{eqn/P11kl.1.4_k=l.1}
\end{align}
and on the other hand, we have
\begin{multline}
2\sum_{\kappab\in\mathbb{N}^{T}}
   \left[
   \left(
   \sum_{u=\ell_{k-1}+1}^{\ell_k}
   \sum_{v=u+1}^{\ell_k}
   \kappa_{u}\kappa_{v}
   \right)
   \prod_{i=1}^{K+1}
   \prod_{t=\ell_{i-1}+1}^{\ell_i}
   \frac{\lambda_i^{\kappa_t}}{\kappa_t!}
   \exp(-\lambda_i)
   \right]
   \\
\begin{aligned}[b]
&= 2\sum_{u=\ell_{k-1}+1}^{\ell_k}
   \sum_{v=u+1}^{\ell_k}
   \left[
   \left(
   \sum_{\kappa_{u}=1}^{+\infty}
   \frac{\lambda_k^{\kappa_{u}}}{(\kappa_{u}-1)!}
   \exp(-\lambda_k)
   \right)
   \left(
   \sum_{\kappa_{v}=1}^{+\infty}
   \frac{\lambda_k^{\kappa_{v}}}{(\kappa_{v}-1)!}
   \exp(-\lambda_k)
   \right)
   \right]
   \\
&= 2\sum_{u=\ell_{k-1}+1}^{\ell_k}
   \sum_{v=u+1}^{\ell_k}
   \lambda_k^2
   \\
&= 2\lambda_k^2
   \left(
   \ell_k(\ell_k - \ell_{k-1})
   - \frac{(\ell_k - \ell_{k-1})(\ell_k + \ell_{k-1} +1)}{2}
   \right)
   \\
&= \lambda_k^2
   (\ell_k - \ell_{k-1})
   (\ell_k - \ell_{k-1}-1).
\label{eqn/P11kl.1.4_k=l.2}
\end{aligned}
\end{multline}
Thus, if $k=l$, gathering (\ref{eqn/P11kl.1.4_k=l}),
(\ref{eqn/P11kl.1.4_k=l.1}) and (\ref{eqn/P11kl.1.4_k=l.2}) together and after
plugging the result into (\ref{eqn/P11kl.1}), we finally obtain
\begin{multline}
\sum_{\kappab\in\mathbb{N}^{T}}
   \frac{\partial \ln f(\xb=\kappab|\lambdab,\tb=\ellb)}
        {\partial \lambda_k}
   \frac{\partial \ln f(\xb=\kappab|\lambdab,\tb=\ellb)}
        {\partial \lambda_l}
   f(\xb = \kappab|\lambdab,\tb=\ellb) \\
\begin{aligned}[b]
&= -(\ell_k - \ell_{k-1})^2
   + \frac{1}{\lambda_k^2}
   \left[
   \lambda_k (\lambda_k +1)(\ell_k - \ell_{k-1})
   + \lambda_k^2
   (\ell_k - \ell_{k-1})
   (\ell_k - \ell_{k-1}-1)
   \right]
   \\
&= \frac{\ell_k -\ell_{k-1}}{\lambda_k}.
\label{eqn/P11kl.1_k=l}
\end{aligned}
\end{multline}

Finally, considering (\ref{eqn/deriv_prior}),
(\ref{eqn/sum_likelihood}), (\ref{eqn/zero_terms_P11kl}),
(\ref{eqn/P11kl.1_kneql}) and (\ref{eqn/P11kl.1_k=l}), and after plugging
them into (\ref{eqn/P11kl_expanded}), we obtain
\begin{equation}
\begin{split}
[\boldsymbol{P}_{11}]_{k,l}
&= \int_{\mathbb{R}_{+}^{K+1}}
   \sum_{\ellb\in\mathbb{Z}^{T}}
   \left[
   \frac{\ell_k-\ell_{k-1}}{\lambda_k}\delta_{k,l}
   \cdot
   \frac{1}{\tau^K}
   \indicf{\mathcal{J}_{0,0}}{\ellb}
   \cdot
   \prod_{i=1}^{K+1}
   \frac{\beta^{\alpha_i}}{\Gamma (\alpha_i)}
   \lambda_i^{\alpha_i-1}
   \exp (-\beta\lambda_i)
   \right]
   \diff\lambdab
   \\
&  \quad  +
   \int_{\mathbb{R}_{+}^{K+1}}
   \sum_{\ellb\in\mathbb{Z}^{T}}
   \left[
   \left(
   \frac{\alpha_k-1}{\lambda_k}-\beta
   \right)
   \left(
   \frac{\alpha_l-1}{\lambda_l}-\beta
   \right)
   \cdot
   \frac{1}{\tau^K}
   \indicf{\mathcal{J}_{0,0}}{\ellb}
   \cdot
   \prod_{i=1}^{K+1}
   \frac{\beta^{\alpha_i}}{\Gamma (\alpha_i)}
   \lambda_i^{\alpha_i-1}
   \exp (-\beta\lambda_i)
   \right]
   \diff\lambdab .
\end{split}
\label{eqn/P11kl_expanded2}
\end{equation}
The first term in the right-hand side of (\ref{eqn/P11kl_expanded2}) can be
developed as
\begin{multline}
\int_{\mathbb{R}_{+}^{K+1}}
\sum_{\ellb\in\mathbb{Z}^{T}}
\left[
\frac{\ell_k-\ell_{k-1}}{\lambda_k}\delta_{k,l}
\cdot
\frac{1}{\tau^K}
\indicf{\mathcal{J}_{0,0}}{\ellb}
\cdot
\prod_{i=1}^{K+1}
\frac{\beta^{\alpha_i}}{\Gamma (\alpha_i)}
\lambda_i^{\alpha_i-1}
\exp (-\beta\lambda_i)
\right]
\diff\lambdab
\\
\begin{aligned}[b]
&= \frac{\delta_{k,l}}{\tau^K}
   \int_{\mathbb{R}_{+}^{K+1}}
   \left[
   \left(
   \sum_{\ellb\in\mathcal{J}_{0,0}}
   \frac{\ell_k-\ell_{k-1}}{\lambda_k}
   \right)
   \prod_{i=1}^{K+1}
   \frac{\beta^{\alpha_i}}{\Gamma (\alpha_i)}
   \lambda_i^{\alpha_i-1}
   \exp (-\beta\lambda_i)
   \right]
   \diff\lambdab
   \\
&= \frac{\delta_{k,l}}{\tau^k}
   \int_{\mathbb{R}_{+}^{K+1}}
   \left[
   \left(
   \frac{1}{\lambda_k}
   \sum_{\ellb_{1:k-2}\in\mathcal{J}_{1:k-2,0,0}}
   \sum_{\ell_{k-1}=\ell_{k-2}+1}^{\ell_{k-2}+\tau}
   \sum_{\ell_{k}=\ell_{k-1}+1}^{\ell_{k-1}+\tau}
   (\ell_k-\ell_{k-1})
   \right)
   \prod_{i=1}^{K+1}
   \frac{\beta^{\alpha_i}}{\Gamma (\alpha_i)}
   \lambda_i^{\alpha_i-1}
   \exp (-\beta\lambda_i)
   \right]
   \diff\lambdab
   \\
&= \frac{\delta_{k,l}}{\tau^k}
   \int_{\mathbb{R}_{+}^{K+1}}
   \left[
   \left(
   \frac{1}{\lambda_k}
   \sum_{\ellb_{1:k-2}\in\mathcal{J}_{1:k-2,0,0}}
   \sum_{\ell_{k-1}=\ell_{k-2}+1}^{\ell_{k-2}+\tau}
   \left(
   \frac{\tau(2\ell_{k-1}+\tau+1)}{2}-\tau\ell_{k-1}
   \right)
   \right)
   \prod_{i=1}^{K+1}
   \frac{\beta^{\alpha_i}}{\Gamma (\alpha_k)}
   \lambda_i^{\alpha_i-1}
   \exp (-\beta\lambda_i)
   \right]
   \diff\lambdab
   \\
&= \delta_{k,l}
   \frac{\tau +1}{2}
   \int_{\mathbb{R}_{+}^{K+1}}
   \left(
   \prod_{\substack{i=1\\
                    i\neq k}}
        ^{K+1}
   \frac{\beta^{\alpha_i}}{\Gamma (\alpha_i)}
   \lambda_i^{\alpha_i-1}
   \exp (-\beta\lambda_i)
   \right)
   \frac{\beta^{\alpha_k}}{\Gamma (\alpha_k)}
   \lambda_k^{\alpha_k-2}
   \exp (-\beta\lambda_k)
   \diff\lambdab
   \\
&= \delta_{k,l}
   \frac{\tau +1}{2}
   \frac{\beta^{\alpha_k}}{\Gamma (\alpha_k)}
   \int_{\mathbb{R}_{+}^{K+1}}
   \lambda_k^{\alpha_k-2}
   \exp (-\beta\lambda_k)
   \diff\lambda_k
   \\
&= \delta_{k,l}
   \frac{\tau +1}{2}
   \frac{\beta^{\alpha_k}}{\Gamma (\alpha_k)}
   \frac{\Gamma (\alpha_k-1)}{\beta^{\alpha_k -1}}
   \\
&= \delta_{k,l}
   \frac{\beta(\tau +1)}{2(\alpha_k -1)}
\end{aligned}
\raisetag{15pt}
\label{eqn/P11kl.1_final}
\end{multline}
in which $\delta_{k,l}$ denotes the Dirac delta. The second term in the 
right-hand side of (\ref{eqn/P11kl_expanded2}) can be
developed, for $k\neq l$, as
\begin{multline}
\int_{\mathbb{R}_{+}^{K+1}}
\sum_{\ellb\in\mathbb{Z}^{T}}
\left[
\left(
\frac{\alpha_k-1}{\lambda_k}-\beta
\right)
\left(
\frac{\alpha_l-1}{\lambda_l}-\beta
\right)
\cdot
\frac{1}{\tau^K}
\indicf{\mathcal{J}_{0,0}}{\ellb}
\cdot
\prod_{i=1}^{K+1}
\frac{\beta^{\alpha_i}}{\Gamma (\alpha_i)}
\lambda_i^{\alpha_i-1}
\exp (-\beta\lambda_i)
\right]
\diff\lambdab
\\
\begin{aligned}[b]
&= \underset{=1}{
   \underbrace{
   \frac{1}{\tau^K}
   \left(
   \sum_{\ellb\in\mathbb{Z}^{T}}
   \indicf{\mathcal{J}_{0,0}}{\ellb}
   \right)
   }
   }
   \int_{\mathbb{R}_{+}^{K+1}}
   \left[
   \left(
   \frac{\alpha_k-1}{\lambda_k}-\beta
   \right)
   \left(
   \frac{\alpha_l-1}{\lambda_l}-\beta
   \right)
   \prod_{i=1}^{K+1}
   \frac{\beta^{\alpha_i}}{\Gamma (\alpha_i)}
   \lambda_i^{\alpha_i-1}
   \exp (-\beta\lambda_i)
   \right]
   \diff\lambdab
   \\
&= \begin{aligned}[t]
   \prod_{\substack{i=1\\
                    i\neq k,i\neq l }}
        ^{K+1}
   \biggl(
   \underset{=1}{
   \underbrace{
   \frac{\beta^{\alpha_i}}{\Gamma (\alpha_i)}
   \int_{\mathbb{R}_{+}}
   \lambda_i^{\alpha_i-1}
   \exp (-\beta\lambda_i)
   \diff\lambda_i
   }
   }
   \biggr)
&  \frac{\beta^{\alpha_k}}{\Gamma (\alpha_k)}
   \int_{\mathbb{R}_{+}}
   \left(
   \frac{\alpha_k-1}{\lambda_k}-\beta
   \right)
   \lambda_k^{\alpha_k-1}
   \exp (-\beta\lambda_k)
   \diff\lambda_k
   \\
{} \times {}
&  \frac{\beta^{\alpha_l}}{\Gamma (\alpha_l)}
   \int_{\mathbb{R}_{+}}
   \left(
   \frac{\alpha_l-1}{\lambda_l}-\beta
   \right)
   \lambda_l^{\alpha_l-1}
   \exp (-\beta\lambda_l)
   \diff\lambda_l
   \end{aligned}
   \\
&= \begin{aligned}[t]
   \frac{\beta^{\alpha_k}}{\Gamma (\alpha_k)}
   \frac{\beta^{\alpha_l}}{\Gamma (\alpha_l)}
&  \left[
   (\alpha_k -1)
   \biggl(
   \int_{\mathbb{R}_{+}}
   \lambda_k^{\alpha_k-2}
   \exp (-\beta\lambda_k)
   \diff\lambda_k
   \biggr)
   -\beta \frac{\Gamma (\alpha_k)}{\beta^{\alpha_k}}
   \right]
   \\
{} \times {}
&  \left[
   (\alpha_l -1)
   \biggl(
   \int_{\mathbb{R}_{+}}
   \lambda_l^{\alpha_l-2}
   \exp (-\beta\lambda_l)
   \diff\lambda_l
   \biggr)
   -\beta \frac{\Gamma (\alpha_l)}{\beta^{\alpha_l}}
   \right]
   \end{aligned}
   \\
&= \frac{\beta^{\alpha_k}}{\Gamma (\alpha_k)}
   \frac{\beta^{\alpha_l}}{\Gamma (\alpha_l)}
   \left[
   (\alpha_k -1)
   \frac{\Gamma (\alpha_k -1)}{\beta^{\alpha_k -1}}
   - \frac{\Gamma (\alpha_k)}{\beta^{\alpha_k -1}}
   \right]
   \cdot
   \left[
   (\alpha_l -1)
   \frac{\Gamma (\alpha_l -1)}{\beta^{\alpha_l -1}}
   - \frac{\Gamma (\alpha_l)}{\beta^{\alpha_l -1}}
   \right]
   \\
&= 0,
\end{aligned}
\label{eqn/P11kl.2_kneql_final}
\end{multline}
provided that $\alpha_k,\alpha_l >2$. Conversely, if $k=l$, we obtain
\begin{multline}
\int_{\mathbb{R}_{+}^{K+1}}
\sum_{\ellb\in\mathbb{Z}^{T}}
\left[
\left(
\frac{\alpha_k-1}{\lambda_k}-\beta
\right)^2
\cdot
\frac{1}{\tau^K}
\indicf{\mathcal{J}_{0,0}}{\ellb}
\cdot
\prod_{i=1}^{K+1}
\frac{\beta^{\alpha_i}}{\Gamma (\alpha_i)}
\lambda_i^{\alpha_i-1}
\exp (-\beta\lambda_i)
\right]
\diff\lambdab
\\
\begin{aligned}[b]
&= \frac{\beta^{\alpha_k}}{\Gamma (\alpha_k)}
   \int_{\mathbb{R}_{+}}
   \left(
   \frac{\alpha_k-1}{\lambda_k}-\beta
   \right)^2
   \lambda_k^{\alpha_k-1}
   \exp (-\beta\lambda_k)
   \diff\lambda_k
   \\
&= \frac{\beta^{\alpha_k}}{\Gamma (\alpha_k)}
   \left[
   (\alpha_k -1)^2
   \biggl(
   \int_{\mathbb{R}_{+}}
   \lambda_k^{\alpha_k-3}
   \exp (-\beta\lambda_k)
   \diff\lambda_k
   \biggr)
   -2\beta (\alpha_k -1)
   \biggl(
   \int_{\mathbb{R}_{+}}
   \lambda_k^{\alpha_k-2}
   \exp (-\beta\lambda_k)
   \diff\lambda_k
   \biggr)
   + \beta^2 \frac{\Gamma (\alpha_k)}{\beta^{\alpha_k}}
   \right]
   \\
&= \frac{\beta^{\alpha_k}}{\Gamma (\alpha_k)}
   \left[
   (\alpha_k -1)^2
   \frac{\Gamma (\alpha_k -2)}{\beta^{\alpha_k -2}}
   -2(\alpha_k -1)
   \frac{\Gamma (\alpha_k -1)}{\beta^{\alpha_k -2}}
   + \frac{\Gamma (\alpha_k)}{\beta^{\alpha_k -2}}
   \right]
   \\
&= \frac{\beta^{\alpha_k}}{\Gamma (\alpha_k)}
   \frac{
   (\alpha_k -1)^2\,\Gamma (\alpha_k -2)
   - \Gamma (\alpha_k )
   }{\beta^{\alpha_k -2}}
   \\
&= \frac{\beta^2 (\alpha_k -1)\, \Gamma (\alpha_k -2) }
        {(\alpha_k -1)(\alpha_k-2)\Gamma (\alpha_k -2)}
   \\
&= \frac{\beta^2}{\alpha_k -2},
\end{aligned}
\label{eqn/P11kl.2_k=l_final}
\end{multline}
provided that $\alpha_k >2$.

Finally, considering (\ref{eqn/P11kl.1_final}),
(\ref{eqn/P11kl.2_kneql_final}) and (\ref{eqn/P11kl.2_k=l_final}), and after
plugging them into (\ref{eqn/P11kl_expanded2}),
we retrieve (\ref{eqn/P11kk}), i.e.,
\begin{equation}
[\boldsymbol{P}_{11}]_{k,l}=
\left(
\frac{\beta \left( \tau +1\right) }
{2\left( \alpha_k -1\right) }+\frac{\beta^2}{\alpha_k - 2}\right)
\delta_{k,l}.
\end{equation}

\subsection{Derivation of $\boldsymbol{P}_{12}$}

\begin{equation}
\left[ \boldsymbol{P}_{12}\right]_{k,l}
= \mathbb{E}_{\xb,\lambdab,\tb}
  \left\{
  \frac{\partial \ln f(\xb =\kappab ,\lambdab ,\tb =\ellb ) }
       {\partial \lambda _{k}}
  \left( 
  \sqrt{
  \frac{f (\xb =\kappab ,\lambdab ,\tb =\ellb +\hb_{l}) }
       {f (\xb =\kappab ,\lambdab ,\tb =\ellb         ) }
       }
  -\sqrt{
   \frac{f (\xb =\kappab ,\lambdab ,\tb =\ellb -\hb_{l}) }
        {f (\xb =\kappab ,\lambdab ,\tb =\ellb         ) }}
  \right)
  \right\}
\label{eqn/P12kl}
\end{equation}

Let us first derive the quantity
\begin{multline}
\mathbb{E}_{\xb,\lambdab,\tb}
\left\{
\frac{\partial \ln f(\xb =\kappab ,\lambdab ,\tb =\ellb ) }
     {\partial \lambda _{k}}
\sqrt{
\frac{f (\xb =\kappab ,\lambdab ,\tb =\ellb +\hb_{l}) }
     {f (\xb =\kappab ,\lambdab ,\tb =\ellb         ) }
     }
\right\}
\\
\begin{aligned}[b]
&= \int_{\mathbb{R}_{+}^{K+1}}
   \sum_{\ellb \in \mathbb{Z}^{K}}
   \sum_{\kappab\in\mathbb{N}^{T}}
   \frac{\partial \ln f(\xb =\kappab ,\lambdab ,\tb =\ellb ) }
        {\partial \lambda _{k}}
   \sqrt{f (\xb =\kappab ,\lambdab ,\tb =\ellb +\hb_{l}) 
         f (\xb =\kappab ,\lambdab ,\tb =\ellb         )}
   \diff\lambdab
   \\
&= \int_{\mathbb{R}_{+}^{K+1}}
   \sum_{\ellb \in \mathbb{Z}^{K}}
   \begin{aligned}[t]
   \left[
   \vphantom{\sum_{\ellb \in \mathbb{Z}^{K}}}
   \right.
   & \sqrt{f (\lambdab,\tb =\ellb +\hb_{l})
           f (\lambdab,\tb =\ellb         )}
     \\
   & \begin{aligned}[t]
     {}\times
     \Biggl(
     & \sum_{\kappab\in\mathbb{N}^{T}}
       \biggl(
       \frac{\partial \ln f(\xb =\kappab |\lambdab ,\tb =\ellb ) }
            {\partial \lambda _{k}}
       \sqrt{f (\xb =\kappab |\lambdab ,\tb =\ellb +\hb_{l}) 
             f (\xb =\kappab |\lambdab ,\tb =\ellb         )}
       \biggr)
       \\
     & + \left.
       \frac{\partial \ln f(\lambdab ,\tb =\ellb ) }
            {\partial \lambda _{k}}
       \sum_{\kappab\in\mathbb{N}^{T}}
       \sqrt{f (\xb =\kappab |\lambdab ,\tb =\ellb +\hb_{l}) 
             f (\xb =\kappab |\lambdab ,\tb =\ellb         )}
       \Biggl)
       \vphantom{\sum_{\ellb \in \mathbb{Z}^{K}}}
       \right]
       \diff\lambdab
     \end{aligned}
   \end{aligned}
   \\
&= \int_{\mathbb{R}_{+}^{K+1}}
   \sum_{\ellb \in \mathbb{Z}^{K}}
   \begin{aligned}[t]
   \left[
   \vphantom{\sum_{\ellb \in \mathbb{Z}^{K}}}
   \right.
   & \sqrt{f (\lambdab,\tb =\ellb +\hb_{l})
           f (\lambdab,\tb =\ellb         )}
     \hspace{28.5em}
     \\
   & \begin{aligned}[b]
     {}\times
     \Biggl(
     & \sum_{\kappab\in\mathbb{N}^{T}}
       \biggl(
       \biggl[
       \ell_{k-1}-\ell_{k}
       + \frac{1}{\lambda_k}
         \tsum_{t'=\ell_{k-1}+1}^{\ell_k} \kappa_{t'}
       \biggr]
       \sqrt{f (\xb =\kappab |\lambdab ,\tb =\ellb +\hb_{l}) 
             f (\xb =\kappab |\lambdab ,\tb =\ellb         )}
       \biggr)
       \\
     & + \left.
       \left[
       \frac{\alpha_k -1}{\lambda_k} - \beta
       \right]
       \sum_{\kappab\in\mathbb{N}^{T}}
       \sqrt{f (\xb =\kappab |\lambdab ,\tb =\ellb +\hb_{l}) 
             f (\xb =\kappab |\lambdab ,\tb =\ellb         )}
       \Biggl)
       \vphantom{\sum_{\ellb \in \mathbb{Z}^{K}}}
       \right]
       \diff\lambdab .
       \label{eqn/P12kl.1_expanded}
     \end{aligned}
     \raisetag{22.5pt}
   \end{aligned}
\end{aligned}
\end{multline}
Note that the quantity
$\sum_{\kappab\in\mathbb{N}^T}
 \sqrt{f (\xb =\kappab |\lambdab ,\tb =\ellb +\hb_{l}) 
 f (\xb =\kappab |\lambdab ,\tb =\ellb )}$ is nothing else than
$\acute{\zeta}(\lambdab ,\ellb ,\hb_{l},\mathbf{0}_{2K+1})$, which is given in
(\ref{eqn/zetacute_D3}).

In (\ref{eqn/P12kl.1_expanded}), we develop
\begin{multline}
\sum_{\kappab\in\mathbb{N}^{T}}
\Biggl(
\biggl[
\ell_{k-1}-\ell_{k}
+ \frac{1}{\lambda_k}
  \sum_{t'=\ell_{k-1}+1}^{\ell_k} \kappa_{t'}
\biggr]
\sqrt{f (\xb =\kappab |\lambdab ,\tb =\ellb +\hb_{l}) 
      f (\xb =\kappab |\lambdab ,\tb =\ellb         )}
\Biggr)
\\
\begin{aligned}[b]
&= (\ell_{k-1}-\ell_{k}) \,
   \rho^{\abs{h_{l}}}(\lambda_{l},\lambda_{l+1})
   +
   \frac{1}{\lambda_{k}}
   \sum_{\kappab\in\mathbb{N}^{T}}
   \begin{aligned}[t]
   \Biggl(
   \biggl[
   \sum_{t'=\ell_{k-1}+1}^{\ell_k} \kappa_{t'}
   \biggr]
   & \Biggl[
     \prod_{i=1}^{K+1}
     \prod_{t=\ell_{i-1}+\delta_{i-1,l}\max (h_l,0)+1}
          ^{\ell_{i}-\delta_{i,l}\max (-h_l,0)}
     \frac{\lambda_i^{\kappa_t}}{\kappa_t!}
     \exp (-\lambda_i)
     \Biggr]
     \\
   & \times
     \prod_{t=\ell_{l}-\max (-h_{l},0)+1}^{\ell_{l}+\max (h_{l},0)}
     \frac{\sqrt{(\lambda_{l}\lambda_{l+1})^{\kappa_t}}}{\kappa_t!}
     \exp \biggl\{
          -\frac{\lambda_l+\lambda_{l+1}}{2}
          \biggr\}
     \Biggr) .
   \end{aligned}
   \label{eqn/P12kl.1/aux1} 
\end{aligned}
\end{multline}
The derivation of the second term in the right hand side of
(\ref{eqn/P12kl.1/aux1}) again depends upon the cases:
\begin{description}
\item[\textit{1)}] $l\neq k$ and $l\neq k-1$, case referred to as ``ULT'' (for
``upper and lower triangles'');

\item[\textit{2)}] $l = k$, case referred to as ``D1'' (for ``1st
diagonal'');

\item[\textit{3)}] $l = k-1$, case referred to as ``D2'' (for ``2nd
diagonal'').
\end{description}
Details for each case are given in the following sections.

\subsubsection{Case ULT ($l\neq k$ and $l\neq k-1$)}

We develop the second term in the right hand side of (\ref{eqn/P12kl.1/aux1}):
\begin{multline}
\sum_{\kappab\in\mathbb{N}^{T}}
\Biggl(
\biggl[
\sum_{t'=\ell_{k-1}+1}^{\ell_k} \kappa_{t'}
\biggr]
\Biggl[
\prod_{i=1}^{K+1}
\prod_{t=\ell_{i-1}+\delta_{i-1,l}\max (h_l,0)+1}
     ^{\ell_{i}-\delta_{i,l}\max (-h_l,0)}
\frac{\lambda_i^{\kappa_t}}{\kappa_t!}
\exp (-\lambda_i)
\Biggr]
\prod_{t=\ell_{l}-\max (-h_{l},0)+1}^{\ell_{l}+\max (h_{l},0)}
\frac{\sqrt{(\lambda_{l}\lambda_{l+1})^{\kappa_t}}}{\kappa_t!}
\exp \biggl\{
     -\frac{\lambda_l+\lambda_{l+1}}{2}
     \biggr\}
\Biggr)
\\
\begin{aligned}[b]
&= \begin{aligned}[t]
   \sum_{\kappab\in\mathbb{N}^{T}}
   \Biggl(
   & \Biggl[
     \prod_{\substack{i=1 \\
                      i\neq k}}
          ^{K+1}
     \prod_{t=\ell_{i-1}+\delta_{i-1,l}\max (h_l,0)+1}
          ^{\ell_{i}-\delta_{i,l}\max (-h_l,0)}
     \frac{\lambda_i^{\kappa_t}}{\kappa_t!}
     \exp (-\lambda_i)
     \Biggr]
     \sum_{t'=\ell_{k-1}+1}^{\ell_k}
     \Biggl[
     \biggl(
     \prod_{\substack{t=\ell_{k-1}+1 \\
                      t\neq t'}}
            ^{\ell_{k}}
     \frac{\lambda_k^{\kappa_t}}{\kappa_t!}
     \exp (-\lambda_k)
     \biggr)\,
     \frac{\lambda_k^{\kappa_{t'}}}{(\kappa_{t'}-1)!}
     \exp (-\lambda_k)
     \Biggr]
     \\
   & \times
     \prod_{t=\ell_{l}-\max (-h_{l},0)+1}^{\ell_{l}+\max (h_{l},0)}
     \frac{\sqrt{(\lambda_{l}\lambda_{l+1})^{\kappa_t}}}{\kappa_t!}
     \exp \biggl\{
          -\frac{\lambda_l+\lambda_{l+1}}{2}
          \biggr\}
     \Biggr)
   \end{aligned}
   \\
&= \Biggl(
   \sum_{t'=\ell_{k-1}+1}^{\ell_k}
   \sum_{\kappa_{t'}=1}^{+\infty}
   \frac{\lambda_k^{\kappa_{t'}}}{(\kappa_{t'}-1)!}
   \exp (-\lambda_k)
   \Biggr)
   \Biggl(
   \sum_{\kappa=0}^{+\infty}
   \frac{\sqrt{(\lambda_{l}\lambda_{l+1})^{\kappa_t}}}{\kappa_t!}
   \exp \biggl\{
        -\frac{\lambda_l+\lambda_{l+1}}{2}
        \biggr\}
   \Biggr)^{\abs{h_l}}
   \\
&= \lambda_k\,
   (\ell_k - \ell_{k-1})\,
   \rho^{\abs{h_l}} (\lambda_{l},\lambda_{l+1})
   \label{eqn/P12kl.1/aux1.2_ULT}
   \raisetag{11.25pt}
\end{aligned}
\end{multline}
Then, plugging (\ref{eqn/P12kl.1/aux1.2_ULT}) into (\ref{eqn/P12kl.1/aux1}),
we simply find
\begin{equation}
\sum_{\kappab\in\mathbb{N}^{T}}
\Biggl(
\biggl[
\ell_{k-1}-\ell_{k}
+ \frac{1}{\lambda_k}
  \sum_{t'=\ell_{k-1}+1}^{\ell_k} \kappa_{t'}
\biggr]
\sqrt{f (\xb =\kappab |\lambdab ,\tb =\ellb +\hb_{l}) 
      f (\xb =\kappab |\lambdab ,\tb =\ellb         )}
\Biggr)
= 0
\label{eqn/P12kl.1/aux1_ULT}
\end{equation}
for $k\neq l$, $k\neq l+1$.

\subsubsection{Case D1 ($l = k$)}

The writing of the second term in the right hand side of
(\ref{eqn/P12kl.1/aux1}) depends upon whether $h_k \gtrless 0$.

\noindent $\bullet$ Let us first assume that $h_k >0$. In this case, the
second term in the right hand side of (\ref{eqn/P12kl.1/aux1}) becomes
\begin{multline}
\sum_{\kappab\in\mathbb{N}^{T}}
\Biggl(
\biggl[
\sum_{t'=\ell_{k-1}+1}^{\ell_k} \kappa_{t'}
\biggr]
\Biggl[
\prod_{i=1}^{K+1}
\prod_{t=\ell_{i-1}+\delta_{i-1,k} h_k +1}
     ^{\ell_{i}}
\frac{\lambda_i^{\kappa_t}}{\kappa_t!}
\exp (-\lambda_i)
\Biggr]
\prod_{t=\ell_{k}+1}^{\ell_{k}+h_{k}}
\frac{\sqrt{(\lambda_{k}\lambda_{k+1})^{\kappa_t}}}{\kappa_t!}
\exp \biggl\{
     -\frac{\lambda_k+\lambda_{k+1}}{2}
     \biggr\}
\Biggr)
\\
\begin{aligned}[b]
&= \begin{aligned}[t]
   \sum_{\kappab\in\mathbb{N}^{T}}
   \Biggl(
   & \Biggl[
     \prod_{\substack{i=1 \\
                      i\neq k, k+1}}
          ^{K+1}
     \prod_{t=\ell_{i-1}+1}
         ^{\ell_{i}}
     \frac{\lambda_i^{\kappa_t}}{\kappa_t!}
     \exp (-\lambda_i)
     \Biggr]
     \sum_{t'=\ell_{k-1}+1}^{\ell_k}
     \Biggl[
     \biggl(
     \prod_{\substack{t=\ell_{k-1}+1 \\
                      t\neq t'         }}
          ^{\ell_{k}}
     \frac{\lambda_k^{\kappa_t}}{\kappa_t!}
     \exp (-\lambda_k)
     \biggr)
     \frac{\lambda_k^{\kappa_{t'}}}{(\kappa_{t'}-1)!}
     \exp (-\lambda_k)
     \Biggr]
     \\
   & \times
     \prod_{t=\ell_{k}+1}^{\ell_{k}+h_{k}}
     \frac{\sqrt{(\lambda_{k}\lambda_{k+1})^{\kappa_t}}}{\kappa_t!}
     \exp \biggl\{
          -\frac{\lambda_k+\lambda_{k+1}}{2}
          \biggr\}
     \prod_{t=\ell_{k}+h_{k}+1}^{\ell_{k+1}}
     \frac{\lambda_{k+1}^{\kappa_t}}{\kappa_t!}
     \exp (-\lambda_{k+1})
     \Biggr)
   \end{aligned}
   \\
&= \lambda_k\,
   (\ell_k - \ell_{k-1})\,
   \rho^{\abs{h_k}} (\lambda_{k},\lambda_{k+1})
   \label{eqn/P12kl.1/aux1.2_D1_hk>0}
\end{aligned}
\end{multline}
in the same way as in the case ULT. We then have, again, for $h_{k}>0$
\begin{equation}
\sum_{\kappab\in\mathbb{N}^{T}}
\Biggl(
\biggl[
\ell_{k-1}-\ell_{k}
+ \frac{1}{\lambda_k}
  \sum_{t'=\ell_{k-1}+1}^{\ell_k} \kappa_{t'}
\biggr]
\sqrt{f (\xb =\kappab |\lambdab ,\tb =\ellb +\hb_{k}) 
      f (\xb =\kappab |\lambdab ,\tb =\ellb         )}
\Biggr)
= 0.
\label{eqn/P12kl.1/aux1_D1.hk>0}
\end{equation}

\noindent $\bullet$ In the converse case $h_k <0$,
(\ref{eqn/P12kl.1/aux1.2_D1_hk>0}) becomes
\begin{multline}
\sum_{\kappab\in\mathbb{N}^{T}}
\Biggl(
\Biggl[
\sum_{t'=\ell_{k-1}+1}^{\ell_k} \kappa_{t'}
\Biggr]
\Biggl[
\prod_{i=1}^{K+1}
\prod_{t=\ell_{i-1}+1}
     ^{\ell_{i}+\delta_{i,k}h_k}
\frac{\lambda_i^{\kappa_t}}{\kappa_t!}
\exp (-\lambda_i)
\Biggr]
\prod_{t=\ell_{k}+h_{k}+1}^{\ell_{k}}
\frac{\sqrt{(\lambda_{k}\lambda_{k+1})^{\kappa_t}}}{\kappa_t!}
\exp \Bigl\{
     -\frac{\lambda_{k}+\lambda_{k+1}}{2}
     \Bigr\}
\Biggr)
\\
\begin{aligned}[b]
&= \begin{aligned}[t]
   \sum_{\kappab\in\mathbb{N}^{T}}
   \Biggl(
   & \Biggl[
     \sum_{t'=\ell_{k-1}+1}^{\ell_k +h_{k}}
     \kappa_{t'}
     +
     \sum_{t'=\ell_{k}+h_{k}+1}^{\ell_k}
     \kappa_{t'}
     \Biggr]
     \Biggl[
     \prod_{\substack{i=1 \\
                      i\neq k}}
          ^{K+1}
     \prod_{t=\ell_{i-1}+1}^{\ell_{i}}
     \frac{\lambda_i^{\kappa_t}}{\kappa_t!}
     \exp (-\lambda_i)
     \Biggr]
     \\
   & \times
     \prod_{t=\ell_{k-1}+1}^{\ell_{k}+h_{k}}
     \frac{\lambda_{k}^{\kappa_t}}{\kappa_t!}
     \exp (-\lambda_{k})
     \prod_{t=\ell_{k}+h_{k}+1}^{\ell_{k}}
     \frac{\sqrt{(\lambda_{k}\lambda_{k+1})^{\kappa_t}}}{\kappa_t!}
     \exp \Bigl\{
          -\frac{\lambda_{k}+\lambda_{k+1}}{2}
          \Bigr\}
     \Biggr)
   \end{aligned}
   \\
&= \sum_{\kappab\in\mathbb{N}^{T}}
   \Biggl(
   \Biggl[
   \prod_{\substack{i=1 \\
                    i\neq k}}
        ^{K+1}
   \tprod_{t=\ell_{i-1}+1}^{\ell_{i}}
   \frac{\lambda_i^{\kappa_t}}{\kappa_t!}
   \exp (-\lambda_i)
   \Biggr]
   \begin{aligned}[t]
   \Biggl\{
   & \Biggl[
     \begin{aligned}[t]
     & \tprod_{t=\ell_{k}+h_{k}+1}^{\ell_{k}}
       \frac{\sqrt{(\lambda_{k}\lambda_{k+1})^{\kappa_t}}}{\kappa_t!}
       \exp \Bigl\{
            -\frac{\lambda_{k}+\lambda_{k+1}}{2}
            \Bigr\}
       \\
     & \times
       \tsum_{t'=\ell_{k-1}+1}^{\ell_k +h_{k}}
       \biggl[
       \biggl(
       \tprod_{\substack{t=\ell_{k-1}+1 \\
                         t\neq t'}}
             ^{\ell_{k}+h_{k}}
       \frac{\lambda_{k}^{\kappa_t}}{\kappa_t!}
       \exp (-\lambda_{k})
       \biggr)
       \frac{\lambda_{k}^{\kappa_{t'}}}{(\kappa_{t'}-1)!}
       \exp (-\lambda_{k})
       \biggr]
       \Biggr]
     \end{aligned}
     \\
   & + \Biggl[
     \begin{aligned}[t]
     & \tprod_{t=\ell_{k-1}+1}
             ^{\ell_{k}+h_{k}}
       \frac{\lambda_{k}^{\kappa_t}}{\kappa_t!}
       \exp (-\lambda_{k})
       \\
     & \times
       \tsum_{t'=\ell_{k}+h_{k}+1}^{\ell_k}
       \begin{aligned}[t]
       \biggl[
       & \biggl(
         \tprod_{\substack{t=\ell_{k}+h_{k}+1 \\
                           t\neq t'}}
               ^{\ell_{k}}
         \frac{\sqrt{(\lambda_{k}\lambda_{k+1})^{\kappa_t}}}{\kappa_t!}
         \exp \Bigl\{
              -\frac{\lambda_{k}+\lambda_{k+1}}{2}
              \Bigr\}
         \biggr)
         \\
       & \times
         \frac{\sqrt{(\lambda_{k}\lambda_{k+1})^{\kappa_{t'}}}}
              {(\kappa_{t'}-1)!}
         \exp \Bigl\{
              -\frac{\lambda_{k}+\lambda_{k+1}}{2}
              \Bigr\}
         \biggr]
         \Biggr]
         \Biggr\}
         \Biggl)
       \end{aligned}
     \end{aligned}
   \end{aligned}
   \\
&= \Biggl(
   \sum_{\kappa =0}^{+\infty}
   \frac{\sqrt{(\lambda_{k}\lambda_{k+1})^{\kappa_t}}}{\kappa_t!}
   \exp \Bigl\{
        -\frac{\lambda_{k}+\lambda_{k+1}}{2}
        \Bigr\}
   \Biggr)^{\!\!\! -h_{k}}
   \,
   \Biggl[
   \sum_{t'=\ell_{k-1}+1}^{\ell_{k}+h_{k}}
   \biggl(
   \sum_{\kappa_{t'}=1}^{+\infty}
   \frac{\lambda_{k}^{\kappa_{t'}}}{(\kappa_{t'}-1)!}
   \exp (-\lambda_{k})
   \biggr)
   \Biggr]
   \\
&  \quad \: +
   \sum_{t'=\ell_{k}+h_{k}+1}^{\ell_{k}}
   \Biggl[
   \biggl(
   \sum_{\kappa =0}^{+\infty}
   \frac{\sqrt{(\lambda_{k}\lambda_{k+1})^{\kappa}}}{\kappa!}
   \exp \Bigl\{
        -\frac{\lambda_{k}+\lambda_{k+1}}{2}
        \Bigr\}
   \biggr)^{\!\!\! -h_{k}-1}
   \,
   \biggl(
   \sum_{\kappa_{t'}=1}^{+\infty}
   \frac{\sqrt{(\lambda_{k}\lambda_{k+1})^{\kappa_{t'}}}}
        {(\kappa_{t'}-1)!}
   \exp \Bigl\{
        -\frac{\lambda_{k}+\lambda_{k+1}}{2}
        \Bigr\}
   \biggr)
   \Biggr]
   \\
&= \rho^{-h_{k}} (\lambda_{k},\lambda_{k+1})
   \,
   \lambda_{k}
   (\ell_{k}+h_{k}-\ell_{k-1})
   +
   \sum_{t'=\ell_{k}+h_{k}+1}^{\ell_{k}}
   \Bigl[
   \rho^{-h_{k}-1} (\lambda_{k},\lambda_{k+1})
   \,
   \sqrt{\lambda_{k}\lambda_{k+1}}
   \,
   \rho(\lambda_{k},\lambda_{k+1})
   \Bigr]
   \\
&= \rho^{-h_{k}} (\lambda_{k},\lambda_{k+1})
   \Bigl[
   \lambda_{k}
   (\ell_{k}-\ell_{k-1})
   +
   h_{k}
   \bigl(
   \lambda_{k} - \sqrt{\lambda_{k}\lambda_{k+1}}
   \bigr)
   \Bigr]
   \\
&= \lambda_{k} \,
   \rho^{-h_{k}} (\lambda_{k},\lambda_{k+1})
   \biggl[
   (\ell_{k}-\ell_{k-1})
   +
   h_{k}
   \biggl(
   1 - \sqrt{\frac{\lambda_{k+1}}{\lambda_{k}}}
   \biggr)
   \biggr].
   \label{eqn/P12kl.1/aux1.2_D1_hk<0}
\end{aligned}
\end{multline}
Then, by plugging (\ref{eqn/P12kl.1/aux1.2_D1_hk<0}) into
(\ref{eqn/P12kl.1/aux1}), we obtain
\begin{multline}
\sum_{\kappab\in\mathbb{N}^{T}}
\Biggl(
\biggl[
\ell_{k-1}-\ell_{k}
+ \frac{1}{\lambda_k}
  \sum_{t'=\ell_{k-1}+1}^{\ell_k} \kappa_{t'}
\biggr]
\sqrt{f (\xb =\kappab |\lambdab ,\tb =\ellb +\hb_{k}) 
      f (\xb =\kappab |\lambdab ,\tb =\ellb         )}
\Biggr)
\\
= (-h_{k})
  \biggl(
  \sqrt{\frac{\lambda_{k+1}}{\lambda_{k}}} - 1
  \biggr)
  \rho^{-h_{k}} (\lambda_{k},\lambda_{k+1}).
  \label{eqn/P12kl.1/aux1_D1_hk<0}
\end{multline}

\subsubsection{Case D2 ($l = k-1$)}

In this case as well, we have to distinguish between the two cases $h_{k-1}
\gtrless 0$.

\noindent $\bullet$ Let us first assume that $h_{k-1} < 0$. This case can be
handled by using exactly the same methodology as in the case D1 with $h_{k}>0$
(see (\ref{eqn/P12kl.1/aux1.2_D1_hk>0})). We then find, for $h_{k-1}<0$
\begin{equation}
\sum_{\kappab\in\mathbb{N}^{T}}
\Biggl(
\biggl[
\ell_{k-1}-\ell_{k}
+ \frac{1}{\lambda_k}
  \sum_{t'=\ell_{k-1}+1}^{\ell_k} \kappa_{t'}
\biggr]
\sqrt{f (\xb =\kappab |\lambdab ,\tb =\ellb +\hb_{k-1}) 
      f (\xb =\kappab |\lambdab ,\tb =\ellb         )}
\Biggr)
=0.
\label{eqn/P12kl.1/aux1_D2_hk<0}
\end{equation}

\noindent $\bullet$ The converse case $h_{k-1}>0$ can be handled also by using
the exact same methodology as that used in the case D1 with $h_{k}<0$
(see (\ref{eqn/P12kl.1/aux1.2_D1_hk<0})), so that we obtain
\begin{multline}
\sum_{\kappab\in\mathbb{N}^{T}}
\Biggl(
\biggl[
\ell_{k-1}-\ell_{k}
+ \frac{1}{\lambda_k}
  \sum_{t'=\ell_{k-1}+1}^{\ell_k} \kappa_{t'}
\biggr]
\sqrt{f (\xb =\kappab |\lambdab ,\tb =\ellb +\hb_{k-1}) 
      f (\xb =\kappab |\lambdab ,\tb =\ellb         )}
\Biggr)
\\
= h_{k-1}
  \biggl(
  \sqrt{\frac{\lambda_{k-1}}{\lambda_{k}}} - 1
  \biggr)
  \rho^{h_{k-1}} (\lambda_{k-1},\lambda_{k}).
  \label{eqn/P12kl.1/aux1_D2_hk>0}
\end{multline}

Finally, we can merge both cases D1 and D2 (i.e., whether $l=k$ or $l=k-1$) by
writing
\begin{multline}
\sum_{\kappab\in\mathbb{N}^{T}}
\Biggl(
\biggl[
\ell_{k-1}-\ell_{k}
+ \frac{1}{\lambda_k}
  \sum_{t'=\ell_{k-1}+1}^{\ell_k} \kappa_{t'}
\biggr]
\sqrt{f (\xb =\kappab |\lambdab ,\tb =\ellb +\hb_{l}) 
      f (\xb =\kappab |\lambdab ,\tb =\ellb         )}
\Biggr)
\\
= \max(\pm h_{l},0)
  \biggl(
  \Bigl(
  \frac{\lambda_{l}}{\lambda_{l+1}}
  \Bigr)^{\pm1/2} - 1
  \biggr)
  \rho^{\abs{h_{l}}} (\lambda_{l},\lambda_{l+1})
  \label{eqn/P12kl.1/aux1_D1_D2}
\end{multline}
where the ``$\pm$'' signs both are ``$+$'' signs if $l=k-1$, and they are 
``$-$'' signs if $l=k$.

\bigskip

\subsubsection*{Completion of the derivation (\ref{eqn/P12kl.1_expanded}) and final expression of $\boldsymbol{P}_{12}$}

From the foregoing, we can now complete the derivation of (\ref{eqn/P12kl.1_expanded}). Let us do it separately, according to the cases ULT, and D1-2.

\paragraph*{i) Case ULT ($l\neq k-1$ and $l\neq k$)}

Carrying on the derivation started in (\ref{eqn/P12kl.1_expanded}), and using
(\ref{eqn/P12kl.1/aux1_ULT}), we have
\begin{multline}
\mathbb{E}_{\xb,\lambdab,\tb}
\left\{
\frac{\partial \ln f(\xb =\kappab ,\lambdab ,\tb =\ellb ) }
     {\partial \lambda _{k}}
\sqrt{
\frac{f (\xb =\kappab ,\lambdab ,\tb =\ellb +\hb_{l}) }
     {f (\xb =\kappab ,\lambdab ,\tb =\ellb         ) }
     }
\right\}
\\
\begin{aligned}[b]
&= \int_{\mathbb{R}_{+}^{K+1}}
   \sum_{\ellb \in \mathbb{Z}^{K}}
   \left[
   \sqrt{f (\lambdab,\tb =\ellb +\hb_{l})
         f (\lambdab,\tb =\ellb         )}
   \left(
   \frac{\alpha_k -1}{\lambda_k} - \beta
   \right)
   \rho^{\abs{h_{l}}} (\lambda_{l},\lambda_{l+1})
   \right]
   \diff\lambdab
   \\
&= \sum_{\ellb\in\mathbb{Z}^{K}}
   \int_{\mathbb{R}_{+}^{K+1}}
   \Biggl[
   \biggl(
   \tprod_{i=1}^{K+1}
   \frac{\beta^{\alpha _{i}}}
        {\Gamma \left( \alpha _{i}\right) }
   \lambda_{i}^{\alpha _{i}-1}\exp \left( -\beta \lambda _{i}\right)
   \biggr)
   \frac{1}{\tau ^{K}}
   \indicf{\mathcal{J}_{h_{l},0}}{\ellb}
   \\
&  \phantom{{} = \sum_{\ellb\in\mathbb{Z}^{K}}
                 \int_{\mathbb{R}_{+}^{K+1}}
                 \Biggl[{}
           }
   \times
   \left(
   \frac{\alpha_k -1}{\lambda_k} - \beta
   \right)
   \exp \biggl\{ 
        -\abs{h_{l}}
        \frac{(\sqrt{\lambda_{l+1}}-\sqrt{\lambda_{l}})^2}{2}
        \biggr\}
   \Biggr]
   \diff \lambdab
   \\
&= \biggl(
   \frac{1}{\tau ^{K}}
   \sum_{\boldsymbol{\ell }\in \mathbb{Z}^{K}}
   \indicf{\mathcal{J}_{h_{l},0}}{\ellb}
   \biggr)
   \Biggl[
   \tprod_{\substack{i=1 \\
                    i\neq k,l,l+1}}
         ^{K+1}
    \int_{\mathbb{R}_{+}^{K-1}}
    \frac{\beta^{\alpha_{i}}}
         {\Gamma \left( \alpha_{i}\right) }
    \lambda_{i}^{\alpha_{i}-1}\exp \left( -\beta \lambda_{i}\right)
    \diff\lambda_{i}
    \Biggr]
    \Phi (h_{l})
    \\
& \quad \: \times
  \frac{\beta^{\alpha_{k}}}
       {\Gamma \left( \alpha_{k}\right) }
  \underset{= \: 0}{
  \underbrace{
  \int_{\mathbb{R}_{+}}
  \left(
  \frac{\alpha_k -1}{\lambda_k} - \beta
  \right)
  \lambda_{k}^{\alpha_{k}-1}\exp \left( -\beta \lambda_{k}\right)
  \diff\lambda_{k}
  }
  }
  \\
&= 0
\label{eqn/P12kl.1_ULT}
\end{aligned}
\end{multline}
by using the same arguments as in (\ref{eqn/P11kl.2_kneql_final}), for
example. Of course, in (\ref{eqn/P12kl.1_ULT}), when replacing $h_{l}$ with
$-h_{l}$, we obtain zero as well. Then, by plugging (\ref{eqn/P12kl.1_ULT})
into (\ref{eqn/P12kl}) twice (once with $+\hb_{l}$ and once with
$-\hb_{l}$), we obtain, for $l\neq k-1$ and $l\neq k$
\begin{equation}
[\boldsymbol{P}_{12}]_{k,l} = 0.
\label{eqn/P12kl_ULT}
\end{equation}

\paragraph*{ii) Cases D1 and D2 ($l = k-1$ or $l=k$)}

By plugging (\ref{eqn/P12kl.1/aux1_D1_D2}) into (\ref{eqn/P12kl.1_expanded}),
we obtain
\begin{multline}
\mathbb{E}_{\xb,\lambdab,\tb}
\left\{
\frac{\partial \ln f(\xb =\kappab ,\lambdab ,\tb =\ellb ) }
     {\partial \lambda _{k}}
\sqrt{
\frac{f (\xb =\kappab ,\lambdab ,\tb =\ellb +\hb_{l}) }
     {f (\xb =\kappab ,\lambdab ,\tb =\ellb         ) }
     }
\right\}
\\
\begin{aligned}[b]
&= \int_{\mathbb{R}_{+}^{K+1}}
   \sum_{\ellb \in \mathbb{Z}^{K}}
   \left[
   \sqrt{f (\lambdab,\tb =\ellb +\hb_{l})
         f (\lambdab,\tb =\ellb         )}
   \Biggl(
   \frac{\alpha_k -1}{\lambda_k} - \beta
   + \max(\pm h_{l},0)
     \biggl(
     \Bigl(
     \frac{\lambda_{l}}{\lambda_{l+1}}
     \Bigr)^{\pm1/2} - 1
     \biggr)
   \Biggr)
   \rho^{\abs{h_{l}}} (\lambda_{l},\lambda_{l+1})
   \right]
   \diff\lambdab
   \\
&= \biggl(
   \frac{1}{\tau ^{K}}
   \sum_{\boldsymbol{\ell }\in \mathbb{Z}^{K}}
   \indicf{\mathcal{J}_{h_{l},0}}{\ellb}
   \biggr)
   \Biggl[
   \tprod_{\substack{i=1 \\
                    i\neq l,l+1}}
         ^{K+1}
   \int_{\mathbb{R}_{+}^{K-1}}
   \frac{\beta^{\alpha_{i}}}
        {\Gamma \left( \alpha_{i}\right) }
   \lambda_{i}^{\alpha_{i}-1}\exp \left( -\beta \lambda_{i}\right)
   \diff\lambda_{i}
   \Biggr]
   \cdot
   \Biggl[
   \frac{\beta^{\alpha_{l}+\alpha_{l+1}}}
        {\Gamma \left( \alpha_{l}\right)
         \Gamma \left( \alpha_{l+1}\right)}
   \\
&  \quad \: \times
   \int_{\mathbb{R}_{+}^{2}}
   \Biggl(
   \frac{\alpha_k -1}{\lambda_k} - \beta
   + \max(\pm h_{l},0)
     \biggl(
     \Bigl(
     \frac{\lambda_{l}}{\lambda_{l+1}}
     \Bigr)^{\pm 1/2} - 1
     \biggr)
   \Biggr)
   \exp \biggl\{
        -\beta (\lambda_{l} +\lambda_{l+1})
        -\abs{h_{l}}
         \frac{\bigl(\sqrt{\lambda_{l+1}}-\sqrt{\lambda_{l}}\bigr)^2}{2}
        \biggr\}
   \diff\lambdab_{l:l+1}
   \Biggr]
   \\
&= u(\tau,h_{l}) \,
   \frac{\beta^{\alpha_{l}+\alpha_{l+1}}}
        {\Gamma \left( \alpha_{l}\right)
         \Gamma \left( \alpha_{l+1}\right)}
   \int_{\mathbb{R}_{+}^{2}}
   \Biggl(
   \frac{\alpha_k -1}{\lambda_k} - \beta
   + \max(\pm h_{l},0)
     \biggl(
     \Bigl(
     \frac{\lambda_{l}}{\lambda_{l+1}}
     \Bigr)^{\pm1/2} - 1
     \biggr)
   \Biggr)
   \varphi_{h_{l}} (\lambdab_{l:l+1})
   \diff\lambdab_{l:l+1}.
   \label{eqn/P12kl.1_D1_D2}
   \raisetag{20.5pt}
\end{aligned}
\end{multline}
When replacing $h_{l}$ with $-h_{l}$ in (\ref{eqn/P12kl.1_D1_D2}), we
obtain, accordingly
\begin{multline}
\mathbb{E}_{\xb,\lambdab,\tb}
\left\{
\frac{\partial \ln f(\xb =\kappab ,\lambdab ,\tb =\ellb ) }
     {\partial \lambda _{k}}
\sqrt{
\frac{f (\xb =\kappab ,\lambdab ,\tb =\ellb -\hb_{l}) }
     {f (\xb =\kappab ,\lambdab ,\tb =\ellb         ) }
     }
\right\}
\\
= u(\tau,h_{l}) \,
  \frac{\beta^{\alpha_{l}+\alpha_{l+1}}}
        {\Gamma \left( \alpha_{l}\right)
         \Gamma \left( \alpha_{l+1}\right)}
  \int_{\mathbb{R}_{+}^{2}}
  \Biggl(
  \frac{\alpha_k -1}{\lambda_k} - \beta
  + \max(\mp h_{l},0)
    \biggl(
    \Bigl(
    \frac{\lambda_{l}}{\lambda_{l+1}}
    \Bigr)^{\pm1/2} - 1
    \biggr)
  \Biggr)
  \varphi_{h_{l}} (\lambdab_{l:l+1})
  \diff\lambdab_{l:l+1}.
   \label{eqn/P12kl.2_D1_D2}
\end{multline}
Thus, we finally obtain, by subtracting both right hand sides of
(\ref{eqn/P12kl.1_D1_D2}) and (\ref{eqn/P12kl.2_D1_D2}), we find, for
$l=k-1$ or $l=k$
\begin{align}
[\boldsymbol{P}_{12}]_{k,l}
&= u(\tau,h_{l}) \,
   \frac{\beta^{\alpha_{l}+\alpha_{l+1}}}
        {\Gamma \left( \alpha_{l}\right)
         \Gamma \left( \alpha_{l+1}\right)}
   \int_{\mathbb{R}_{+}^{2}}
   \Biggl(
   \max(\pm h_{l},0)
   - \max(\mp h_{l},0)
     \biggl(
     \Bigl(
     \frac{\lambda_{l}}{\lambda_{l+1}}
     \Bigr)^{\pm 1/2} - 1
     \biggr)
   \Biggr)
   \varphi_{h_{l}} (\lambdab_{l:l+1})
   \diff\lambdab_{l:l+1}
   \notag \\
&= \pm h_{l} \,
   u(\tau,h_{l}) \,
   \frac{\beta^{\alpha_{l}+\alpha_{l+1}}}
        {\Gamma \left( \alpha_{l}\right) \Gamma \left( \alpha_{l+1}\right)}
   \int_{\mathbb{R}_{+}^{2}}
   \biggl(
   \Bigl(
   \frac{\lambda_{l}}{\lambda_{l+1}}
   \Bigr)^{\pm 1/2} - 1
   \biggr)
   \varphi_{h_{l}} (\lambdab_{l:l+1})
   \diff\lambdab_{l:l+1}
   \label{eqn/P12kl_D1_D2}
\end{align}
where both ``$\pm$'' signs are ``$+$'' signs if $l=k-1$, and they are
``$-$'' signs if $l=k$. More explicitly, for $k=1,\ldots ,K$,
\begin{equation}
[\boldsymbol{P}_{12}]_{k,k}
= -h_{k} \,
   u(\tau,h_{k}) \,
   \frac{\beta^{\alpha_{k}+\alpha_{k+1}}}
        {\Gamma \left( \alpha_{k}\right) \Gamma \left( \alpha_{k+1}\right)}
   \int_{\mathbb{R}_{+}^{2}}
   \biggl(
   \sqrt{\frac{\lambda_{k+1}}{\lambda_{k}}}
   - 1
   \biggr)
   \varphi_{h_{k}} (\lambdab_{k:k+1})
   \diff\lambdab_{k:k+1}
   \label{eqn/P12kl_D1}
\end{equation}
and
\begin{equation}
[\boldsymbol{P}_{12}]_{k+1,k}
= h_{k} \,
  u(\tau,h_{k}) \,
  \frac{\beta^{\alpha_{k}+\alpha_{k+1}}}
       {\Gamma \left( \alpha_{k}\right) \Gamma \left( \alpha_{k+1}\right)}
  \int_{\mathbb{R}_{+}^{2}}
  \biggl(
  \sqrt{\frac{\lambda_{k}}{\lambda_{k+1}}}
  - 1
  \biggr)
  \varphi_{h_{k}} (\lambdab_{k:k+1})
  \diff\lambdab_{k:k+1}
  \label{eqn/P12kl_D2}
\end{equation}

Finally, considering (\ref{eqn/P12kl_ULT}) and (\ref{eqn/P12kl_D1_D2}),
we obtain the structure of the matrix $\boldsymbol{P}_{12}$ as given by
(\ref{eqn/P12}) and its elements as given by (\ref{eqn/def_Akl}).

We have thus completed the derivation of the bound given in Section
\ref{sect/bound_results}.

\pagebreak

\bibliographystyle{IEEEtran}
\bibliography{Alex,ICASSP17}

\end{document}